\documentclass{article}
\usepackage{epsfig,epsf}
\newcounter{fixy}
 \newenvironment{fixy}[1]{\setcounter{figure}{#1}}
{\addtocounter{fixy}{1}}

\begin{document} 
\begin{center}
{\Large\bf STATIC AXIALLY SYMMETRIC SOLUTIONS OF EINSTEIN-YANG-MILLS EQUATIONS
 WITH A NEGATIVE COSMOLOGICAL CONSTANT:  BLACK HOLE SOLUTIONS }
\vspace{0.5cm}

{\bf Eugen Radu}\footnote{E-mail:
{\tt radu@heisenberg1.thphys.may.ie}}
{\bf and Elizabeth Winstanley}\footnote{E-mail:
{\tt E.Winstanley@shef.ac.uk}}

\vspace*{0.2cm}
{\it $^1$ Department of Mathematical Physics, National University of Ireland,
Maynooth, Ireland}

{\it $^2$ Department of Applied Mathematics, The University of Sheffield, 
Hicks Building,
\\
Hounsfield Road, Sheffield, S3 7RH, United Kingdom.}
\vspace{0.5cm}
\end{center}
\begin{abstract}
We investigate static axially symmetric black hole solutions
in a four-dimensional Einstein-Yang-Mills-SU(2) theory
with a negative cosmological constant $\Lambda$.
These solutions approach asymptotically the anti-de Sitter spacetime
and possess a regular event horizon.
A discussion of the main properties
of the solutions and the differences with respect to the asymptotically
flat case is presented.
The mass of these configurations is computed by using a counterterm method.
We note that the $\Lambda=-3$ configurations have an higher dimensional
interpretation in context of
$d=11$ supergravity.
The existence of axially symmetric monopole and dyon solutions
in a fixed Schwarzschild-anti-de Sitter background is also discussed.
An exact solution of the Einstein-Yang-Mills equations is presented in Appendix.
\end{abstract}
\section{INTRODUCTION}
There has been much interest in recent years 
in the study of black hole solutions
in gravitational theories with a negative cosmological constant $\Lambda$.
The initial interest in asymptotically anti-de Sitter (AAdS) solutions was
due to the result that (sufficiently large) Schwarzschild-AdS  (SAdS)
black holes are thermodynamically stable \cite{Hawking:1982dh}.
A renewed interest in the study of AAdS black
holes has appeared following the  AdS/conformal field theory conjecture, 
which proposes a correspondence  between physical effects
associated with gravitating fields propagating in 
AdS spacetime and those of a conformal
field theory (CFT) on the boundary of AdS spacetime 
\cite{Witten:1998qj, Maldacena:1997re}.
The AAdS black hole solutions would offer  the posibility  
of studying the nonperturbative structure of some conformal field theories.
For example, the  AdS$_5$ Hawking-Page phase transition \cite{Hawking:1982dh} is interpreted
as a thermal phase transition 
from a confining to a deconfining phase in the dual $D = 4$, $N = 4$ super Yang-Mills 
theory \cite{Witten:1998zw}, while the 
phase structure of Reissner- Nordstr\"om-AdS black holes resembles that of a 
van der Waals-Maxwell liquid-gas system \cite{Chamblin:1999tk}.
Also, when $\Lambda<0$, the so-called topological black holes, whose topology
of the event horizon is no longer the two-sphere $S^2$ may appear
(see \cite{Mann:1997iz} for reviews of the subject).

These results motivate at least partially attempts to
find new black hole solutions in AAdS spacetimes.
Particularly interesting  are  solutions violating the no hair conjecture.
This conjecture states that the only allowed 
characteristics  of a stationary black hole are those associated 
with the Gauss law, such as mass angular momentum and U(1) charge.
Black holes with hair may  be useful for probing not only quantum 
gravity, but also may be interesting tests of the AdS/CFT correspondence, 
particularly since we can find such objects which are classically stable. 
For AAdS black holes, it has been shown recently that conformally 
coupled scalar field can be painted as hair \cite{Winstanley:2002jt}.
The existence of long range Nielson-Olesen vortex as hair 
for asymptotically AdS black holes has been investigated 
in \cite{Dehghani:2001nz,Dehghani:2002qc} for SAdS black hole
and charged black strings.

AAdS black holes solutions with SU(2) nonabelian fields 
have been presented in
\cite{Winstanley:1998sn,Bjoraker:2000qd}.
The obtained results are strikingly different from
those valid in the asymptotically flat case 
(nontrivial solutions exist for all values of $\Lambda<0$).
For example, regular black hole solutions exist 
for continuous intervals of the parameter 
space, rather than discrete points
and there are configurations for which the gauge field has no zeros.
Regular and black hole dyon solutions without a
 Higgs field have also been found \cite{Bjoraker:2000qd}.
Insofar as the no-hair theorem is concerned, it has been shown that
there exist stable black hole solutions in 
SU(2) Einstein-Yang-Mills (EYM) theory that are AAdS 
\cite{Winstanley:1998sn,Bjoraker:2000qd}.
However, the same system in the presence of a Higgs scalar field
presents only configurations 
with very similar properties to the asymptotically-flat space counterparts
\cite{Lugo:1999fm, Lugo:1999ai,  vanderBij:2002sq}.

The literature on AAdS solutions with nonabelian fields is continuously growing,
including stability analyses \cite{Sarbach:2001mc,Breitenlohner:2003qj}, 
the study of configurations with a NUT charge \cite{Radu:2002hf},
higher dimensional counterparts \cite{Okuyama:2002mh} and 
axially symmetric generalizations \cite{Radu:2002rv,Radu:2001ij}.
Here we remark that the EYM black hole solutions with $\Lambda<0$ 
discussed in the literature
are spherically symmetric or correspond to topological
black holes \cite{VanderBij:2001ia} presenting the same amount of symmetry.
However, for a vanishing cosmological constant,
the EYM-SU(2) theory is known to possess also
static axially symmetric finite energy black hole solutions,
as exhaustively discussed in \cite{Kleihaus:1997ic,Kleihaus:1997ws}. 
The situation for a nonabelian field is very different from 
the Einstein-Maxwell theory,
where the static black hole solution is spherically symmetric.
Representing generalizations of the well known 
spherically symmetric asymptotically flat 
black hole solutions \cite{Volkov:sv}, 
these static axially symmetric solutions are characterized by 
two integers, for a given event
event horizon radius. 
These are the node number $k$ of the gauge field functions and 
the winding number $n$ with respect to the azimuthal angle $\varphi$ . 
The static spherically symmetric solutions have winding number $n = 1$. 
Similar to the case of regular configurations \cite{Kleihaus:1997mn}, 
winding numbers $n >1$ leads to axially symmetric black hole
solutions.
Outside their regular event horizon, the static $\Lambda=0$ axially symmetric black hole solutions 
possess non-trivial magnetic gauge field configurations, 
but they carry no global magnetic charge.
In a remarkable development, these asymptotically flat black hole solutions
have been generalized to include the effects of rotation, leading to nonabelian 
counterparts of the Kerr-Newmann solution \cite{Kleihaus:2002ee}.

The regular $n>1$ axially symmetric EYM configurations of Ref. \cite{Kleihaus:1997mn} have 
been generalized for $\Lambda<0$ in Ref. \cite{Radu:2001ij}.
Although  some common features are present,
the results in the AAdS case are rather different
from those valid in the $\Lambda=0$ limit, in particular presenting 
arbitrary values for the mass and
magnetic charge.
These distinctions arise from differences that
already exist in the spherically symmetric case. 

It is reasonable to suppose that the AAdS axially symmetric regular solutions discussed
in  Ref. \cite{Radu:2001ij}
can be generalized to include a black hole event horizon inside them.
The purpose of this paper is to present numerical arguments for the
existence of this type of configurations and to analyze the properties 
of the axially symmetric EYM-SU(2) black hole
solutions, in light of
the existing results for the asymptotically flat case, discussing
the points where the differences are relevant.
The numerical methods used here are similar to the methods succesfully
employed for $\Lambda<0$ axially symmetric regular solutions.

The mass of these solutions is computed by using a counterterm
method inspired by the AdS/CFT correspondence.
Although further research is clearly necessary, at least some of these solutions,
emerging as consistent reduction of $d=11$ supergravity 
on a seventh dimenensional sphere
\cite{Pope:1985bu},
may have a relevance in AdS/CFT context.
In this case the four-dimensional cosmological constant is completely fixed
by the gauge coupling constant and exact solutions for configurations 
preserving some amount of supersymmetry are 
likely to exist.

The  paper is structured as follows: in the next Section we explain the model 
and derive the basic equations, while in Section 3 we present a detailed computation
of the physical quantities of the solutions such as mass, magnetic charge, temperature and entropy
noting the relevance of a special class of solutions in $d=11$ supergravity.
In Section 4 solutions of YM equations in a 
fixed SAdS background are discussed.
Some features of the axially symmetric solutions possessing a net YM electric charge
in a fixed SAdS black hole background
are also presented in this Section.
The general properties of the axially symmetric gravitating solutions are
 presented in Section 5 where we show results obtained by numerical calculations.
We give our conclusions and remarks in the final section.
The Appendix contains a derivation of an exact solution of the 
EYM equations with negative cosmological constant with a 
planar symmetry.

\section{GENERAL FRAMEWORK AND EQUATIONS OF MOTION}
\subsection{Einstein-Yang-Mills action}
We will follow in this work most of the conventions and notations 
used by Kleihaus and Kunz in their papers on asymptotically flat  EYM black holes
\cite{Kleihaus:1997ws}, \cite{Volkov:sv}-\cite{Kleihaus:2002ee}.

The equations for a static, axially symmetric SU(2) gauge field 
coupled to Einstein gravity  with a cosmological term  
have been presented in  Ref. \cite{Radu:2001ij}.
The starting point is the EYM action
\begin{equation} 
\label{lag0}
I=\int d^{4}x\sqrt{-g}[\frac{1}{16\pi G}(\mathcal{R} - 2 \Lambda)
-\frac{1}{2}Tr(F_{\mu \nu }F^{\mu \nu })]
-\frac{1}{8\pi G}\int_{\partial\mathcal{M}} d^{3}x\sqrt{-h}K,
\end{equation}
where the field strength tensor is
\begin{equation}
F_{\mu \nu} = 
\partial_\mu A_\nu -\partial_\nu A_\mu + i e \left[A_\mu , A_\nu \right] 
\ , \label{fmn} \end{equation}
and the gauge field
$A_{\mu} = \frac{1}{2} \tau^a A_{\mu}^{(a)},$
where $e$ and $G$ are the Yang-Mills (YM) and the gravity coupling constants.
The last term in  (\ref{lag0}) is the Hawking-Gibbons surface term \cite{Gibbons:1976ue}, 
where $K$ is the trace 
of the extrinsic curvature for the boundary $\partial\mathcal{M}$ and $h$ is the induced 
metric of the boundary. Of course, this term does not affect the equations of motion 
but it is relevant for the discussion in Section 3 of the solutions' mass and boundary stress tensor.

Variation of the action (\ref{lag0})
 with respect to the metric $g^{\mu\nu}$ leads to the Einstein equations
\begin{equation}
\label{einstein-eqs}
R_{\mu\nu}-\frac{1}{2}g_{\mu\nu}R +\Lambda g_{\mu\nu}  = 8\pi G T_{\mu\nu},
\end{equation}
where the YM stress-energy tensor is 
\begin{eqnarray}
T_{\mu\nu} = 2{\rm Tr}
    ( F_{\mu\alpha} F_{\nu\beta} g^{\alpha\beta}
   -\frac{1}{4} g_{\mu\nu} F_{\alpha\beta} F^{\alpha\beta}). 
\end{eqnarray}
Variation with respect to the gauge field $A_\mu$ 
leads to the YM equations
\begin{equation}
\label{YM-eqs}
\nabla_{\mu}F^{\mu\nu}+ie[A_{\mu},F^{\mu\nu}]=0.
\end{equation}
\subsection{Static axially symmetric ansatz and gauge condition}
As for the globally regular static solutions, we consider a line element on the form
\begin{equation} 
\label{metric}
ds^2=g_{\mu \nu}dx^{\mu}dx^{\nu}=- fN dt^2 +  \frac{m}{f} ( \frac{d r^2}
{N}+ r^2 d \theta^2 )
           +  \frac{l}{f} r^2 \sin ^2 \theta d\varphi^2,
\end{equation}
where the metric functions
$f$, $m$ and $l$ are only functions of 
$r$ and $\theta$ and we note $N=1-\Lambda r^2/3$. Here $r$ is the radial coordinate 
(we are interested in the region $r_h\leq r < \infty$),
$t$ is a global time coordinate, 
$(\theta,\varphi)$ being the usual coordinates on the sphere,
with $0\leq \theta\leq\pi,0\leq \varphi<2 \pi$.
The metric ansatz (\ref{metric}) generalizes for a nonzero $\Lambda$
the isotropic axisymmetric line element considered 
by Kleihaus and Kunz in \cite{Kleihaus:1997ws} 
(a slightly modified version of this metric form has been considered in
\cite{Dehghani:2001nz} for asymptotically AdS  black holes
with abelian Higgs hair).
The line element (\ref{metric}) has two Killing vectors: $\partial_t$
coresponding to time translation  symmetry and $\partial_{\varphi}$
corresponding to rotation symmetry around the $z-$   axis
(with $z=r \cos \theta,~\rho=r \sin \theta$).
The event horizon resides at a surface of constant radial coordinate
$r=r_h$ and is characterized by the condition $f(r_h)=0$,
while the metric coefficients $g_{rr}, g_{\theta \theta}$ 
and $g_{\varphi \varphi}$  take nonzero and finite values.

The construction of an axially symmetric YM ansatz has been discussed by many authors 
starting with the pioneering papers by Manton \cite{Manton:1977ht} 
and Rebbi and Rossi \cite{Rebbi:1980yi}. 
The most general purely magnetic axially symmetric YM-SU(2) ansatz contains  nine magnetic
 potentials and can be easily obtained in cylindrical coordinates 
$x^{\mu}=(\rho,\varphi,z$)
\begin{eqnarray} 
\label{A-gen-cil}
A_{\mu}=\frac{1}{2}A_{\mu}^{(\rho)}(\rho,z)\tau_{\rho}^n
        +\frac{1}{2}A_{\mu}^{(\varphi)}(\rho,z)\tau_{\varphi}^n
        +\frac{1}{2}A_{\mu}^{(z)}(\rho,z)\tau_{z}^n,
\end{eqnarray}
where the only $\varphi$-dependent terms are the $SU(2)$ matrices
(composed of the standard $(\tau_1,~\tau_2,~\tau_3)$ Pauli matrices)
\begin{eqnarray} 
\label{u-cil}
\tau_{\rho}^n=~~\cos n\varphi~\tau_1+\sin n\varphi~\tau_2,~~
\tau_{\varphi}^n=-\sin n\varphi~\tau_1+\cos n\varphi~\tau_2,~~
\tau_{z}^n=\tau_3.
\end{eqnarray}
Transforming to spherical coordinates, it is convenient to introduce,
without any loss of generality, a new  SU(2)  basis 
$(\tau_{r}^n,\tau_{\theta}^n,\tau_{\varphi}^n)$,
with
\begin{eqnarray} 
\label{u-sph}
\tau_{r}^n=\sin \theta~\tau_{\rho}^n+\cos \theta~\tau_z^n,
~~
\tau_{\theta}^n=\cos \theta~\tau_{\rho}^n-\sin \theta~\tau_z^n.
\end{eqnarray}
The general expression (\ref{A-gen-cil}) takes the following form in spherical coordinates
\begin{eqnarray} 
\label{A-gen-sph}
A_{\mu}=\frac{1}{2}A_{\mu}^{(r)}(r,\theta)\tau_{r}^n
        +\frac{1}{2}A_{\mu}^{(\theta)}(r,\theta)\tau_{\theta}^n
        +\frac{1}{2}A_{\mu}^{(\varphi)}(r,\theta)\tau_{\varphi}^n,
\end{eqnarray}
where
$A_\mu^{(a)} dx^\mu=A_r^{(a)} dr + A_\theta^{(a)} d\theta + A_\varphi^{(a)} d\varphi$.
This ansatz is axially symmetric in the sense that a rotation around the $z-$axis 
can be compensated by a gauge rotation 
${\mathcal{L}}_\varphi A=D\Psi$ \cite{Forgacs:1980zs}, 
with $\Psi$ being a Lie-algebra valued gauge function. 
For the ansatz (\ref{A-gen-sph}),
$\Psi=n \cos \theta  \tau^n_{r}/2 - n \sin \theta  \tau^n_{\theta}/2$.
Therefore we find
$ F_{\mu \varphi} = D_{\mu}W, $
where $W=A_{\varphi}-\Psi$.

Searching for axially symmetric solutions within the most general ansatz
is a difficult task, and, to our knowledge there are no 
analytical or numerical results
in this case.
We use in this paper a reduced ansatz, employed also 
in all previous studies on EYM solutions,
with five of the gauge potentials taken identically zero
\begin{eqnarray}
\nonumber
A_{r}^{(r)}=A_{r}^{(\theta)}~=~A_{\theta}^{(r)}~=~A_{\theta}^{(\theta)}
~=~A_{\varphi}^{(\varphi)}=0.
\end{eqnarray}
A suitable parametrization of the four nonzero
components of $A_\mu^{(a)}$ which factorizes the trivial $\theta$-depencence is
\begin{eqnarray}
\label{ansatz}
A_r^{(r)}=\frac{H_1(r,\theta)}{r},~A_{\theta}^{(r)}=1-H_2(r,\theta),
~A_{\varphi}^{(r)}=-n \sin \theta H_3(r,\theta),~
A_{\varphi}^{(\theta)}=-n \sin \theta (1-H_4(r,\theta)),
\end{eqnarray}
This  consistent reduction of the general ansatz
satisfies also some additional discrete symmetries \cite{Rebbi:1980yi}, \cite{ymh}
(in particular the parity reflection symmetry).
The explicit expression of the  field strength tensor for this ansatz 
is given in \cite{Kleihaus:1997mn}.
To fix the residual abelian gauge invariance we choose the usual
gauge condition \cite{Kleihaus:1997ws}-\cite{Kleihaus:2002ee}
\begin{eqnarray}
\label{gauge}
\nonumber
r \partial_r H_1 - \partial_\theta H_2 = 0. 
\end{eqnarray}
The ansatz (\ref{A-gen-sph}) contains an integer $n$, representing the winding number with 
respect to the 
azimutal angle $\varphi$.
While $\varphi$ covers the trigonometric circle once, 
the fields wind $n$ times around.
Note that the usual spherically symmetric static ansatz   
corresponds to $n=1$,
$H_1=H_3=0$, $H_2=H_4=w(r)$.
Similar to $n=1$ solutions, we define the node number $k$ by the number of nodes
of the gauge fields $H_2$ and $H_4$.
From (\ref{einstein-eqs}) and (\ref{YM-eqs}) we obtain a set of seven
nonlinear elliptical partial differential equations for $(f,l,m,H_i)$ which 
should  be solved numerically.
These equations are presented in \cite{Radu:2001ij}.

Within this ansatz, the energy density of the matter 
fields $\epsilon=-T_t^t$ is given by
\begin{eqnarray} 
\label{energy} 
-T^t_t=\frac{f^2}{{2e^2 r^4 m}}
\Bigg [  
\frac{N}{m}
 (r\partial_r H_2+\partial_{\theta}H_1)^2+
\frac{n^2 N}{l}
\Big((r\partial_r H_3-H_1H_4)^2+
(r\partial_r H_4+H_1(H_3+\cot \theta))^2\Big)
\\
\nonumber
+\frac{n^2}{l}
\Big(
(\partial_{\theta}H_3-1+\cot \theta H_3+H_2H_4)^2
+(\partial_{\theta}H_4+\cot \theta (H_4-H_2)-H_2H_3)^2
\Big)
\Bigg ].
\end{eqnarray}
\subsection{Boundary conditions}
To obtain asymptotically AdS solutions
with a regular event horizon and with the proper symmetries,
we must impose the appropriate boundary conditions,
the boundaries being the horizon and radial infinity,
the $z$-axis and, because of parity reflection symmetry satisfied by the matter fields,
the $\rho$-axis.
The boundary conditions at infinity and along the  $z-$ and the $\rho$-axis
(i.e for $\theta=0,\pi/2$)
agree with those of the globally regular solutions \cite{Radu:2001ij}.

Similar to the $\Lambda=0$ case, we impose  the horizon of
the black hole to reside at a surface of constant radial coordinate $r=r_h$.

The boundary conditions satisfied by the metric functions are
\begin{equation} 
\label{b1}
f|_{r=r_h}=  m|_{r=r_h}=  l|_{r=r_h}= 0,
\end{equation}
\begin{equation}\label{b2}
f|_{r=\infty}= m|_{r=\infty}= l|_{r=\infty}=1.
\end{equation}
We impose also the  condition
\begin{equation}
m|_{\theta=0}=l|_{\theta=0},
\end{equation}
which is implied by the assumption of regularity on the $z-$axis \cite{kramer}.
The boundary conditions satisfied by the matter functions are
very similar to those satisfied in the $\Lambda=0$ case
\begin{equation}
\label{b3}
H_{1}|_{r=r_h}=0,~~\partial_r H_{2}|_{r=r_h}=
\partial_r H_{3}|_{r=r_h}=\partial_r H_{4}|_{r=r_h}=0,
\end{equation}
at the event horizon and 
\begin{equation}
\label{b4}
H_2|_{r=\infty}=H_4|_{r=\infty}= \omega_0, \ \ \ 
H_1|_{r=\infty}=H_3|_{r=\infty}=0,
\end{equation}
at infinity, where this time  $w_0$ is a constant 
(with $w_0 = \pm 1$ for $\Lambda =0$).
The relations (\ref{b3}) come from the condition 
$F_{r \varphi}|_{r=r_h}=F_{r \theta}|_{r=r_h}=0$
imposed by the field equations, together with a special gauge choice on the event horizon
\cite{Kleihaus:2002ee,Hartmann:2001ic}.
For a solution with parity reflection symmetry,
the boundary conditions along the axes are
\begin{equation}\label{b5}
\begin{array}{lllllllllll}
H_1|_{\theta=0,\pi/2}&=&H_3|_{\theta=0,\pi/2}&=&0,
\\
\partial_\theta H_2|_{\theta=0,\pi/2} 
&=& \partial_\theta H_4|_{\theta=0,\pi/2}
&=& 0,
\end{array}
\end{equation}
\begin{equation}\label{b6}
\begin{array}{lllllll}
\partial_\theta f|_{\theta=0,\pi/2} 
&=& \partial_\theta m|_{\theta=0,\pi/2} &=&
\partial_\theta l|_{\theta=0,\pi/2} &=&0. 
\end{array}
\end{equation}
Therefore we need to consider the solutions only in the region $0\leq \theta \leq \pi/2$. 
Regularity on the $z-$axis requires also
\begin{equation}
H_2|_{\theta=0}=H_4|_{\theta=0}.
\end{equation}
Dimensionless quantities are obtained by using the following rescaling
\begin{eqnarray} 
\label{resc} 
r \to (\sqrt{4\pi G}/e) r,~~\Lambda \to (e^2/4 \pi G) \Lambda,~~
~~M \to (eG/\sqrt{4\pi G}) M,
\end{eqnarray}
where $M$ is the mass of the solutions 
(defined in the next section).

\section{PHYSICAL QUANTITIES OF THE SOLUTIONS}
\subsection{Mass and boundary stress tensor}

At spatial infinity, the line element (\ref{metric}) can be written as
\begin{equation}\label{mass1}
ds^2=ds_0^2+c_{\mu \nu}dx^{\mu}d x^{ \nu},
\end{equation}
where $c_{\mu \nu}$ are deviations from the background AdS metric $ds_0^2$.
Similar to the asymptotically flat case, one expects 
the value of mass to be encoded in the functions $c_{\mu \nu}$.
The construction of the conserved quantities for an asymptotically AdS spacetime 
was addressed for the first time in the eighties, with various aproaches 
(see for instance 
Ref. \cite{Abbott:1982ff, Henneaux:1985tv}).
The generalization of Komar's formula in this case is not 
straightforward and 
requires the further subtraction of a background configuration in order 
to render a finite result.
The mass of the axially symmetric regular solutions discussed in \cite{Radu:2001ij} 
has been computed in this way,
by using the Hamiltonian formalism of Henneaux and Teitelboim \cite{Henneaux:1985tv}.

Here we prefer to use a different approach
and to follow the general 
procedure proposed by Balasubramanian and Kraus \cite{Balasubramanian:1999re}
to compute conserved quantities
for a spacetime with a negative cosmological constant.
This technique was inspired by AdS/CFT correspondence and consists in 
adding suitable counterterms 
$I_{ct}$
to the action. These counterterms are built up with
curvature invariants of a boundary $\partial \cal{M}$ (which is sent to 
infinity after the integration)
and thus obviously they do not alter the bulk equations of motion.

The following counterterms are sufficient to cancel divergences in four dimensions,
for vacuum solutions with a negative cosmological constant
\begin{eqnarray}
\label{ct}
I_{\rm ct}=-\frac{1}{8 \pi G} \int_{\partial {\cal M}}d^{3}x\sqrt{-h}\Biggl[
\frac{2}{\tilde l}+\frac{\tilde l}{2}\rm{R}
\Bigg]\ .
\end{eqnarray}
Here $\rm{ R}$ is the Ricci scalar for the boundary metric $h$. 
In this section we will define also $\tilde l^2=-3/\Lambda$. 

Using these counterterms one can
construct a divergence-free stress tensor from the total action
$I{=}I_{\rm bulk}{+}I_{\rm surf}{+}I_{\rm ct}$ by defining 
\begin{eqnarray}
\label{s1}
 T_{ab}&=& \frac{2}{\sqrt{-h}} \frac{\delta I}{ \delta h^{ab}}
=\frac{1}{8\pi G}(K_{ab}-Kh_{ab}-\frac{2}{\tilde l}h_{ab}+\tilde lE_{ab}),
\end{eqnarray}
where $E_{ab}$ is the Einstein tensor of the intrinsic metric $h_{ab}$.
The efficiency of this approach has been demonstrated in a broad range of examples,
the counterterm subtraction method being developed on its own interest and applications.
If there are matter fields on $\cal{M}$ additional counterterms 
may be needed to regulate the action (see $e.g.$ \cite{Taylor-Robinson:2000xw}).
However, we find that for a SU(2) nonabelian matter content
in four dimensions, the prescription  (\ref{ct}) removes all divergences
(a different situation is found for the five dimensional AAdS
nonabelian solutions where the counterterm method fails and
logarithmic divergences are presented in the
total action and the expression of mass \cite{Okuyama:2002mh}).

In deriving the explicit expression of $T_{ab}$ for our solutions, we make use of
the asymptotic behavior of the metric functions
\begin{eqnarray}
\label{as1}
f=1+\frac{f_1+f_2 \sin^2 \theta}{r^3}+O(\frac{1}{r^5}),~~
m=1+\frac{m_1+m_2 \sin^2 \theta}{r^3}+O(\frac{1}{r^5}),~~
l=1+\frac{l_1+l_2 \sin^2 \theta}{r^3}+O(\frac{1}{r^5}),
\end{eqnarray}
where $f_1,~f_2$ are undetermined constants, while
\begin{eqnarray}
\label{test}
l_1=m_1=\frac{2f_1}{3},~~~l_2=\frac{6f_2}{17},~~~m_2=\frac{14f_2}{17},
\end{eqnarray}
(note that we find a different power decay as compared with the $\Lambda=0$ case \cite{Kleihaus:1997mn}).

The nonvanishing components of the boundary stress 
tensor at large $r$ are 
\begin{eqnarray}
\label{BD4}
\nonumber
T_{\theta \theta} &=&-\frac{1}{8 \pi G \tilde l r}
\left(\frac{2f_1}{3}+\frac{6f_2}{17}\sin^2 \theta \right)+O\left(\frac{1}{r^2}\right),
\\
T_{\varphi \varphi}&=&-\frac{\sin^2 \theta}{8 \pi G \tilde l r}
\left(\frac{2f_1}{3}+\frac{18f_2}{17}\sin^2 \theta \right)
+O\left(\frac{1}{r^2}\right),
\\
\nonumber
T_{tt}&=&-\frac{1}{8 \pi G \tilde l^3 r}
\left(\frac{4f_1}{3}+\frac{24f_2}{17}\sin^2 \theta \right)
+O\left(\frac{1}{r^2}\right).
\end{eqnarray}
Direct computation shows that this stress tensor is traceless.
This result is expected from the AdS/CFT correspondence, since even dimensional
bulk theories with $\Lambda<0$ are dual to odd dimensional CFTs which have a 
vanishing trace anomaly.

We can use this formalism to assign a mass to our EYM solutions
by writing the boundary metric in the following form \cite{Balasubramanian:1999re}
\begin{eqnarray}
\label{b-AdS}
h_{\mu \nu}dx^{\mu} dx^{\nu}=-N_{\Sigma}^2dt^2
+\sigma_{ab}(dx^a+N_{\sigma}^a dt) (dx^b+N_{\sigma}^b dt).
\end{eqnarray}
If $\xi^{\mu}$ is a Killing vector generating an isometry of the boundary geometry,
there should be an associated conserved charge. The conserved charge associated with
time translation is the mass of spacetime
\begin{eqnarray}
\label{mass}
M=\int_{\partial \Sigma}d^{2}x\sqrt{\sigma}N_{\Sigma}\epsilon
\end{eqnarray}
(in the absence of an electric part of the YM potential, 
$N_{\sigma}^a=0$ and the solutions carry no angular momentum).
Here $\Sigma$ is a spacelike hypersurface from event horizon to infinity with timelike unit normal 
$n^{\mu}$ and $\epsilon=n^{\mu}n^{\nu}T_{\mu \nu}$ is the proper energy density.
By using this relation, we  find that the mass of our solutions is given by
\begin{eqnarray}
\label{ct-mass}
M= \frac{\Lambda}{3G}\left(\frac{2f_1}{3}+\frac{8f_2}{17}\right).
\end{eqnarray}

The metric restricted to the boundary $h_{ab}$ diverges due to an infinite
conformal factor $r^2/\tilde l^2$. The background metric upon which the dual field
theory resides is
\begin{eqnarray}
\nonumber
\gamma_{ab}=\lim_{r \rightarrow \infty} \frac{\tilde l^2}{r^2}h_{ab}.
\end{eqnarray}
For the asymptotically AdS solutions considered here, the  
boundary metric is 
\begin{equation}
\label{l1} 
\gamma_{ab}dx^a dx^b=\tilde l^2(d \theta^2+\sin ^2 \theta d \varphi^2)-dt^2,
\end{equation}
which is just the line-element of the $(2+1)$ Einstein universe.
Corresponding to the boundary metric (\ref{l1}), the stress-energy tensor $<\tau_{ab}>$
for the dual theory can be calculated using the following relation \cite{Myers:1999qn}
\begin{eqnarray}
\label{r1}
\sqrt{-\gamma}\gamma^{ab}<\tau_{bc}>=
\lim_{r \rightarrow \infty} \sqrt{-h} h^{ab}T_{bc}.
\end{eqnarray}
By using this prescription, we find that the stress tensor 
of the field theory has the curious form
\begin{eqnarray}
\label{st}
<\tau^{a}_b>=A~{\rm diag}(1,1,-2)+B~{\rm diag}(1,3,-4)\sin^2 \theta,
~~{\rm with~}A=\frac{1}{8 \pi \tilde l^2}(M+\frac{8f_2}{17}\frac{1}{G \tilde l^2})
,~B=-\frac{1}{8 \pi G \tilde l^4}\frac{6f_2}{17},
\end{eqnarray}
where $x^1=\theta,~x^2=\varphi,~x^3=t$. Here $M,f_2$ are continuous variables
which encode the bulk parameters.
This tensor is covariantly conserved and manifestly traceless.
A winding number $n>1$ of the bulk configurations implies  $f_2 \neq 0$ and thus
a $\theta-$dependence
of the dual theory stress tensor (although the boundary metric is spherically symmetric),
which is a unique property of AAdS gravitating nonabelian configurations.
This also suggests the dual  theory should also contain the integer $n$.

From the AdS/CFT correspondence, we expect the nonabelian hairy black holes to be described by 
some thermal states
in a dual theory formulated in a $(2+1)$ Einstein Universe background.
The spherically and axially symmetric solitons will correspond to zero-temperature states 
in the same theory. 
However, for the EYM action (\ref{lag0}) we do not know the underlying boundary CFT.
In particular, we do not know what the SU(2) field corresponds 
to in CFT language.
The bulk YM fields have nothing to do, of course, 
with the nonabelian fields of the dual 
gauge theory.
Further work in this direction will be of great interest.

\subsection{From $\Lambda=-3$, $d=4$ EYM to $d=11$ supergravity}
However, the EYM solutions with a negative cosmological constant $\Lambda=-3$ 
may have some relevance in AdS/CFT context
(we'll work in this section with rescaled quantities as defined by (\ref{resc})).
As proven in \cite{Pope:1985bu}, for this value of the cosmological constant,
an arbitrary solution $(g_{\mu \nu},A_{\mu}^{(a)})$ of the four dimensional
EYM equations (\ref{einstein-eqs}), (\ref{YM-eqs}) gives
a solution of the equations of motion of the $d=11$ supergravity
(note also that the EYM system with $\Lambda>0$ can also 
be embedded in $d=11$ supergravity
\cite{Lu:2003dm}). 

Since this reduction of the eleven dimensional theory
 has been   explained  in detail in \cite{Pope:1985bu},
we present here only the basic results.
The Lagrangian of the bosonic sector of $d=11$
supergravity comprises the eleven-dimensional metric tensor $\hat{g}_{MN}$
and a 4-index antisymmetric tensor field $\hat{F}_{MNPQ}$.
Following \cite{Pope:1985bu}, 
the eleven dimensional metric ansatz reads
(with $\mu,~\nu\dots $ denoting indices of the four dimensional space, $M,N \dots$ 
indices of the eleventh dimensional solution, 
while $m,n,\dots$ are indices on the internal space)
\begin{eqnarray}
\label{d11-1}
d\hat{s}^2=\hat{g}_{MN}dx^Mdx^n
=g_{\mu \nu}dx^{\mu}dx^{\nu}+g_{mn}(y)(dy^m-K^{(a) m}A_{\mu}^{(a)}(x)dx^{\mu})
(dy^n-K^{(a) n}A_{\nu}^{(a)}(x)dx^{\nu}),
\end{eqnarray}
where
$g_{\mu \nu}dx^{\mu} dx^{\nu}=e^{\alpha}\otimes e^{\beta} \eta_{\alpha \beta}$
is the four-dimensional metric (\ref{metric}), 
$g_{mn}dy^mdy^n=e^{p}\otimes e^{q} \delta_{pq}$ is the metric on a round 7-sphere,
$(e^{\alpha},e^{p})$ being the orthonormal bases on these subspaces. 
$K^{(a) m}$ are three Killing vectors of the seventh dimensional internal space generating the
SU(2) Lie algebra.

The antisymmetric tensor field $\hat{F}$ can be read from
($F^{(a) \mu \nu}$ being a solution of the four dimensional YM equations)
\begin{eqnarray}
\label{d11-3}
\hat{F}=\frac{3}{2} \varepsilon_4 -\frac{1}{8}\tilde{G}^{(a)}_{\alpha \beta} M^{(a)}_{cd}
\hat{e}^{\alpha}\land \hat{e}^{\beta} \land \hat{e}^{c} \land \hat{e}^d,
\end{eqnarray}
where $\varepsilon_4=(4!)^{-1}\varepsilon_{\alpha \beta \gamma \delta}
\hat{e}^{\alpha}\land \hat{e}^{\beta}\land \hat{e}^{\gamma}\land \hat{e}^{\delta}$,
$\tilde{G}^{(a)}_{\alpha \beta}=1/2 \varepsilon_{\alpha \beta \mu \nu}F^{(a) \mu \nu}$,
while
$M^{(a)}_{mn}=-2\nabla_m K^{(a)}_n$.

This embedding suggests the possible existence of 
BPS configurations with $\Lambda =-3$, satisfying first order differential equations
which may lead to exact solutions.
An exact solution of the field equations where 
the round 
two-sphere $(\theta, \varphi)$ is replaced by a two-dimensional space  
of  vanishing curvature is presented in Appendix.

\subsection{Temperature and entropy}
The zeroth law of black hole physics states that 
the surface gravity $\kappa$ 
is constant at the horizon of the black hole solutions,
where 
\begin{equation}
\label{kappa} 
\kappa^2 =
-(1/4)g^{tt}g^{ij}(\partial_i g_{tt})(\partial_j g_{tt})\Big|_{r=r_h}. 
\end{equation}
To evaluate $\kappa$, we 
use the following expansions of the metric functions at the horizon
\begin{eqnarray}
\label{expan-h}
\nonumber
f(r,\theta)&=&f_2(\theta)\left(\frac{r-r_h}{r_h}\right)^2 
 + O\left(\frac{r-r_h}{r_h}\right)^3 \ ,
\\
m(r,\theta)&=&m_2(\theta)\left(\frac{r-r_h}{r_h}\right)^2 
 + O\left(\frac{r-r_h}{r_h}\right)^3\,
\\
\nonumber
l(r,\theta)&=&l_2(\theta)\left(\frac{r-r_h}{r_h}\right)^2 
 + O\left(\frac{r-r_h}{r_h}\right)^3. 
\end{eqnarray}
Since from general arguments the temperature $T$ is proportional
to the surface gravity
$\kappa $, $
T=\kappa /(2 \pi),
$
we obtain the relation
\begin{equation}
\label{temp}
T=\frac{f_2(\theta) (1- \Lambda r_h^2/3)}{2 \pi r_h \sqrt{m_2(\theta)} } 
\ . 
\label{tempe} 
\end{equation}
Similar to the $\Lambda=0$ case, we can show, with help of the $(r~\theta)$
Einstein equation which implies
\begin{eqnarray}
\label{con}
f_2  m_{2, \theta}=2m_2 f_{2,\theta},
\end{eqnarray}
that the temperature $T$, as given in (\ref{temp}), is indeed constant.

For the line element (\ref{metric}),
the  area $A$ 
of the event horizon  is given by 
\begin{equation}
\label{area}
A = 2 \pi \int_0^\pi  d\theta \sin \theta
\frac{\sqrt{l_2(\theta) m_2(\theta)}}{f_2(\theta)} r_h^2.
\end{equation}
According to the usual thermodynamic arguments, the entropy $S$ is proportional 
to the area $A$ \cite{wald}
\begin{equation}
\label{entropy}
S = \frac{A}{4G},  
\end{equation}
leading to the product
\begin{equation}
TS = \frac{r_h(1- \Lambda r_h^2/3)}{4G} \int_0^\pi  d\theta \sin \theta
{\sqrt{l_2(\theta) }}. 
\end{equation}
Unfortunatelly, for $\Lambda<0$ we could not derive a simple Smarr-type relation, 
similar to that valid for asymptotically flat hairy black holes \cite{Kleihaus:1997ws},
which relates asymptotic quantities  to quantities defined on the event horizon
(this  relation can also  be used to test the accuracy of the numerical results).
When integrating the Killing identity
\begin{equation}
\label{killing}
\nabla^a\nabla_b K_a=R_{bc}K^c,  
\end{equation}
for the Killing field $K^a=\delta^a_t$ 
over a spacelike hypersurface $\Sigma$, we find the expression
\begin{equation}
\label{smarr1}
- \frac{1}{2}\int_0^{\pi} d \theta \sin \theta \frac{r^2 \sqrt{l}}{f}
\big(N\partial_r f-\frac{2\Lambda r}{3}f\big) \Big|_{r_h}^{\infty}
= \Lambda \int_{rh}^{\infty} dr \int_0^{\pi} d \theta \sin \theta \frac{r^2 m\sqrt{l}}{f} 
+8\pi G \int_{rh}^{\infty} dr \int_0^{\pi} d \theta \sin \theta \frac{r^2 m\sqrt{l}}{f} T_t^t,
\end{equation}
(here we use $R=4\Lambda$ and $T_{\mu}^{\mu}=0$).
To make sense on this expression we must regularize it, 
by extracting a background contribution $(g_{\mu \nu}^0)$.
However, this gives useful results for very simple configurations only (with
$\sqrt{-g}=\sqrt{-g^0}$).
This background subtraction is of no practical use for numerical solutions like ours.
%
\subsection{A computation of the Euclidean action}
The expression (\ref{entropy}) for the entropy can be derived in a more rigorous way by using
Euclidean quantum gravity arguments.
Here we start by constructing the path integral \cite{Gibbons:1976ue}
\begin{eqnarray}
\label{Z1}
Z=\int D[g]D[\Psi]e ^{-iI[g,\Psi]}
\end{eqnarray}
by integrating over all metrics and matter fields between  some given initial and final
hypersurfaces, $\Psi$ corresponding here to the SU(2) potentials.
By analytically continuing the time coordinate $t \to i\tau$,
the path integral formally converges, and in the leading order one obtains
\begin{eqnarray}
\label{Z2}
Z \simeq e^{-I_{cl}}
\end{eqnarray}
where $I_{cl}$ is the classical action evaluated on the equations of motion
of the gravity/matter system.
We note that the considered Lorentzian solutions of the EYM equations extremize also the
Euclidean action, $t \to i\tau$ having no effects at the level of the equations of motion 
\footnote{Note that this analytical continuation becomes  problematic for nonabelian solutions
with an electric potential.}.
The value of $\beta$ is found here by demanding regularity of the Euclideanized manifold as $r \to r_h$,
which together with the expansion (\ref{expan-h}) and the condition (\ref{con})
gives $\beta=1/T$. The physical interpretation of this formalism is that
the class of regular stationary metrics forms an ensemble of
thermodynamic systems at equilibrium temperature
$T$ \cite{Mann:2002fg}.
$Z$ has  the interpretation of partition function and we can
define the free energy of the system
$F=-\beta^{-1} \log Z$.
Therefore
\begin{eqnarray}
\label{i1}
\log Z=-\beta F=S-\beta M,
\end{eqnarray}
or
\begin{eqnarray}
\label{i2}
S=\beta M-I_{cl},
\end{eqnarray}
straightforwardly follows.

To compute $I_{cl}$, we make use of the Einstein equations, replacing the $R-2\Lambda$ volume term with
$2R_t^t-16\pi G T_t^t$.
For our purely magnetic ansatz, the term $T_t^t$  exactly cancels the matter field
lagrangean in the bulk action
 $L_m=-1/2Tr(F_{\mu \nu }F^{\mu \nu })$.
The divergent contribution given by the surface integral term at infinity in $R_t^t$ is also canceled by 
$I_{\rm{surface}}+I_{ct}$ and we arrive at the simple finite expression
\begin{eqnarray}
\label{itot}
I_{cl}=\beta\left(\frac{\Lambda}{3G}\bigg(\frac{2f_1}{3}+\frac{8f_2}{17}\bigg)
-\frac{r_h}{4G}\bigg(1-\frac{\Lambda r_h^2}{3}\bigg)\int_0^{\pi} 
d \theta \sin \theta \sqrt{l_2(\theta)}
\right)
\end{eqnarray}
Replacing now in (\ref{i2}) (where $M$ is the mass-energy computed in Section 3.1), we find
\begin{eqnarray}
\label{i3}
S= \frac{\pi }{2G}\int_0^\pi  d\theta \sin \theta
\frac{\sqrt{l_2(\theta) m_2(\theta)}}{f_2(\theta)} r_h^2,
\end{eqnarray}
which is one quarter of the event horizon area, as expected.

\subsection{Nonabelian charges}
Solutions of the field equations are also classified by the
nonabelian electric and magnetic charges $Q_E$ and $Q_M$.  
For  purely magnetic configurations, the usual definition used by various
authors (and employed also in this paper) is 
\begin{equation}
\label{charge}
Q_M
= {e\over 4\pi} \int dS_k \, \sqrt{-g} \, 
Tr  \{ \tilde F^{k0}  \tau_r \},
\end{equation}
where  $\tilde F^{\mu \nu}$ is the dual field strength tensor.
For the boundary conditions at infinity 
(\ref{b4}), we find $Q_M=n(1-\omega_0^2)$.
However, the integral (\ref{charge}) does not intrinsically characterize a field 
configuration, as it is gauge dependent.
A gauge invariant definition for the nonabelian charges has been proposed in  
\cite{Corichi:1999nw}  
and used in \cite{Kleihaus:2002ee} for asymptotically flat nonabelian solutions
\begin{eqnarray}
\label{charges}
Q_E={e\over 4\pi} \oint d \theta d \varphi |\tilde F_{\theta \varphi}|,~~~
Q_M={e\over 4\pi} \oint d \theta d \varphi | F_{\theta \varphi}|,
\end{eqnarray}
where the vertical bars denote the Lie-algebra norm and the integrals 
are evaluated as $r \to \infty$.
One can verify that this definition yields the absolute value
of the magnetic charge as defined in (\ref{charge}).
%
%

\section{YANG-MILLS FIELDS IN FIXED BLACK HOLE BACKGROUND}
For asymptotically AdS geometries, 
it has been proven useful in a number of cases to consider first
the matter system in a fixed background, as a first step 
toward the study of the gravitating configurations.
These solutions preserve the basic features of the gravitating counterparts
and can be obtained much more easily.
This situation resembles the well-known case of
(asymptotically-) flat monopoles and dyons, where the inclusion of gravity
does not change the essential picture.

In this section we present numerical arguments 
for the existence of nontrivial monopoles and dyons solutions of pure YM equations in a 
four dimensional SAdS black hole background, 
gravity being regarded as a fixed  external field.
Although being very simple, nevertheless this model appears to contain all 
the essential features of the gravitating black hole solutions.

The background metric is given by
\begin{eqnarray} 
\label{SAdS}
ds^2&=&\frac{dr^2}{F}+
r^2(d \theta ^2+\sin^2 \theta d \varphi^2)-F dt^2,
\\
\nonumber
F&=& 1-\frac{2\tilde M}{r}- \frac{\Lambda r^2}{3}, 
\end{eqnarray}
where $\tilde M$ is the black hole mass, 
 and the black hole horizon $r_h$ is located at the largest root of the equation
$r_h-2\tilde M-\Lambda r_h^3/3=0$.
Note that this line element is not on the form (\ref{metric}); 
unfortunately, the SAdS metric does not present a 
simple form in "isotropic" coordinates
and we should make use of the usual Schwarzschild coordinates.
\subsection{Monopole solutions}
\subsubsection{Spherically symmetric configurations}
We start by briefly discussing the solutions obtained for $n=1$ 
within the magnetic  ansatz (\ref{ansatz}). 
In this case $H_1=H_3=0$,~$H_2=H_4=\omega(r)$ and  the YM equations have  the simple form
\begin{eqnarray} 
\label{weq}
\omega''=\frac{1}{F} \left [\frac{\omega(\omega^2-1)}{r^2}
-(\frac{2\tilde M}{r^2}-\frac{2\Lambda r}{3}) \omega'\right ],
\end{eqnarray}
where the prime denotes derivative with respect to $r$.
This  problem  possesses two nontrivial limits, 
with known exact solutions, 
noticed 
for the first time in \cite{Boutaleb-Joutei:1979va}.
The first one is found for $M \to 0$
\begin{eqnarray} 
\label{exact1}
\omega=\frac{1}{\sqrt{1-\Lambda/3 r^2}},
\end{eqnarray} 
describing a monopole in AdS spacetime with unit magnetic charge. 
Apart from this exact solution, a continuous family of $k=0$ finite energy 
configurations is found \cite{Radu:2001ij, Hosotani:2001iz}.
Even more remarkable, the second one valid as $\Lambda \to 0$ describes a nonabelian hair 
for a Schwarzschild black hole
\begin{eqnarray} 
\label{exact2}
\omega=\frac{c-r}{r+3(c-1)},
\end{eqnarray} 
where $c=M(3+\sqrt{3})$. 
In this case, 
we find an infinite sequence of excited solutions, indexed by the node number $k$ 
(the solution (\ref{exact2}) corresponds to $k=1$), with
similar properties to the well known gravitating counterparts
\cite{Bizon:1994dh}).

In the $\Lambda<0$ case, the numerical results show the existence of a one-parameter
family of solutions regular at $r=r_h$ with the following  behavior
as $r \to r_h$
\begin{eqnarray}
\label{w-hor}
 \omega(r) &=& \omega_h+\frac{\omega_h(\omega_h^2-1)}{r_h(1-\Lambda r_h^2)}(r-r_h)+O(r-r_h)^2,
\end{eqnarray}
and the asymptotic expansion at large $r$
\begin{eqnarray}
\label{w-infinity}
\omega = \omega_0+\frac{w_1}{r} + O(\frac{1}{r^2}),
\end{eqnarray}
where $\omega_0$, $\omega_1$ are constants to
be determined by numerical calculations and there are no restrinctions on the
value of $\omega_0$.
Since the field equations are invariant under the transformation $\omega \to -\omega$, only
values of $\omega_h>0$ are considered.
The  boundary condition (\ref{w-infinity}) permits a non-vanishing magnetic 
charge $Q_M=1-\omega_0^2$.

The overall picture we find for the general case 
is rather similar to 
the one described in \cite{Bjoraker:2000qd} where gravity is taken into account.
By varying the parameter $\omega_h$, a continuum of monopole solutions is obtained.
The same general behavior is noticed for the 
gauge function $\omega$, with very similar 
branch structure.
As expected, the properties of the solutions depends esentially on the 
values of $\Lambda,~r_h$.
For $|\Lambda|$ sufficiently large, there exist finite energy solutions for 
which the gauge function has no nodes.
Values of $\omega_h>1$ are always allowed, leading to nodeless solutions.
For a fixed value of $\Lambda$, we obtain finite energy solutions 
for a finite number of intervals in parameter space.
We noticed always the existence of  at least two such interval around the 
trivial solutions $\omega_h=1,~0$.
For large enough values of $|\Lambda|$, 
the $\omega$ function remain close to 
its initial value on the event horizon, and we find
only one branch of solutions.
For small enough $|\Lambda|$, we have found  
solutions where $\omega$ crosses the $r$ axis and a large number of branches.
The arguments presented in \cite{Bjoraker:2000qd} for the (linear) stability of the 
nodeless $n=1$ monopole 
solutions apply directly to the nongravitating case.

\subsubsection{Axially symmetric configurations}
The next step is to consider the $n>1$ axially symmetric generalizations
of these configurations.
Subject to the  boundary conditions (\ref{b1})-(\ref{b6}),
we solve numerically  for a fixed SAdS background 
the set of four YM equations
implied by (\ref{YM-eqs}).

The algorithm we have used for all axially symmetric solutions presented in this paper
is only a slightly modified version of the 
approach employed by Kleihaus and Kunz in their studies of
asymptotically flat nonabelian configurations.
The field equations are first discretized on a ($r,~\theta$) grid with $N_r\times N_{\theta}$ points.
The angular coordinate $\theta$ runs
from $0$ to $\pi/2$ and the radial coordinate goes from $r_h$ to some large enough value 
$r_{max}$ (this value is not fixed apriori,
depending on the parameters $r_h,\Lambda$; typically $r_{max}\simeq 3\times 10^3 \div 6\times 10^3 r_h$).
We tested for a number of solutions that the relevant quantities are insensitive to the
cut off value $r_{max}$.
The grid spacing in the $r-$direction is non-uniform, while the values of the grid points
in the angular direction are given by $\theta_k=(k-1)\pi/(2(N_{\theta}-1))$.
Essentially, our numerical problems  come from the region near the event horizon,
where a carefull grid choice is necessary.

In this scheme, a new radial variable is introduced
which maps the semi infinite region $[r_h,\infty)$ to the closed region $[0,1]$.
Our choice for this transformation was $ x=(r-r_h)/(r+c)$,
 where $c$ is a properly chosen constant in order to minimize the numerical errors.
A $c \neq 0$ proven to be useful for higher mass solutions
(however in most of the computations $c$ takes a value smaller than 20).
Typical grids have sizes $150 \times 30$, 
covering the integration region 
$0\leq{x}\leq 1$ and $0\leq\theta\leq\pi/2$.
The resulting system is solved iteratively until convergence is achieved.
All numerical calculations for axially symmetric configurations 
are performed by using the program FIDISOL (written in Fortran), based on the iterative Newton-Raphson
method. 
Details of the FIDISOL code are presented in \cite{FIDISOL}.
This code requests the system of nonlinear partial differential equations to be written in the
form 
$
F(r,\theta,u,u_{r},u_{\theta}
,u_{r \theta},u_{rr},u_{\theta \theta})=0,
$
(where $u$ denotes the unknown functions) subject to a set of boundary conditions on a rectangular domain.
The user must deliver to FIDISOL the equations, the boundary conditions, and the Jacobian matrices 
for the equations
and the boundary conditions.
Starting with a guess solution, small corrections are computed until a desired accuracy is reached.
FIDISOL automatically provides also an error  estimate for each unknown function, 
which is the maximum of the discretization error divided by the
maximum of the function.

The output of the code was analysed and visualised mainly with MATHEMATICA.

To obtain  axially symmetric solutions, 
we start with the $n=1$ solution 
 as initial guess and increase the value of $n$ slowly.
The iterations converge, and repeating the procedure one obtains
in this way solutions for arbitrary $n$.
The physical values of $n$ are integers.
For some of the configurations, we use these configurations as a starting guess on a finer grid. 
For magnetic monopoles solutions
in a fixed SAdS background, the typical  numerical error 
for the functions is estimated to be lower than $10^{-3}$.

The energy density of these nongravitating
solutions is given by the
$tt$-component of the energy momentum tensor $T_{\mu}^{\nu}$; integration
over all space yields their total mass-energy
\begin{equation}
\label{mass-monopole}
M = \int{\left\{\frac{1}{4} F_{ij}^a F^{aij}
\right\} \sqrt{-g}d^3x} 
=\pi
\int_{r_h}^{\infty} \int_{0}^{\pi/2} dr d \theta~r^2 \sin \theta
F_{ij}^a F^{aij}.
\end{equation}  
We have obtained higher winding number generalizations
for every spherically symmetric configuration. 
Also, the branch structure noticed for $n=1$
is retained for higher winding number solutions.
These solutions have very similar properties with 
the corresponding EYM counterparts
and the general picture we present here applies also in the Section 5.

We have studied a large number of configurations with $n=2,~3,~4$ and node number 
$k=0,~1,~2$
in SAdS backgrounds with $r_h=1$ and various $\Lambda$ between $10^{-3}$ and $10^2$.
It is not easy to extract some general characteristic properties of the solutions, 
valid for every choice of the parameters $(\Lambda,~\omega_0,~n,~k)$.
However, we have found that
the functions $H_1$ and $H_3$ present always a considerable angle-dependence.
With increasing $\omega_0$, the maximal value of the $H_1$ and $H_3$ increase. 
The functions $H_2$ and $H_4$ present a small $\theta$ dependence,  
although the angular dependence generally increases with $Q_M$. 
The qualitative behavior of the gauge functions does not change by changing the 
value of $\Lambda$ (the picture presented in 
Figure 10 for gravitating gauge functions applies in this case, too).

As a typical example, in Figure 1
the gauge functions $H_i$ 
and the energy density $\epsilon$ are
shown  for three different solutions with node number $k=0,1$ 
as a function of the radial coordinate
$r$ at angles $\theta=0,~\pi/4,~\pi/2$.
Here the winding number is $n=3$ and $\Lambda=-0.01,~r_h=1$; 
also the energy density is given in units $4 \pi /e^2$.
We notice that the gauge field function $H_1$ 
remains nodeless and for every solution with $w_0>1$ it takes only negative values
($H_1$ and $H_3$ are zero on the axes in Fig.~1a and 1c).
The behavior of the energy density strongly depends on the
value of $\omega_0$.
For example, as seen in Figure 1e, for the $k=1$ solution, 
the global maximum of the energy density resides on the $\rho$ axis, for a $r>r_h$.
Surfaces of constant energy density for nongravitating solutions 
have a similar form to 
those presented in Section 5 in the presence of gravity and we will not present them here.

The monopole spectrum is presented in Figure 2 for  a
SAdS black hole with $\Lambda=-3,~r_h=1$.
In this case, the existence of one branch of solutions was noticed,
and the monopole mass strongly increases with the magnetic charge.

\subsection{Nongravitating dyon solutions}
Following \cite{Radu:2001ij}, we present here numerical arguments 
for the existence of  finite energy nongravitating YM dyon solutions
in a fixed SAdS background.
The existence of dyon solutions without a Higgs field
is a new feature for AdS spacetime \cite{Bjoraker:2000qd}.
This is a consequence of the 
different asymptotic behavior of the AdS geometries  
as compared to the asymptotically flat case.
If $\Lambda \ge 0$ the electric part of the gauge fields
is forbidden for static configurations \cite{Bjoraker:2000qd,bizon}. 
In order for the boundary conditions at infinity to permit the electric 
fields and maintain a finite ADM mass 
we have to add scalar fields to the theory. 

The YM axially symmetric ansatz (\ref{ansatz}) can be 
generalized to include an electric part   
(see $e. g.$ \cite{Kleihaus:2002ee,Hartmann:2000ja})
\begin{eqnarray} 
\label{electric-pot}
A_t = H_{5}(r,\theta) \frac{\tau_r^n}{2e} +
H_{6}(r,\theta) \frac{\tau_\theta^n}{2e} .
\end{eqnarray}
For the time translation symmetry, we choose a natural gauge 
such that $\partial A/\partial t$=0.

For $A_t \neq 0$, the total energy  (\ref{mass-monopole})
should be supplemented by an electric contribution
\begin{eqnarray} 
\label{em}
E_e &=&\frac{1}{4} \int{ F_{it}^a F^{ait} \sqrt{-g}d^3x}
\\
\nonumber
&=&\pi
\int_{r_h}^{\infty} \int_{0}^{\pi/2} dr d \theta\sin \theta 
\Big(
(r \partial_rH_5+\frac{H_1 H_6}{r})^2+
(r \partial_rH_6-\frac{H_1 H_5}{r})^2
\\
\nonumber
&&+\frac{1}{r^2 F}
\left((\partial_{\theta}H_5-H_2 H_6)^2+
(\partial_{\theta}H_6+H_2 H_5)^2 +n^2(H_6(H_3+\cot \theta)+H_4H_5)^2\right)
\Big).
\end{eqnarray}
We can use  the existence of the Killing vector $\partial/ \partial t$ which implies
$F_{i t} = D_{i}A_t $,
and the YM equations (\ref{YM-eqs}),
to express the "electric mass" (\ref{em})
as a difference of two surface integrals \cite{wald}
\begin{eqnarray} 
\label{electric-mass1}
-E_e=
Tr( \int \{ D_{i}(A_t F^{i t} \sqrt{-g}) -
A_t D_{i}(F^{it} \sqrt{-g}) d^3x \})
=
 \oint_{\infty} Tr \{A_t F^{\mu t} \} dS_{\mu}-
 \oint_{r_h} Tr \{A_t F^{\mu t} \} dS_{\mu}.
\end{eqnarray}
From the conditions of local finiteness of the energy density, we  find  that
the electric potentials necessarily vanish on the event horizon of a static 
black hole (for a rotating background, the situation may be different).
Thus, the electric mass retains only the asymptotic contribution and,
similar to the regular case,
a vanishing magnitude of the electric potentials at infinity $|A_t|$ 
implies a purely magnetic solution.

Since these axially symmetric solutions have a nonzero electric potential,
we may expect them to possess a nonvanishing angular momentum.
The total angular momentum of a nongravitating solution  is given by
\begin{eqnarray}
\label{J}
J=\int T_{\varphi}^{t}\sqrt{-g} d^{3}x
= \int 2Tr\{F_{r \varphi} F^{r t}
+F_{\theta \varphi} F^{\theta t}\} \sqrt{-g} d^{3}x,
\end{eqnarray}
and can be expressed again as a difference of two surface integrals  
\cite{VanderBij:2001nm}
\begin{eqnarray}
\label{totalJ}
J =\oint_{\infty}2Tr\{WF^{\mu t} \} dS_{\mu}
-\oint_{r_h} 2Tr\{WF^{\mu t} \} dS_{\mu}.
\end{eqnarray}
The relations (\ref{em})-(\ref{totalJ}) provide also an 
useful tests to verify the accuracy 
of the numerical calculations.
The electric charge of the dyon solutions is evaluated by using the definition (\ref{charges}). 

\subsubsection{Static dyon solutions}
A possible set of boundary conditions for the electric potentials $H_5, H_6$ is
\begin{eqnarray} 
\label{st1}
  H_{5}|_{r=r_h}= H_{6}|_{r=r_h}=0,
\ \ H_5|_{r=\infty}=u_0,
\ \ H_{6}|_{r=\infty}=0,
\end{eqnarray}
and 
\begin{eqnarray}
\label{st2}
\partial_\theta H_5|_{\theta=0,\pi/2}=H_6|_{\theta=0,\pi/2}=0
\end{eqnarray}
for a solution with parity reflection symmetry.
The magnetic potentials satisfy the same set of boundary conditions valid in 
purely magnetic case.

These solutions can be considered the counterparts of the regular YM dyons 
in a fixed AdS geometry discussed in \cite{Radu:2001ij},
since they satisfy the same boundary conditions at infinity and on the axes.
The presence of a SAdS black hole will affect the properties
of these solutions in the event horizon region only. 
We remark that, although locally $T_{\varphi}^t  \neq 0$, the configurations 
with the symmetries implying the above 
boundary conditions along the axes have always $J=0$. We use the fact that, 
as $\theta \to \pi-\theta$ we have $H_1\to -H_1,~H_2\to H_2,
~H_3\to -H_3,~H_4\to H_4,~H_5\to H_5,~H_6\to -H_6$
which implies the vanishing of the
integral (\ref{J}).

Spherically symmetric dyon solutions (discussed in \cite{Bjoraker:2000qd} for the gravitating case)
are found by taking $n=1$, $H_5=u(r),~H_2=H_4=\omega(r)$ and $H_1=H_3=H_6=0$.
The behavior of the electric potential in a fixed SAdS geometry is
\begin{eqnarray}
\label{static-eh}
u(r) = u_h(r-r_h)+O(r^3),
\end{eqnarray}
at the event horizon and 
\begin{eqnarray}
\label{infinity-e}
u = u_0+\frac{u_1}{r} + O(\frac{1}{r^2})
\end{eqnarray}
at large $r$, where $u_h,~u_0,~u_1$ are constants.
The expansion (\ref{w-hor}),(\ref{w-infinity}) for the magnetic gauge function
$\omega(r)$ is still valid.
These boundary conditions permit non-vanishing charges $Q_M$ and $Q_E$.

As expected, the general properties of these solutions are rather similar to 
the gravitating counterparts.
There are two adjustable shooting parameters defined on the event horizon
 $(u_h,~\omega_h)$.
At the horizon, $u$ starts at zero and monotonically increases asymptotically
to a finite value.
Solutions are found for a continuous set of parameters $u_h$ and $\omega_h$; 
for some limiting values of these parameters, solutions blow up.
Given  $(u_h,~\omega_h)$, the general behavior of the gauge functions $w, u$ 
is similar to 
the gravitating case; there are also solutions with $Q_M=0$ but $Q_E \ne 0$.
Similar to the soliton solutions discussed in \cite{Bjoraker:2000qd},
purely magnetic solutions are obtained by setting $u_h=0$.

When studying dyon solutions we notice the existence always
of higher node ($k>1$) configurations, for suitable values of $(u_h,~\omega_h)$. 
Also, there are  solutions where $\omega$ does not cross the $r$ axis. 
For a fixed value of $\omega_h$, the number of nodes is determined by 
the value of the parameter $u_1$.
Typical spherically symmetric 
solutions are displayed in Fig. 3.

The same numerical method described above is used to obtain
higher winding number static dyon solutions in a fixed SAdS black hole background
(the typical relative error is estimated to be 
of the order of $10^{-3}$).
The spherically symmetric $n=1$ solution gives the input data in the 
numerical iteration 
and it fixes the values $\omega_0,~u_0$ in the boundary conditions at infinity.
Axially symmetric $n>1$ generalizations seem to exist for every
spherically symmetric configuration. 
Here the inclusion of the electric potential 
does not seem to change qualitatively the 
general picture found in the purely magnetic case.
As expected, for a given value of the magnetic charge, the total mass of the
solutions increases with the electric charge.
The shape of the energy density for the monopole solutions is retained 
for the $n>1$ dyon solutions.

As typical axially symmetrical static dyon configurations, we show 
in Figure 4 the radial dependence of 
the gauge functions $H_i$
and the energy density $\epsilon$ (in units $4\pi /e^2$)
for three solutions with the same magnetic charge and magnitude of the electric potential
at infinity and $n=1,~2,~3$.
The results are given for three different angles.
The value of the cosmological constant is $\Lambda=-3$ 
while the event horizon radius
of the SAdS black hole is $r_h=1$.
These solution have been obtained starting from a spherically symmetric 
configurations with the shooting parameters $\omega_h=0.9$ and $u_h=0.03$.
As seen in Fig. 4a-f, 
 the gauge functions $H_2, ~H_4, ~H_5$
do not exhibit a strong angular dependence, 
while the functions $H_1$, $H_3$ and $H_6$ remain nodeless for $\omega_0>0$.
The angular dependence of the matter functions
increases with $n$ and the location of the
biggest angular spliting slightly moves further outward with $n$.
The maximal values of the functions  $H_1,~H_3$ and $H_6$ increase also with $n$, depending
on the value of $\omega_0$.
Similar results have been obtained for other values of $\Lambda$.

\subsubsection{Rotating dyon solutions}
However, 
the  boundary conditions for the electric potentials $H_5$ and $H_6$  (\ref{st1})-(\ref{st2})
do not exhaust all possibilities even for $n=1$. 
Inspired by \cite{Radu:2002rv}, we have searched for solutions with electric potentials
satisfying a
different set of boundary conditions
(the magnetic potentials  satisfiying the same boundary conditions 
as in the purely magnetic case)
\begin{eqnarray} 
\label{cond-rot}
H_{5}|_{r=r_h}&=& H_{6}|_{r=r_h}=0,~~H_5|_{r=\infty}=V \cos \theta,
~~~H_{6}|_{r=\infty}=V \sin \theta,
\\
\nonumber
\partial_\theta H_5|_{\theta=0}&=&H_6|_{\theta=0}=0,
~~H_{5}|_{\theta=\pi/2}= \partial_\theta H_6|_{\theta=\pi/2}=0,
\end{eqnarray}
where $V$ is a constant corresponding to the magnitude of the electric potential 
at infinity.
The value of the electric charge can be obtained from the asymptotics of the
electric potentials, since as $r \to \infty$, 
$H_5\sim\cos \theta \big(V+(c_1 \sin^2 \theta +c_2)/r\big),~
H_6\sim\sin \theta \big(V+(c_3 \sin^2 \theta +c_4)/r\big)$.
We remark that, similar to the regular case, 
dyon solutions with these symmetries  possess a nonzero angular momentum
(since the integral (\ref{J}) is nonvanishing in this case).

As initial guess in the iteration procedure, we use the $n=1$ static 
YM solution in fixed SAdS background discussed above and slowly increased the value of $V$.
The typical relative error for the gauge functions is estimated to be 
on the order of $10^{-3}$, while the relations 
(\ref{electric-mass1}) and (\ref{totalJ}) are satisfied with a very good accuracy. 
All the solutions we present here have winding number $n=1$ and
have been obtained for a SAdS black hole with $\Lambda=-3,~r_h=1$.
However, a similar general behavior has been found
for some other negative values of $\Lambda$.
Several solutions
with $n=2$ have been also studied, possessing very similar properties.

The situation here presents a number of similarities 
to rotating YM dyons in an AdS background \cite{Radu:2002rv}.
For a given SAdS background, we found nontrivial rotating solutions
for every spherically symmetric configuration we have considered.
The solutions depend on two continous parameters: the values $\omega_0$
of the magnetic potentials $H_2,~H_4$ at infinity and the magnitude of the
electric potential at infinity $V$.
A nonvanishing $V$ leads to rotating configurations, with nontrivial functions
$H_1,~H_3,~H_5,~H_6$ and nonzero $Q_E$.
As we increase $V$ from zero while keeping
$\omega_0$ fixed, a branch of solutions forms.
This branch extends up to a maximal value of $V$, which depends on $\omega_0$.
Along this branch, the total energy, electric charge, 
electric part of energy and the absolute value of the angular momentum  
increase continuously with $V$.
This can be seen in Figure 5, where
we present the properties of a typical branch of solutions in a SAdS background, 
for a fixed value of $\omega_0$.
In this picture, the total energy $E$ and the angular momentum $J$ (in units $4\pi/e^2$)
 as well as the electric charge $Q_E$ and the ratio
$E_{e}/E$ are shown as a function of the parameter $V$.

Depending on $V$, the energy of a rotating dyon can be several orders of
magnitude greater than the energy of
the corresponding monopole solution. 
We find that both $E_{e}/E$ and $J/E^2$  tend to constant values as $V$ is increased.
At the same time, the numerical errors start to increase, we obtain large values for
both $Q_E$ and $E$, and for some $V_{max}$ the numerical
iterations fail to converge. 
In this limit, the total energy and the electric charge diverge, while the magnetic charge
takes a finite value.
It is difficult to find an  accurate value for $V_{max}$, 
especially for large values of $\omega_0$.
Alternatively, we may keep fixed the magnitude of the electric potential at infinity
and vary the parameter $\omega_0$. 

There are also solutions where $Q_M=0$ and $Q_E \neq 0$.
A vanishing $Q_E$ implies a nonrotating, purely magnetic configuration.
However, we find  dyon solutions with vanishing total angular momentum
($J=0$ for some $\omega_0,~V$) which are
not static (locally $T_{\varphi}^t \neq 0$).
For the considered configurations,
we have found that most of the angular momentum in (\ref{totalJ})  
comes from the surface integral at infinity
contribution, the event horizon integral taking small values.

As a typical axially symmetrical configuration, we show 
in Figure 6 three-dimensional plots of  
the gauge functions $H_i$, the magnitude of 
the electric potential $|A_t|=(H_5^2+H_6^2)^{1/2}$,
the local angular momentum $T_{\varphi}^t$
and the energy density $\epsilon$
for a solution with $n=1$, total energy $E=6.72$ (in units $4 \pi /e^2$), 
nonabelian charges 
$Q_{M}=-3.622$ and $Q_{E}=6.833$ 
as a function of the coordinates $\rho$ and $z$.
We can see that both  local energy and angular momentum are localized in a small region 
around the event horizon without being possible to distinguish
any individual component.
For the gauge function, the relevant oscillating region extends at least one order of magnitude
beyond the event horizon.

For some of the configurations we have considered, the surfaces of constant energy
density reveal a curious shape, which is different from the other static solutions considered 
in this paper
or other cases exhibited in the literature (see Figure 6g-i).

In the presence of gravity, the field equations for a rotating dyon
configuration are considerably more complicated, 
the metric ansatz  (\ref{metric}) being supplemented with at least one
extradiagonal term.
However, we expect the gravitating counterparts of these solutions to retain the basic features 
discussed here.

\section{AXIALLY SYMMETRIC SOLUTIONS IN THE PRESENCE OF GRAVITY}
Now we come to the main subject of this paper, by including the backreaction
of the YM fields on the black hole geometry.
In this case, the equations of motion are considerably more complicated
than in the pure YM case.
This is a difficult numerical problem even for asymptotically flat geometries, 
where no exact solution is known yet.

\subsection{Spherically symmetric EYM solutions}
The spherically symmetric EYM black hole solutions presented in  
\cite{Winstanley:1998sn,Bjoraker:2000qd},  have been obtained for a 
Schwarzschild-like line element
\begin{equation}
\label{metric3}
ds^2=- \frac{H(\tilde r)}{p(\tilde r)^2}dt^2 +  \frac{1}{H(\tilde r)} d \tilde r^2 
  + \tilde r^2 \left( d \theta^2 + \sin^2 \theta d\varphi^2 \right),
\end{equation}
where 
\begin{eqnarray}
\nonumber
H(\tilde r) = 1 - \frac{2 \tilde m(\tilde r)}{\tilde r}-
\frac{\Lambda}{3} \tilde r^2 .
\end{eqnarray}
The spherically symmetric horizon resides at radial coordinate
$\tilde r_h$, solution of the equation 
$\tilde r_h-2m(\tilde r_h)-\Lambda \tilde r_h^3/3=0$.
Since we need  the spherically symmetric $n=1$ solution 
as the starting point for the calculation of axially symmetric
configurations, it is useful to reobtain these solutions by using 
a different set of coordinates required by the metric ansatz (\ref{metric}).

By imposing  $l=m$ and the metric functions  $f$ and $m$
to be functions only of the coordinate $r$,
the axially symmetric ansatz  (\ref{metric})
reduces to the spherically symmetric form
\begin{equation}
\label{metric4}
ds^2= -f(1-\frac{\Lambda}{3}r^2) dt^2 
+ \frac{m}{f}  \big(\frac{dr^2}{1-\frac{\Lambda}{3}r^2} + r^2 \left( d \theta^2 
+\sin^2 \theta d\varphi^2 \right) \big). 
\end{equation}
Of course, for the same configuration,
the values of the event horizon radius will differ for 
these different metric parametrizations (\ref{metric3}), (\ref{metric4}).
Comparation of the metric in Schwarzschild coordinates (\ref{metric3})
with the "isotropic" metric (\ref{metric4}) yields
\begin{eqnarray}
\label{co1}
\frac{f(r)}{m(r)} = \frac{r^2}{\tilde r^2},~~~
\frac{dr}{r\sqrt{1-\frac{\Lambda }{3}r^2}} =
\frac{1}{\sqrt{H(\tilde r)}} \frac{d \tilde r}{\tilde r},
~~f(1-\frac{\Lambda}{3}r^2)=\frac{H(\tilde r)}{p(\tilde r)^2}.
\end{eqnarray}
Since the mass function $\tilde m(\tilde r)$ is only known numerically,
we have to numerically integrate the above relations
to obtain the coordinate function  $r(\tilde r)$ and the metric functions $m(r),~f(r)$.
This is a direct generalization of the method used 
in \cite{Kleihaus:1997ws} for $\Lambda=0$ EYM solutions 
and employed also in \cite{Radu:2001ij} for regular EYM 
regular with negative cosmological constant.

However, we  have found more convenient to solve directly the $n=1$ EYM equations
by using the line element (\ref{metric4}).
The field equations read for this parametrization
\begin{eqnarray}
\nonumber
\omega''
-\omega'\left(\frac{m'}{2m}-\frac{f'}{f}+\frac{2\Lambda}{3}\frac{r}{N}\right)
-\frac{\omega(\omega^2-1)}{r^2N}&=&0,
\\
\label{eyms}
f''-\frac{f'^2}{f}+\frac{f'm'}{2m}+\frac{2f'}{r}
-\frac{4f^2}{mr^2}\left(\omega'^2+\frac{1}{2N}\frac{(\omega^2-1)^2}{r^2}\right)
-\frac{\Lambda}{N}\left(2f-2m+\frac{2r}{3}(f'+\frac{m'f}{2m})\right)&=&0,
\\
\nonumber
m''-\frac{m'^2}{2m}+\frac{3m'}{r}-\frac{4\omega'^2f}{r^2}
-\frac{\Lambda}{N}\left(r m'+4m-\frac{4m^2}{f}\right)&=&0,
\end{eqnarray}
and are solved with a set of boundary conditions on the event horizon
\begin{eqnarray}
\label{c1}
m=m_2(r-r_h)^2+O(r-r_h)^3,~~
f=f_2(r-r_h)^2+O(r-r_h)^3,~~
\omega=\omega_0+O(r-r_h)^2.
\end{eqnarray}
Asymptotically we find 
\begin{eqnarray}
m=1+\frac{m_1}{r^3}+O(\frac{1}{r^5}),~~
f=1+\frac{f_1}{r^3}+O(\frac{1}{r^5}),~~
\omega=\omega_0+\frac{\omega_1}{r}+O(\frac{1}{r^2}).
\end{eqnarray}
Using these boundary conditions, the equations (\ref{eyms})
were integrated  for $r_h=1$  and various values of $\Lambda$.
Again, only values of $\omega_h>0$ have been considered.
The value $\omega_h=0$ corresponds to a Reissner-Nordstr\"om-AdS solutions,
while $\omega_h=1$ is the vacuum SAdS solution (although these solutions
do not have a simple form in this coordinate system).

As expected, the solutions obtained for the metric ansatz 
(\ref{metric4}) present
similar properties to those found for the Schwarzschild-like 
line-element (\ref{metric3}).
Also, all the basic features of the spherically symmetric 
nongravitating solutions discussed in Section 4.1.
survive in the presence of gravity.

\subsection{Properties of the axially symmetric EYM  solutions}

To find axially symmetric solutions of
the purely magnetic EYM equations, we have used 
the same numerical algorithm as for the YM solutions 
in a fixed SAdS background  presented above.
We have started always with a $n=1$ EYM solution as initial 
guess and increase the value of $n$ slowly (for a fixed $\omega_0$), 
until approaching the integer physical values. 
For every $n$, corrections are calculated successively,
until the numerical solutions satisfies a given tolerance.
The typical  numerical error for the gravitating solutions is 
of the order of $10^{-3}$, except for some $n=3$  or $k=2$ solutions and most of those with $n=4$,
where we found an error on the order of $10^{-2}$.
This error depends on the magnetic charge and mass of the solutions.
Axially symmetric generalizations of the $n=1$ solutions with a large ratio $M(\omega_0)/M(\omega_0=1)$
are difficult to obtain, with large numerical errors. 
Also, for some of the higher node solutions the convergence is slowed down 
and the errors are not small enough for the results to be reliable.
A set of $\Lambda=0$ test runs was carried out, 
primarily designed to evaluate the code's ability to reproduce the Kleihaus-Kunz results.
In this case, we have obtained a very good agreement with the results of
\cite{Kleihaus:1997ws}.

The axially symmetric solutions depend on three continuous
parameters $(r_h,~\Lambda,~\omega_0)$
as well as two integers: the winding number $n>1$ and the node number $k$.
We have considered solutions with $r_h=1$, several negative values of $\Lambda$, 
the node number $n=2,3,4$ and $k \leq 2$.
Special care has been paid to the properties of the
nodeless solutions  because these configurations are likely 
to be stable against linear perturbations.

Once we have a solution, the horizon variables such as $T_H,A$ are calculated in a 
straightforward way from (\ref{temp}),~(\ref{area}). 
The mass of the solution is computed by using the relation (\ref{ct-mass}), extracting the 
values of the coefficients $f_1,f_2$ from the asymptotics of the metric functions
(we used also the asymptotic relation (\ref{test})
to test the accuracy of our results).
 
The behavior of the solutions is 
in many ways similar to that of the axially symmetric solitons
discussed in \cite{Radu:2001ij}.
Again, starting from a spherically symmetric configuration 
we obtain higher winding number generalizations 
with many similar properties.
Axially symmetric generalizations seem to exist for every
spherically symmetric black hole solution. 
For a fixed winding number, the solutions can also be indexed 
in a finite number of branches classified
by the mass and the non-Abelian magnetic charge.
These branches generally follow the picture found for $n=1$ 
(with higher values of mass, however).
This is in sharp contrast to the $\Lambda=0$ case,
where only a discrete set of solutions is found \cite{Kleihaus:1997ws}.
Also, the Kretschman scalar $K=R_{ijkl}R^{ijkl}$
remains finite for every $(r \geq r_h,~\theta)$.

It is rather difficult to find some general pattern valid
for every considered configuration.
Qualitatively,  the YM field behavior is 
similar to that corresponding
to solutions in a fixed SAdS background.
We notice a similar shape for the functions $H_i$ and also for the 
energy density. 
Similar to $n=1$ case, as long as the YM energy contribution is small enough
(the integral (\ref{mass-monopole}) 
is much smaller than the total mass),
the YM solutions in a fixed SAdS background give a good approximation
to a solution of the EYM equations.
Again, for large enough values of $|\Lambda|$, the derivation of the functions
$H_2$ and $H_4$ from the event horizon value is very small.

The gauge functions $H_2,~H_3,~H_4$  start always at (angle dependent)
nonzero values on the event horizon.
The general picture presented in Fig.~2
is valid in this case too.
Starting with suitable $n=1$ configurations,
we find both solutions where $H_2,~H_4$ cross the $r$ axis and nodeless solutions.
The gauge function $H_2$ is always almost spherically symmetric,
while the gauge functions  $H_1$ and $H_3$ are about one order of magnitude smaller 
than the functions $H_2$ and $H_4$.

For the considered solutions, 
the metric functions $m, ~f, ~l$
do not exhibit a strong angular dependence.
These functions start with a zero value on the event horizon and
approach rapidly the asymptotic values.
We have also observed little dependence of the metric functions on the node number.
The functions  $m$ and $l$ have a rather similar shape, while the ratio $m/l$ indicating
the deviation from spherical symmetry is typically close to one,
except in a region near the horizon.

However, we found that, similar to the asymptotically flat case, the horizon is deformed.
This deformation is parametrized by the ratio $L_e/L_p$, where $L_e$ is the
circumference measured along the equator and $L_p$ is the circumference
measured along the poles
\begin{eqnarray} 
\label{circ}
L_e = \int_0^{2 \pi} { d \varphi 
\left.
\sqrt{ \frac{l}{f}} r \sin\theta \right|_{r=r_{h}, \theta=\pi/2} }
 = 2 \pi r_{h} \left. \sqrt{\frac{l_2}{f_2}}
  \right|_{\theta=\pi/2} 
\ ,
\\
\label{lp}
L_p = 2 \int_0^{ \pi} { d \theta \left.
 \sqrt{ \frac{m  }{f}} r
 \right|_{r=r_{h}, \varphi=const.} }
 = 2 r_{h} \int_0^\pi { d \theta 
 \sqrt{\frac{m_2(\theta)}{f_2(\theta)}} }
\ .   
\end{eqnarray}
For most of the considered solutions, 
this derivation from spherical symmetry is very small, however,
at the limit of the numerical errors.
We have found typically $L_e/L_p \simeq 0.99$ (although the situation
may be different for other values of the event horizon radius).
Similar to the static asymptotically flat case, it remains a challenge to construct
static solutions with a large horizon deformation.

To see the the winding number dependence for a fixed magnetic charge,
we present in Fig.~7 three solutions with $\omega_0=1.73$
and $n=1$, $2$ and $3$. 
In Figs.~7a-d the gauge field functions are shown,
in Figs.~7e-g the metric functions,
and in Fig.~7h the energy density of the matter fields.
These two-dimensional plots exhibit the $r$ dependence
for three angles $\theta=0$, $\pi/4$ and $\pi/2$.
Note that for $\omega_0>1$ the $H_1$, $H_3$ functions remain nodeless
($H_1$ and $H_3$ are zero on the axes in Figs.~7a,c 
as required by the boundary conditions (\ref{b5})).
As expected, the angular dependence of the matter functions
increases with the winding number $n$ and the location of the
biggest angular splitting moves further outward.
Also, we can see  that for every $n$, the global maximum of the energy density
for these black hole solutions resides on the $z-$axis at the event horizon,
while a pronounced minimum develops
on the $\rho-$axis at the horizon.
However, this is not a generic property and we expect the situation to be different
for smaller values of $r_h$ and the same $\Lambda,~\omega_0$.
In fact, even for $r_h=1$ we have found configurations with a maximum of the energy
density along the $\rho$ axis.

In Figure 8 the mass $M$ of this branch of black hole solutions
is plotted as a function of the nonabelian magnetic charge $Q_M$ 
for various winding numbers.
These solutions have been obtained for a cosmological constant $\Lambda=-0.1$.
For an event horizon radius $r_h=1$, this value is of interest 
because the corresponding  solutions combine basic features of the small $|\Lambda|$ case 
(with a large number of branches and node number $k \geq 0$) and large $|\Lambda|$ case
(one branch and nodeless solutions only).
For $\Lambda=-0.1$ and  $n\geq 1$ and we find 
one branch of solutions with both nodeless ($\omega_0>0$) and one-node configurations
(with $\omega_0$ taking negative values up to $\omega_0^{min}\simeq -0.155$ and
extending backwards to zero).
Note also that for the physically more interesting value $\Lambda=-3$, the black hole
configurations with $r_h=1$ present only one branch of nodeless solutions, with 
very small variations 
of the gauge fields outside the event horizon.
A detailed analysis of the static  black hole solutions
as well as rotating regular solutions
with $\Lambda=-3$,
will be presented elsewhere in connection to the $d=11$ supergravity embedding.

Figure 9a shows a typical three-dimensional plot of the  $T_t^t$ component of the
energy-momentum tensor as a function
of the coordinates $\rho$ and $z$,
for a black hole solution with $r_h=1$, $\Lambda=-1$, $Q_M=1.52$, $n=2$ and $k=0$.
In Figures 9b-d surfaces of constant energy density are presented for the same solution.
As expected,
the energy density is not constant at the horizon but angle dependent.
Similar to the asymptotically flat case, 
the surfaces of constant energy density appear ellipsoidal
for small values of $\epsilon$,  being flatter at the poles than in the equatorial
plane.
With increasing values of $\epsilon$, a torus-like shape appears,
with the horizon seen at the center of the torus.
We did not observe a strong dependence of this picture on the value 
of the cosmological constant.

To illustrate the  dependence of the black hole solutions on the value of the cosmological constant,
we show in  Fig.~10 three $n=2$ solutions  
with the same magnetic charge $Q_M=1.52$, obtained for different values of $\Lambda$.
In Figs.~10a-d the gauge field functions are presented,
in Figs.~10e-g the metric functions,
and in Fig.~10h the energy density $-T_t^t$.
We can notice that, for increasing $|\Lambda|$,
the metric functions approach the asymptotic values faster than
in the  limit of small cosmological constant, 
as the magnetic field concentrates near the event horizon.
Given $n,~\omega_0,~r_h$, by decreasing the cosmological 
constant from $0$ to $-\infty$, 
the field variables decrease for fixed $r$.
However, the angular dependence of the metric and matter functions
does not change significantly with $|\Lambda|$
(although the angular dependence of the matter variables on the event 
horizon increases).
The  peak of the energy density 
shifts inward with increasing $|\Lambda|$ and increases in height.

The mass-temperature diagrams for black hole monopole solutions at 
$\Lambda=-0.1,~r_h=1$ and three winding numbers are plotted in Figure 11.
The magnetic potential at infinity $\omega_0$ 
varies along each curve.
The higher branches in this picture corresponds to values of $\omega_0<1$, 
while $\omega_0>1$ for the lower branches.
The common point of these curves corresponds to the SAdS solution with $\omega_0=1$
\footnote{
For any value of $n$, the solution with  $\omega_0=1$
corresponds to a trivial gauge configuration with $H_1=H_3=0$,
$H_2=H_4=1$ $i.e.~F_{\mu \nu}=0$ and a SAdS geometry.}.
Thus, it seems that the Hawking temperature of a EYM system appears to be suppressed relative
to that of a vacuum black hole of equal horizon area
(this result can be proven analytically for $n=1$ only \cite{VanderBij:2001ia}),
while the mass always increases.

We don't address here the problem of limiting solutions,
which is still unclear even in the spherically symmetric case.
Axially symmetric 
generalizations of the spherically
symmetric configurations near the critical solutions
are difficult to obtain, 
with large errors for the gauge functions $H_1$ and $H_3$.

\section{CONCLUSIONS}
In this paper we have presented numerical arguments that
EYM theory with a negative cosmological constant possesses
black hole static axially symmetric solutions.
These static configurations
are asymptotically AdS, possesing a regular event horizon.
They generalize to higher winding number 
the known spherically symmetric solutions,
presenting angle-dependent fields at the horizon.

We started by presenting arguments that YM-SU(2) theory possess
solutions with nonvanishing magnetic and 
electric charges and arbitrary winding number
for a SAdS black hole fixed background.
The spherically symmetric solutions 
we found have  properties similar to the lower branch of
their known gravitating counterparts.
Axially symmetric YM configurations with nonzero electric potentials 
have been considered as well, some of them
presenting a nonvanishing total angular momentum.

When including gravity,
we have presented results suggesting 
the existence of axially symmetric EYM black hole solutions
with negative cosmological constant.
Like their regular counterparts,
these black hole configurations have continuous values of mass and non-Abelian 
magnetic charge and present a branch structure.

We have not  considered  the question of stability for higher winding number solutions.
However, since some of the static spherically symmetric black hole solutions 
are stable \cite{Winstanley:1998sn},
there is all reason to believe, that there are also static axially symmetric black 
 stable against small time-dependent perturbations.
A rigorous proof is however desirable, 
analogous to the proof given for the spherically symmetric case.

Axially symmetric EYM solutions with a different topology of the event horizon, generalizing for
$n>1$ the known topological black holes \cite{VanderBij:2001ia} are also likely to exist.
Rotating configurations should exist as well,
leading to nonabelian counterparts of the Kerr-Newman-AdS abelian solution. 
Here we expect the situation to be more complicated
as compared to the asymptotically flat case.
For $\Lambda \neq 0$ the boundary solutions   in the asymptotic region
are less restrictive allowing for a nonvanishing
value of the electric potential at infinity  
which complicates the general picture.

In fact we expect all known asymptotically flat configurations to generalize for a
negative cosmological constant.
It would be interesting to construct AAdS nonabelian solutions
which possess only discrete symmetries \cite{Kleihaus:2003tn} 
and to find the corresponding 
boundary stress tensor.
Their interpretation in AdS/CFT context is a challange.


\section*{Acknowledgements}
The work of ER was supported by Graduiertenkolleg of the Deutsche
Forschungsgemeinschaft (DFG): Nichtlineare Differentialgleichungen;
Modellierung, Theorie, Numerik, Visualisierung and Enterprise--Ireland Basic
Science Research Project SC/2003/390 of Enterprise-Ireland.
The work of EW was partially supported by the Nuffield Foundation, reference NUF-NAL/02.


\vspace{1.cm}
\appendix

\textbf{\Large Appendix}
\\
\\
We consider spacetimes whose metric can be written locally in the form
\begin{eqnarray}
\label{metrica}
ds^{2}=e^{2C(r)}dr^2+e^{2A(r)-2B(r)}d \Omega_k^2-
e^{2B(r)} dt^{2},
\end{eqnarray}
where $d \Omega_k^2=d\theta^{2}+f^{2}(\theta) d\varphi^{2}$
is the metric on a two-dimensional surface  of constant curvature $2k$.
The discrete parameter $k$ takes the values $1, 0$ and $-1$ 
and implies the form of the function $f(\theta)$
\begin{eqnarray}
f(\theta)=\left \{
\begin{array}{ll}
\sin\theta, & {\rm for}\ \ k=1 \\
\theta , & {\rm for}\ \ k=0 \\
\sinh \theta, & {\rm for}\ \ k=-1.
\end{array} \right.
\end{eqnarray}
In these solutions, the topology of the two-dimensional space $t=const.$, $r=const.$
depends on the value of $k$.
When $k=1$, the metric takes on the familiar spherically symmetric form,
for $k=-1$
the $(\theta, \varphi)$ sector is a space with constant negative curvature
(also known as a hyperbolic plane),
while for $k=0$, this is a flat surface.
For any value of $k$, the metric (\ref{metrica})  
has four Killing vectors, one timelike and three spacelike.

The most general expression for the appropriate  SU(2)  connection 
is obtained by using the standard rule 
for calculating the gauge potentials for any spacetime group 
\cite{Forgacs:1980zs}.
Taking into account 
the symmetries of the line element (\ref{metrica}) we find \cite{VanderBij:2001ia}
\begin{eqnarray} \label{A}
A=\frac{1}{2g} \left\{ u(r,t) \tau_3 dt+ \chi(r,t) \tau_3 dr+
\left( \omega(r,t) \tau_1 +\tilde{\omega}(r,t) \tau_2\right) d \theta
+\left(\frac{d \ln f}{d \theta} \tau_3
+ \omega(r,t) \tau_2-\tilde{\omega}(r,t)\tau_1  \right) f d \varphi \right \}.
\end{eqnarray}
 
For purely magnetic, static configurations ($i.e.$ $u=0$) 
it is convenient to take the $\chi=0$ gauge
and eliminate $\tilde{\omega}$ by using a residual gauge freedom. 
The remaining function $\omega$ depends only on the coordinate $r$.
As a result, we obtain the YM curvature  
\begin{eqnarray} 
F=\frac{1}{2g}\left (
\omega' \tau_1 dr\wedge d\theta +
f \omega' \tau_2 dr\wedge d\varphi +
(w^2-k)f \tau_3 d\theta \wedge d\varphi \right ),
\end{eqnarray}
where a prime denotes a derivative with respect to $r$.

Inserting this ansatz into the
action (\ref{lag0}), integrating and dropping the surface term,
we find that the equations of motion can be derived 
from an effective action whose Lagrangian is given by
(the dependence on the coupling constants $G,e$ is eliminated by taking the
rescaling 
(\ref{resc}))
\begin{eqnarray}                   
\label{lag}
L=G_{ik}(y){ dy^i\over  dr} { dy^k\over  dr} - U(y),
\end{eqnarray}
where $y^i= (A,B,\omega)$ and
$G_{ik}=e^{2A-B-C} {\rm diag}(1,-1,-2 e^{-2A+2B})$
and 
\begin{eqnarray}                   
\label{U}
U= -ke^{B+C}
+\Lambda e^{2A-B+C}+ (\omega^2-k)^2 e^{-2A+3B+C}.
\end{eqnarray}
The field equation for the variable $C$ implies the contraint
\begin{eqnarray}    
-{\rm e}^{2A-B-C} (A'^2-B'^2)+ke^{B+C}+2\omega'^2e^{B-C}
-\Lambda e^{2A-B+C}- (w^2-k)^2 e^{-2A+3B+C}=0.
\end{eqnarray}
We remark that (\ref{lag}) allows for the reparametrization
$r \to \tilde{r}(r)$ which is unbroken by our ansatz.

We are interested in finding a superpotential $W$ such that  $U$ to satisfies the relation
\begin{eqnarray} 
\label{UW} 
U=-G^{ik}\frac{\partial W}{\partial y^i}\frac{\partial W}{\partial y^k},
\end{eqnarray}
which allows the first order Bogomol'nyi equations 
\begin{eqnarray}
\label{BPS}
\frac{dy^i}{dr}=G^{ik}\frac{\partial W}{\partial y^k},
\end{eqnarray}
solving also the second-order EYM equations (\ref{einstein-eqs}), (\ref{YM-eqs}).

In the absence of the YM fields, the superpotential $W$ has the simple form
\begin{eqnarray}
\label{Wg}
W=\sqrt{ke^{2A}+e^{4A-2B}},
\end{eqnarray}
leading to pure-AdS vacuum solutions.

For $\Lambda=-3$ and vanishing curvature ($k=0$) 
of the two-dimensional space $r=const.,~t=const.$,
it is possible to find a simple form of the superpotential
\begin{eqnarray}
\label{Wk=0}
W_0= e^{B}w^2+ e^{2A-B},
\end{eqnarray}
leading to a simple solution of the EYM equations.
The line element reads 
\begin{equation}
\label{solk=0;1}
ds^{2}=dr^2+\frac{e^{-2r}}{2}(B_0e^{4 r}- \omega_0^2)d\Omega_0^2-
\frac{2e^{6 r}}{B_0e^{4 r}- \omega_0^2}dt^2  ,  
\end{equation}
(where $\omega_0,B_0$ are arbitrary real constants; $B_0>0$ for a solution with the right signature),
while the gauge potential is
\begin{eqnarray}
\label{solk=0;2}
\omega(r)=\omega_0e^{-gr}.
\end{eqnarray}
However,  we find that, for every choice of the integration constants,
the line element  (\ref{solk=0;1})  presents 
some unphysical properties.
A direct computation reveals that the point $r=r_0$ 
(with $e^{4 r_0}= \omega_0^2/B_0$) is a curvature singularity
(this can easily be seen by computing the invariant $R_{\mu \nu}R^{\mu \nu})$,
which is not hidden by an event horizon. As $r \to r_0$, the metric potential 
$g_{tt}$ diverges, while 
the other metric functions as well as the gauge field remain finite.
Also, the line element (\ref{solk=0;1}) approaches asymptotically the AdS background.

Following the rules in \cite{Pope:1985bu}, this configuration can be 
uplifted to become a solution
of the equations of motion for the $d=11$ supergravity.
Similar solutions are very likely to exist for $k=\pm 1$.
%
%
%
%
%
%
%
%
%
%
%
%
%
%
%
%
%
%
%
%
%
%
%
%
%
%
%
%
%
%
%
%

\newpage
\setcounter{fixy}{1}

\begin{fixy} {0}

\begin{figure}\centering
{\large Fig. 1a} \vspace{0.0cm}\\
\epsfysize=8.9cm
\mbox{\epsffile{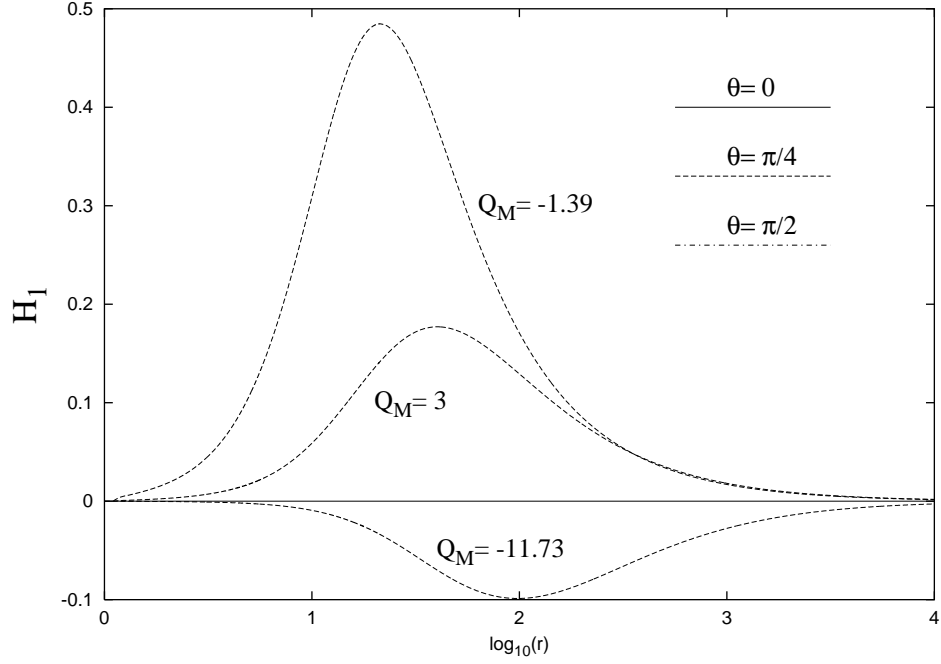}}
\caption{
The gauge function $H_1$ 
is shown as a function of the radial coordinate
$r$ for the angles $\theta=0$, $\pi/4$ and $\pi/2$ for three $n=3$
axially symmetric monopole solutions
in a fixed SAdS background with $\Lambda=-0.01,~r_h=1$.
The magnetic charge and total mass of these YM solutions are
($Q_M=-1.39$, $M=1.09$), ($Q_M=3$, $M=0.33$) and ($Q_M=-11.73$, $M=0.91$).
} 
\end{figure}

\begin{figure}
\centering
{\large Fig. 1b} \vspace{0.0cm}
\\
\epsfysize=8.9cm
\mbox{\epsffile{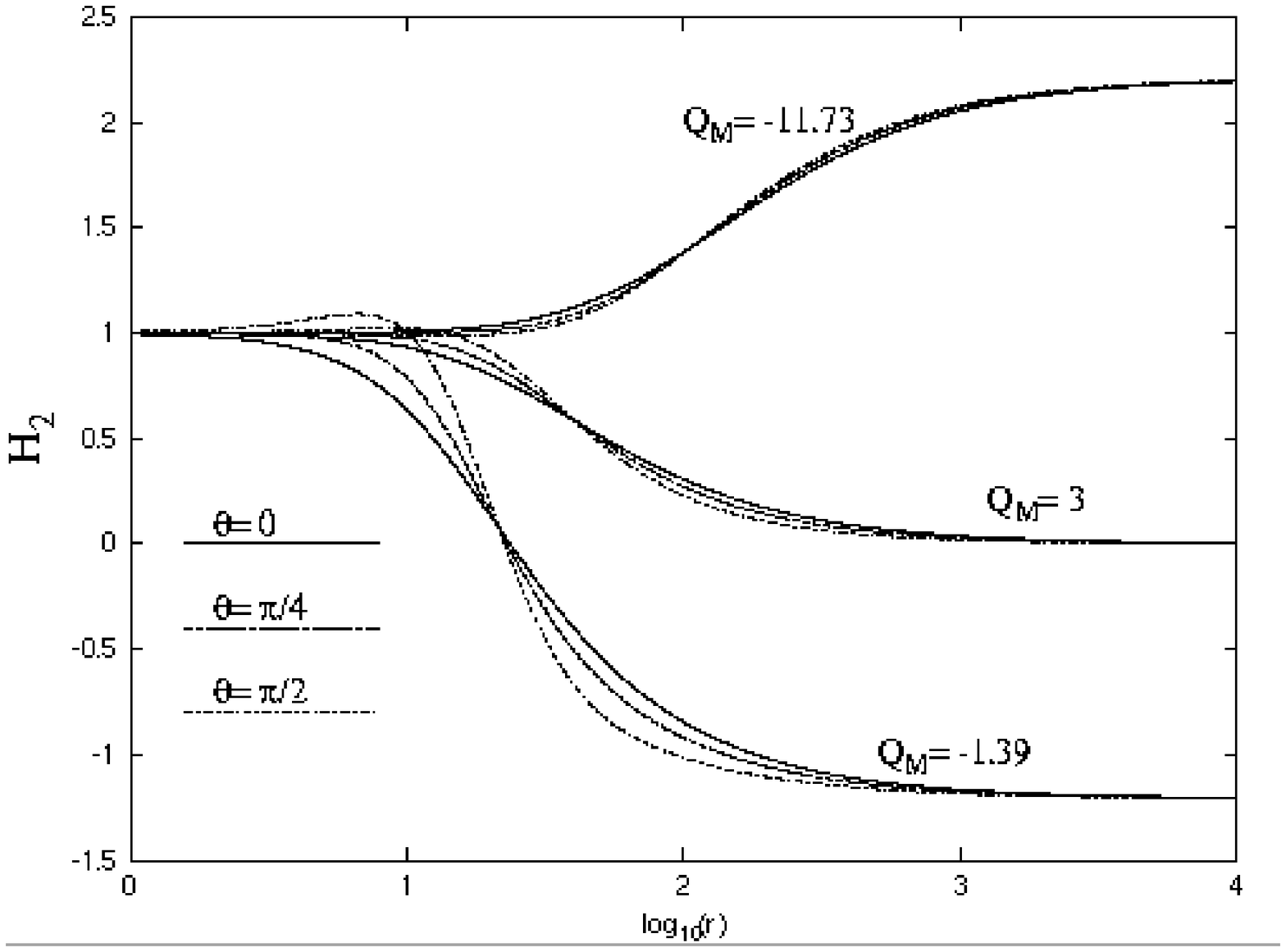}}
\caption
{
Same as Fig.~1a for the gauge function $H_2$.
} 
\end{figure}

\clearpage
\newpage

\begin{figure}
\centering
{\large Fig. 1c} \vspace{0.0cm}
\\
\epsfysize=9.5cm
\mbox{\epsffile{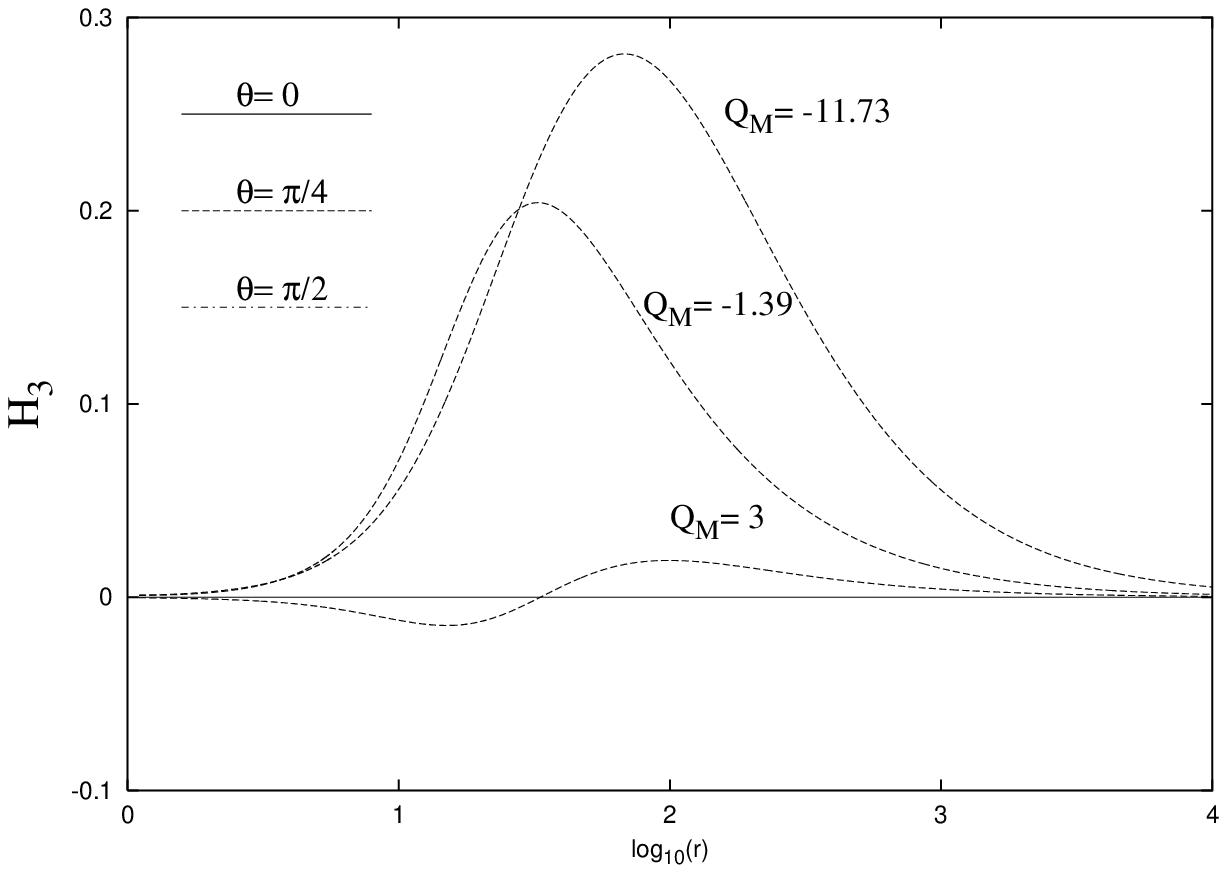}}
\caption
{
Same as Fig.~1a for the gauge function $H_3$.
} 
\end{figure}

\begin{figure}
\centering
{\large Fig. 1d} \vspace{0.0cm}
\\
\epsfysize=9.5cm
\mbox{\epsffile{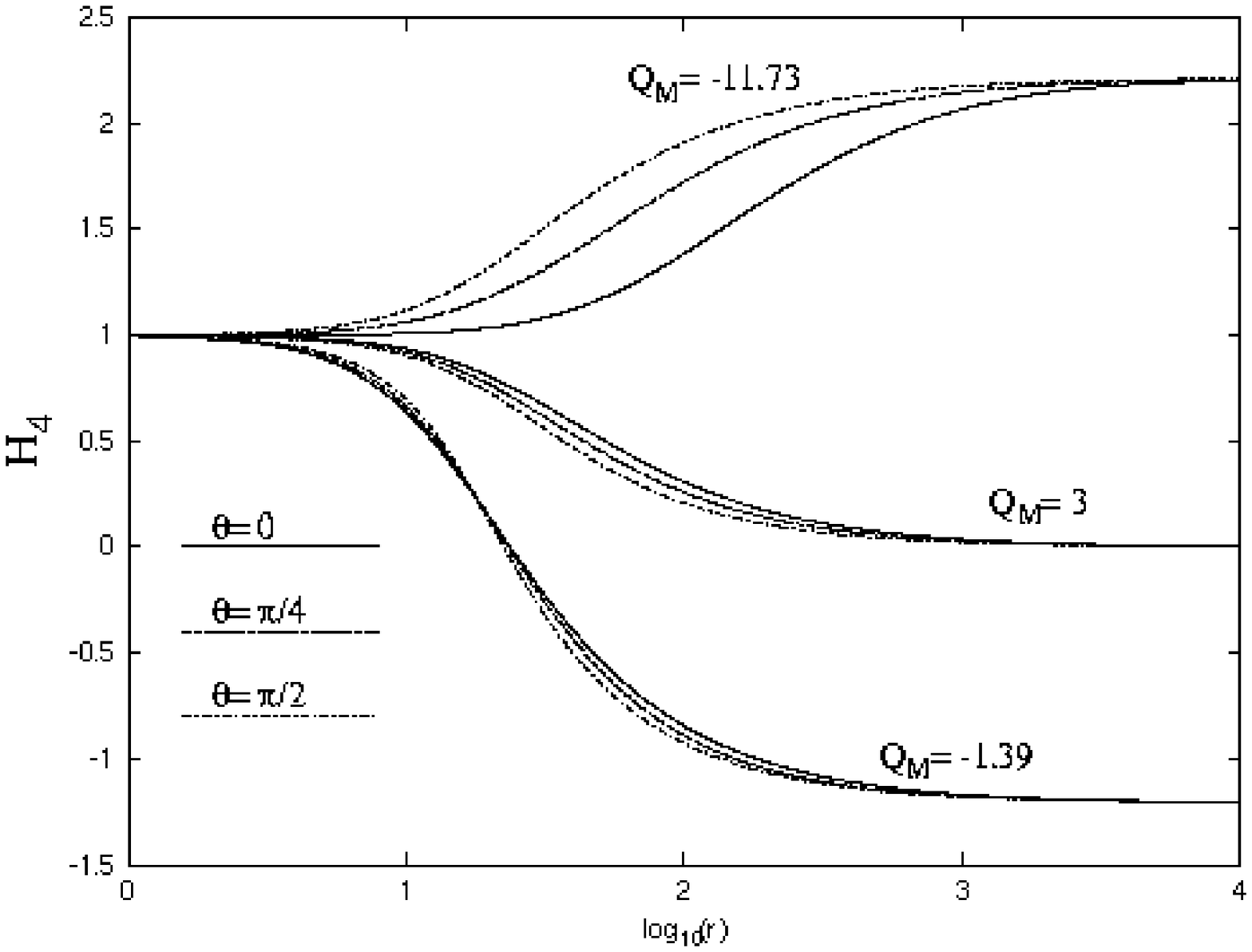}}
\caption
{
Same as Fig.~1a for the gauge function $H_4$.
}
\end{figure}

\clearpage
\newpage

\begin{figure}
\centering
{\large Fig. 1e} \vspace{0.0cm}
\\
\epsfysize=9.3cm
\mbox{\epsffile{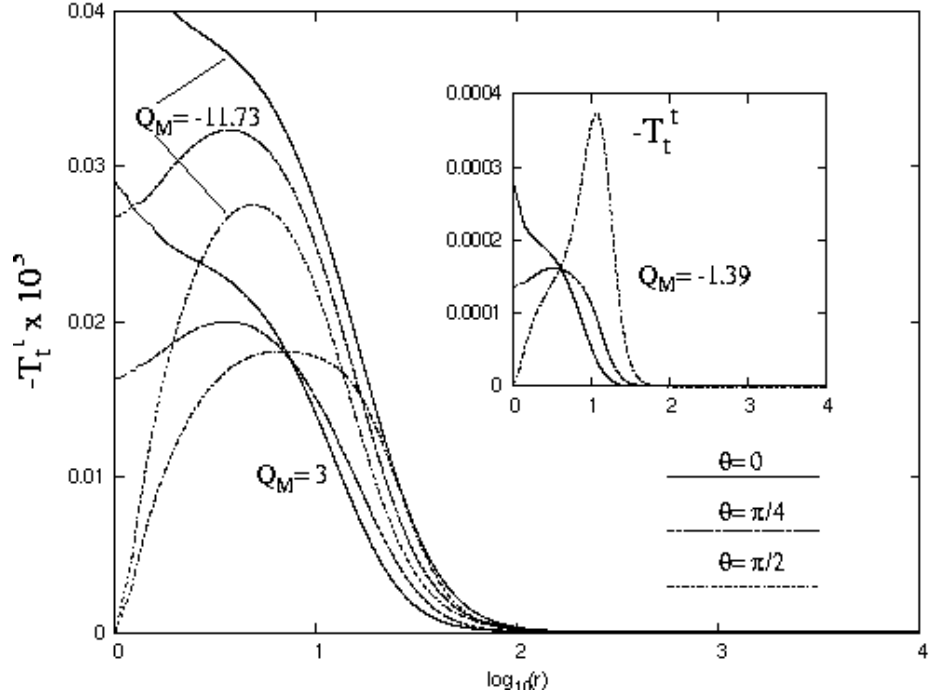}}
\caption
{
Same as Fig.~1a for the energy density $-T_t^t$ (in units $4 \pi /e^2$).
} 
\end{figure}

\end{fixy}


 \begin{fixy}{-1}
\begin{figure}
\centering
{\large Fig. 2} \vspace{0.0cm}
\\
\epsfysize=9.cm
\mbox{\epsffile{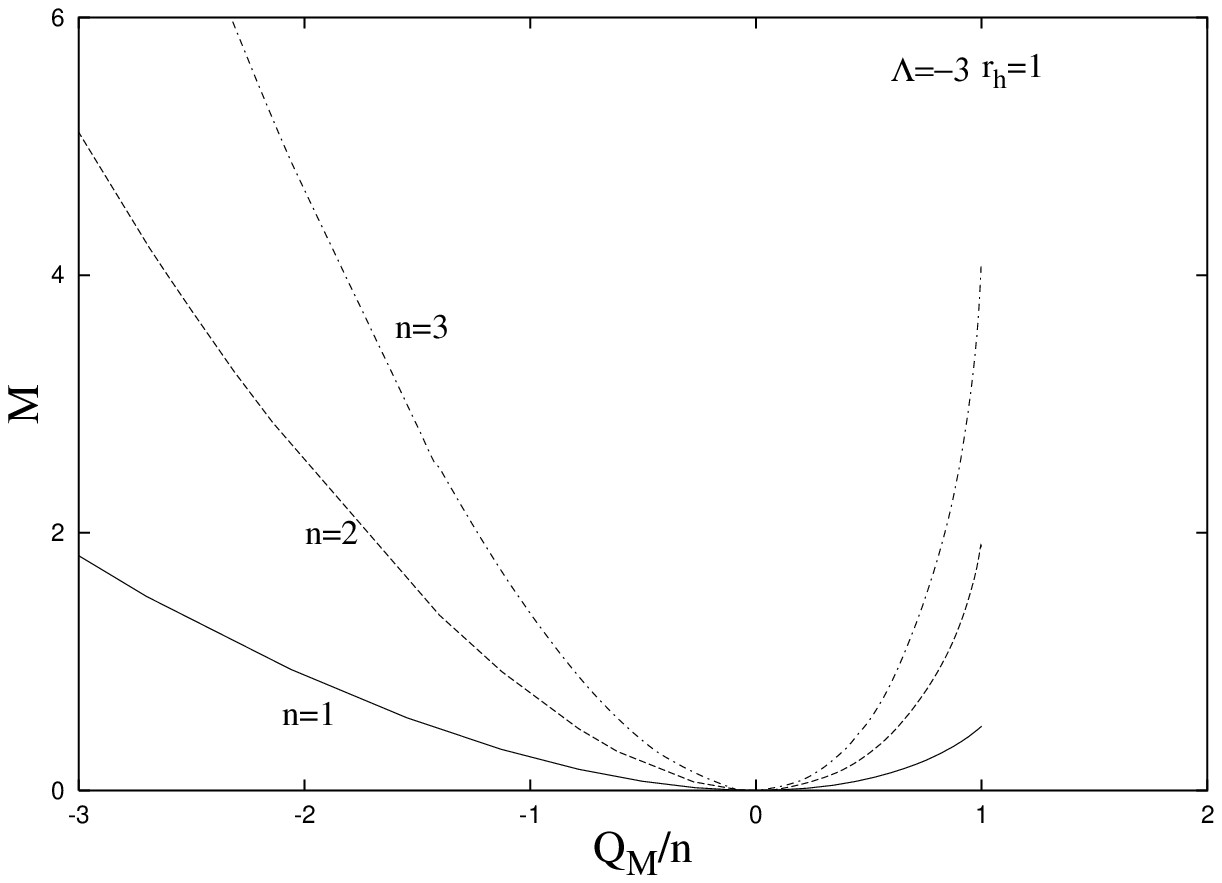}}
\caption
{
The total mass $M$ (in units $4 \pi /e^2$) is plotted as a function of magnetic charge $Q_M$
for monopole solutions in a fixed Schwarzschild-anti de Sitter
background with $\Lambda=-3,~r_h=1$.
The winding number $n$ is also  marked.
} 
\end{figure}
\end{fixy}


\clearpage
\newpage

\begin{fixy}{-1}
\begin{figure}
\centering
{\large Fig. 3} \vspace{-0.0cm}
\\
\epsfysize=8.5cm
\mbox{\epsffile{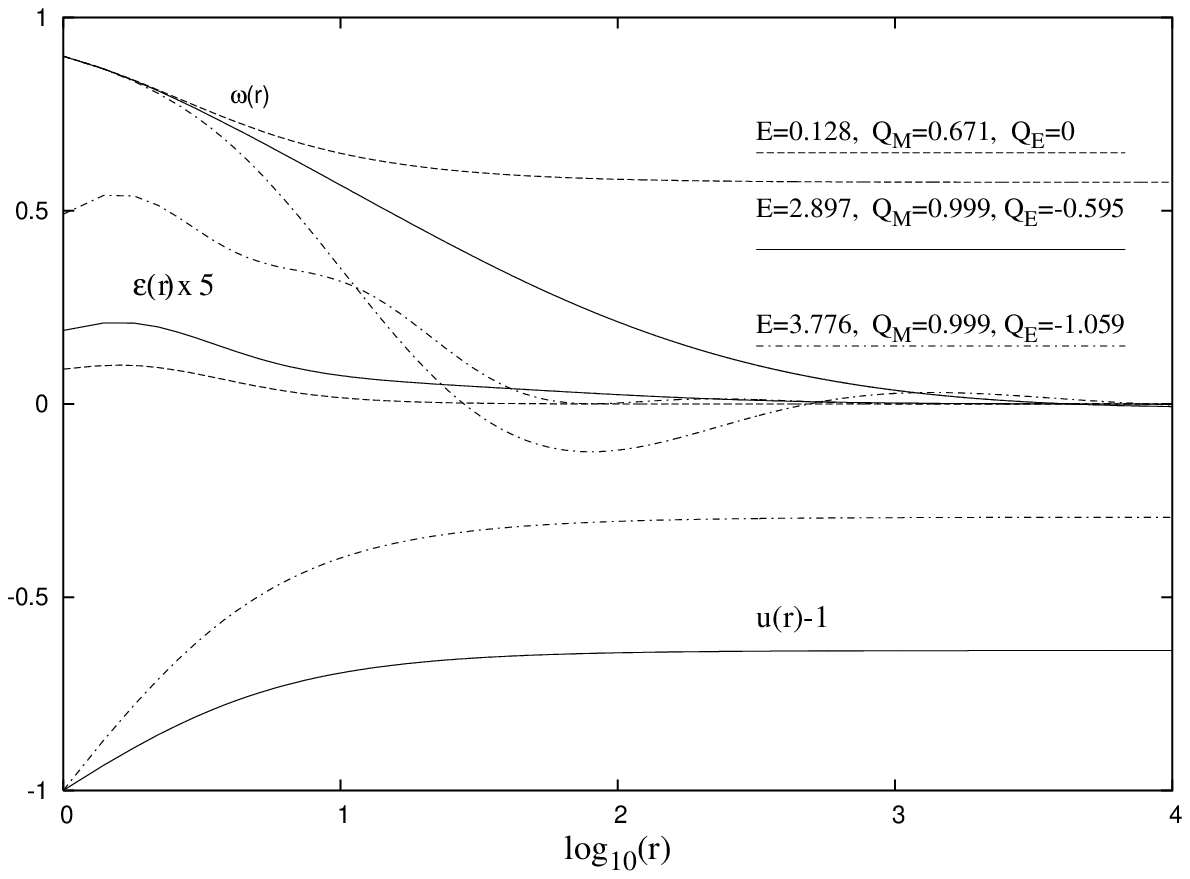}}
\caption
{
Typical nongravitating spherically symmetric dyon
solutions in a SAdS
background, for $\Lambda=-1,~r_h=1$,
the same value of $\omega_h=0.9$ and $u_h=0,~0.2,~0.4$.
The solution with $u_h=0$ corresponds to a magnetic monopole. 
The energy density $\epsilon(r)$ is given in units  $4 \pi /e^2$.
} 
\end{figure} 
\end{fixy}


\begin{fixy}{0}
\begin{figure}
\centering
{\large Fig. 4a} \vspace{0.0cm}
\\
\epsfysize=8.5cm
\mbox{\epsffile{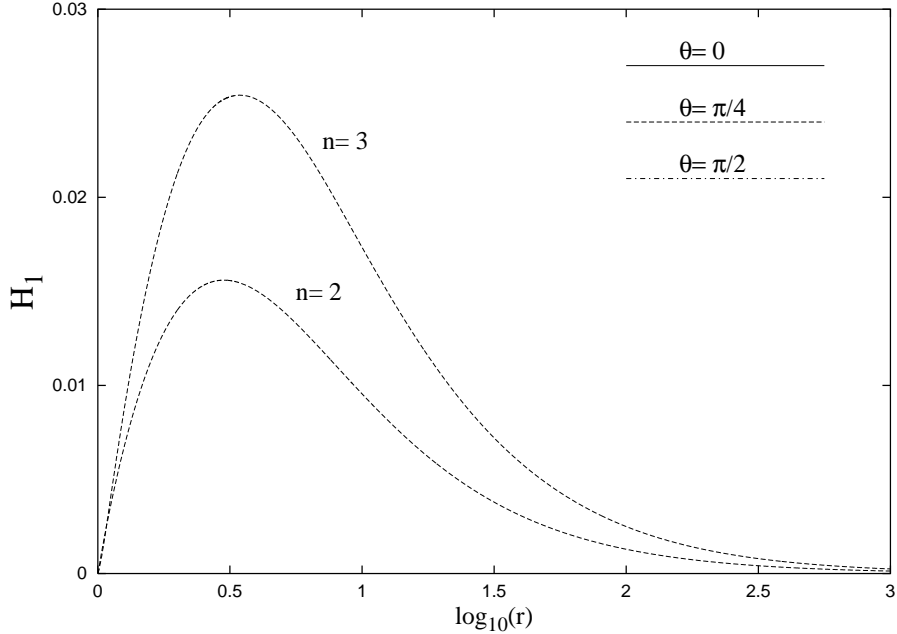}}
\caption
{
The gauge function $H_1$ 
is shown as a function of the radial coordinate
$r$ at $\theta=0$, $\pi/4$ and $\pi/2$ 
for three different static dyon solutions.
Here the solutions have winding numbers $n=1$,~2, 
and 3, the same ratio $Q_M/n=-0.391$,
the same value of the electric potential at 
infinity $u_0=0.04$ and masses
$M(n=1)= 1.541$, $M(n=2)= 1.714$ and  $M(n=3)=2.071$.
} 
\end{figure} 

\clearpage
\newpage

\begin{figure}
\centering
{\large Fig. 4b} \vspace{0.0cm}
\\
\epsfysize=9.5cm
\mbox{\epsffile{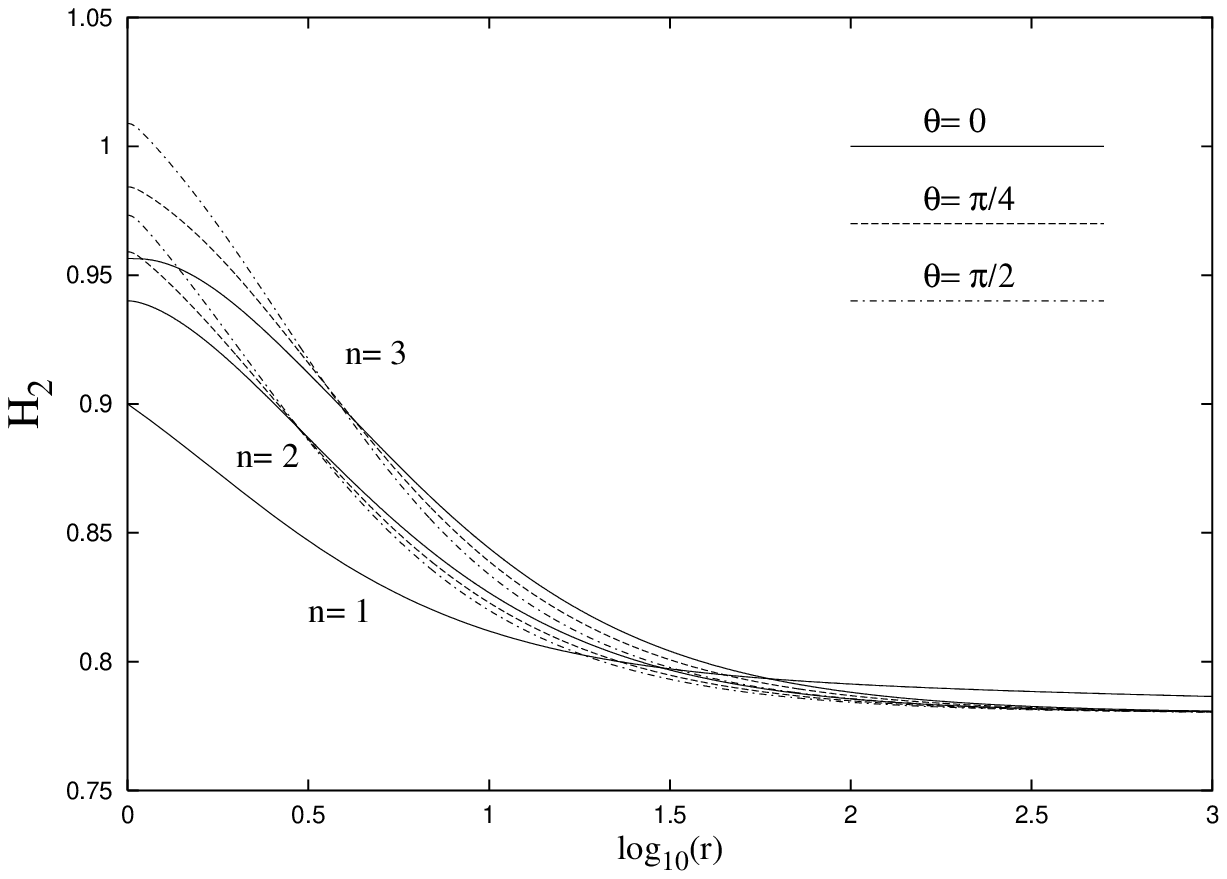}}
\caption
{
Same as Fig.~4a for the gauge function $H_2$.
} 
\end{figure} 

\begin{figure}
\centering
{\large Fig. 4c} \vspace{0.0cm}
\\
\epsfysize=9.5cm
\mbox{\epsffile{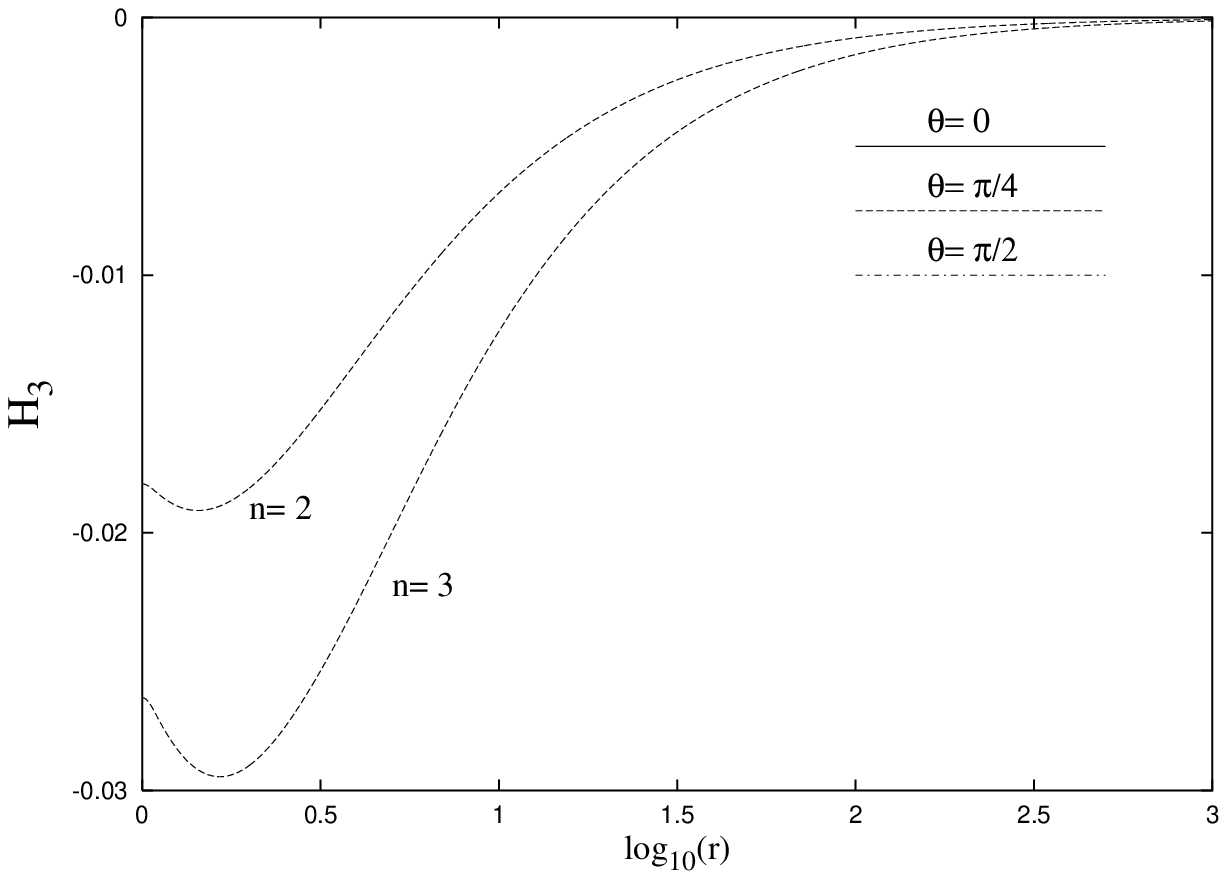}}
\caption
{
Same as Fig.~4a for the gauge function $H_3$.
}
\end{figure} 

\clearpage
\newpage
\begin{figure}
\centering
{\large Fig. 4d} \vspace{0.0cm}
\\
\epsfysize=9.5cm
\mbox{\epsffile{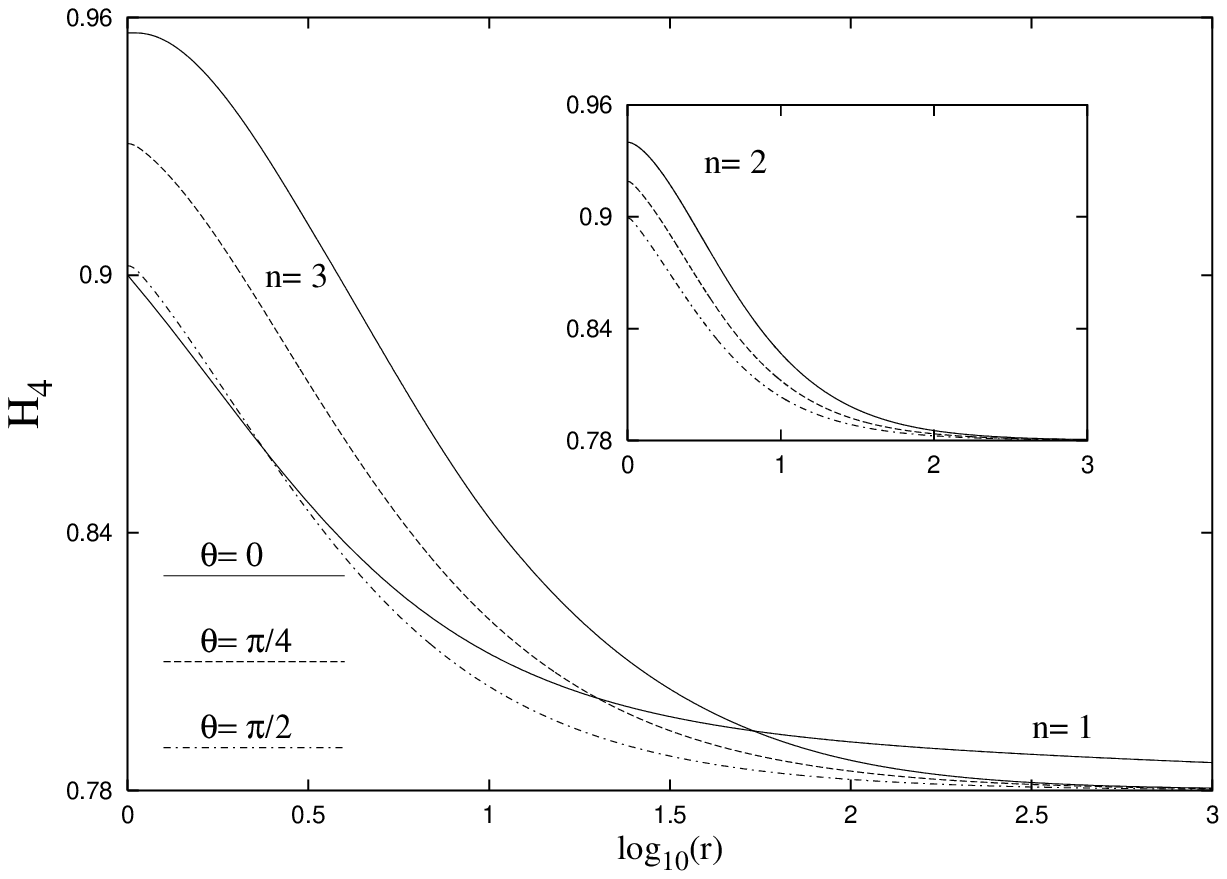}}
\caption
{
Same as Fig.~4a for the gauge function $H_4$.
}
\end{figure} 

\begin{figure}
\centering
{\large Fig. 4e} \vspace{0.0cm}
\\
\epsfysize=9.5cm
\mbox{\epsffile{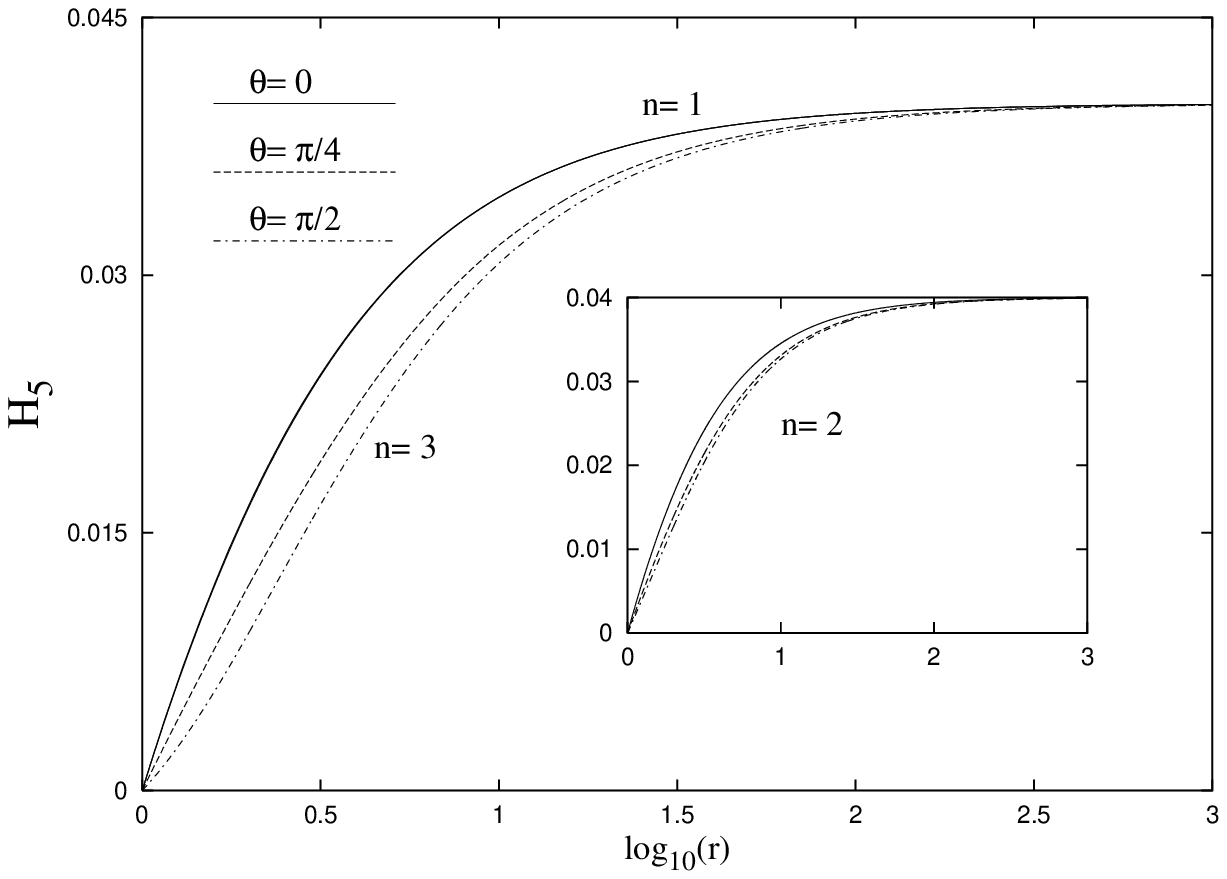}}
\caption
{
Same as Fig.~4a for the gauge function $H_5$.
} 
\end{figure} 

\clearpage
\newpage
\begin{figure}
\centering
{\large Fig. 4f} \vspace{0.0cm}
\\
\epsfysize=9.5cm
\mbox{\epsffile{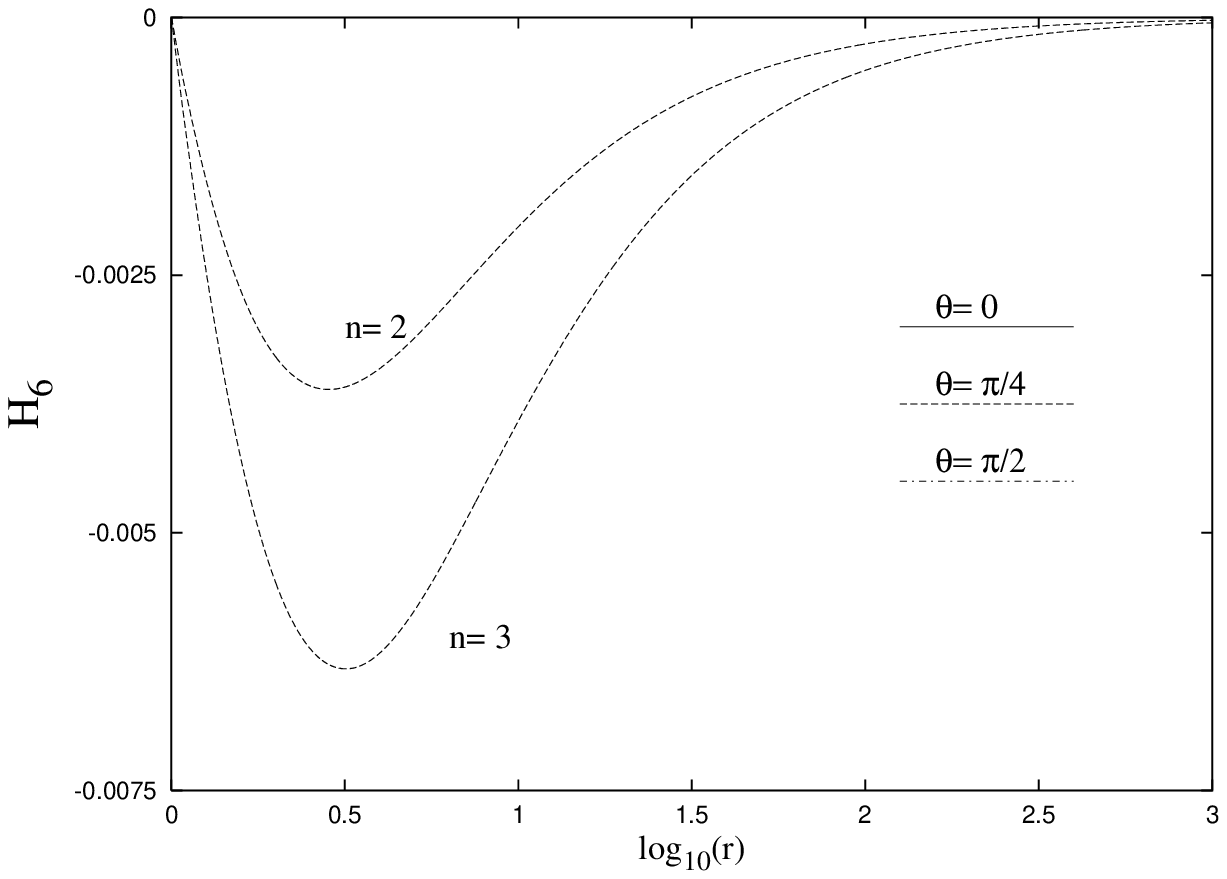}}
\caption
{
Same as Fig.~4a for the gauge function $H_6$.
} 
\end{figure} 

\begin{figure}
\centering
{\large Fig. 4g} \vspace{0.0cm}
\\
\epsfysize=9.5cm
\mbox{\epsffile{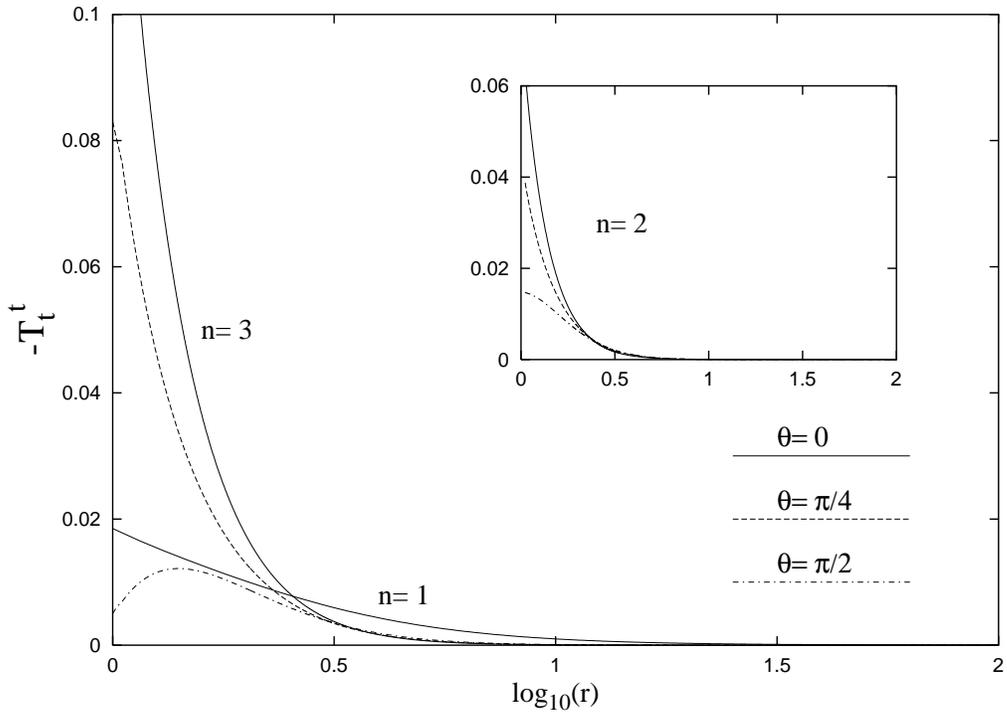}}
\caption
{
Same as Fig.~4a for  the energy density $\epsilon=-T_t^t$
(in units $4 \pi /e^2$).
} 
\end{figure}
\end{fixy}


\clearpage
\newpage

\begin{fixy}{-1}
\begin{figure}
\centering
{\large Fig. 5} \vspace{0.0cm}
\\
\epsfysize=8.5cm
\mbox{\epsffile{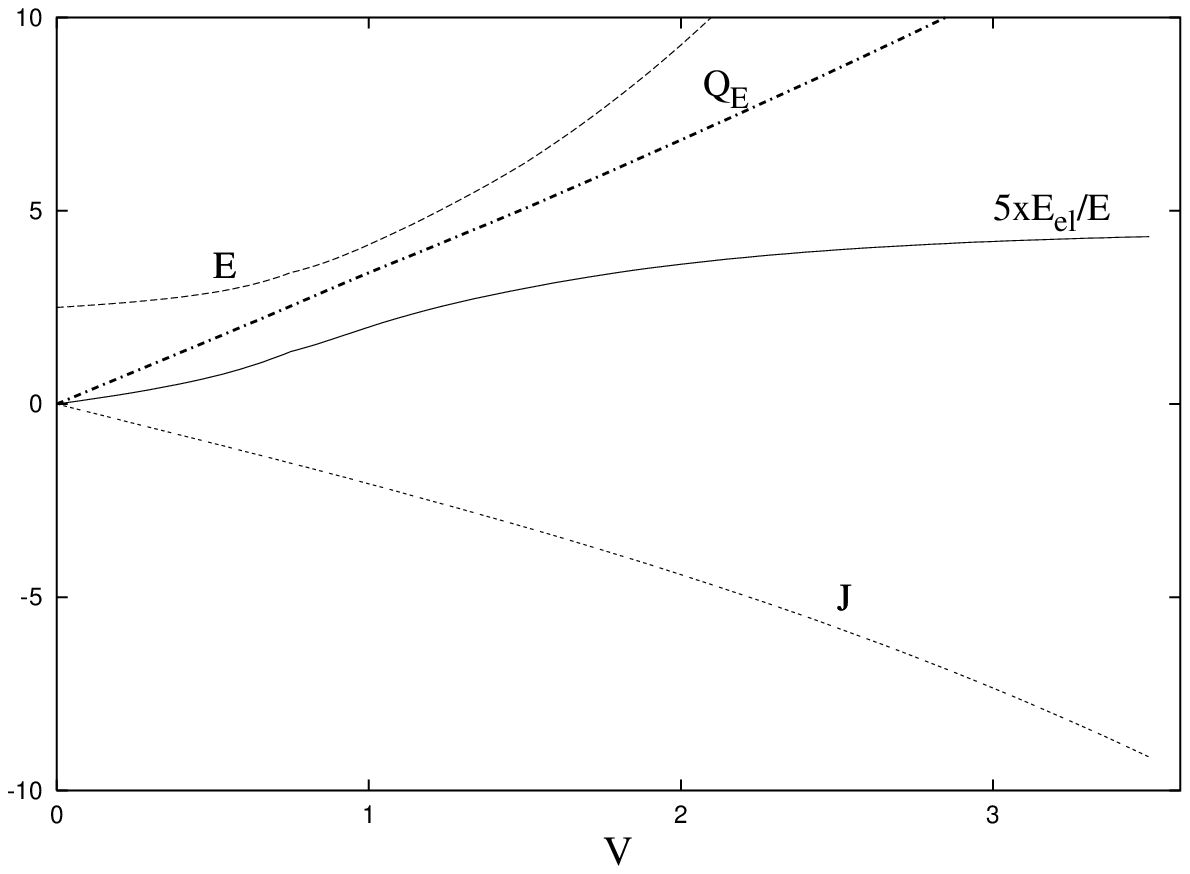}}
\caption
{
The energy $E$ and the angular momentum $J$ (in units $4\pi/e^2$)
for rotating YM dyon solutions in fixed SAdS black hole geometry
with $\Lambda=-3,~r_h=1$
are shown as a function on the parameter $V$  for a fixed $\omega_0=2.15$.
Also shown are  the electric charge $Q_E$ and the ratio
$E_{e}/E$.
} 
\end{figure}
\end{fixy}
\begin{fixy}{0}
\begin{figure}
\centering
{\large  Fig. 6a} \vspace{0.0cm}
\\
\epsfysize=8.5cm
\mbox{\epsffile{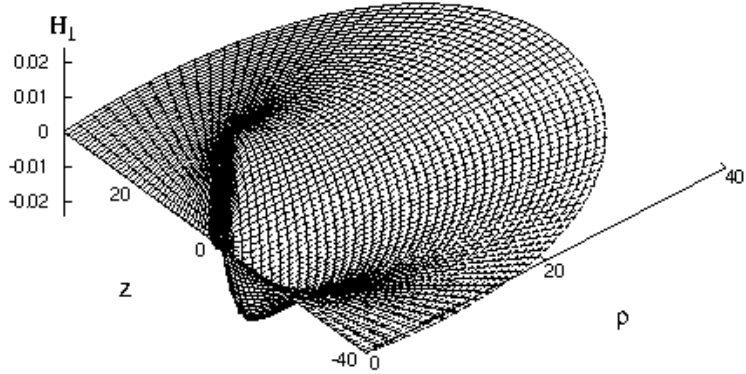}}
\caption
{
The magnetic gauge function $H_1$, 
is shown as a function of the 
coordinates $\rho=r \sin \theta,~z=\rho \cos \theta$ for a rotating dyon solution
with $\omega_0=2.15,~V=2,~n=1$.
The energy and the angular momentum of this configuration (in units $4\pi/e^2$)
are $E=9.29$ and $J=4.414$ respectively, while $Q_E=6.833$.
The results are obtained in a SAdS background with $\Lambda=-3,~r_h=1$.
}
\end{figure} 

\clearpage
\newpage

\begin{figure}
\centering
{\large Fig. 6b} \vspace{0.0cm}
\\
\epsfysize=9.5cm
\mbox{\epsffile{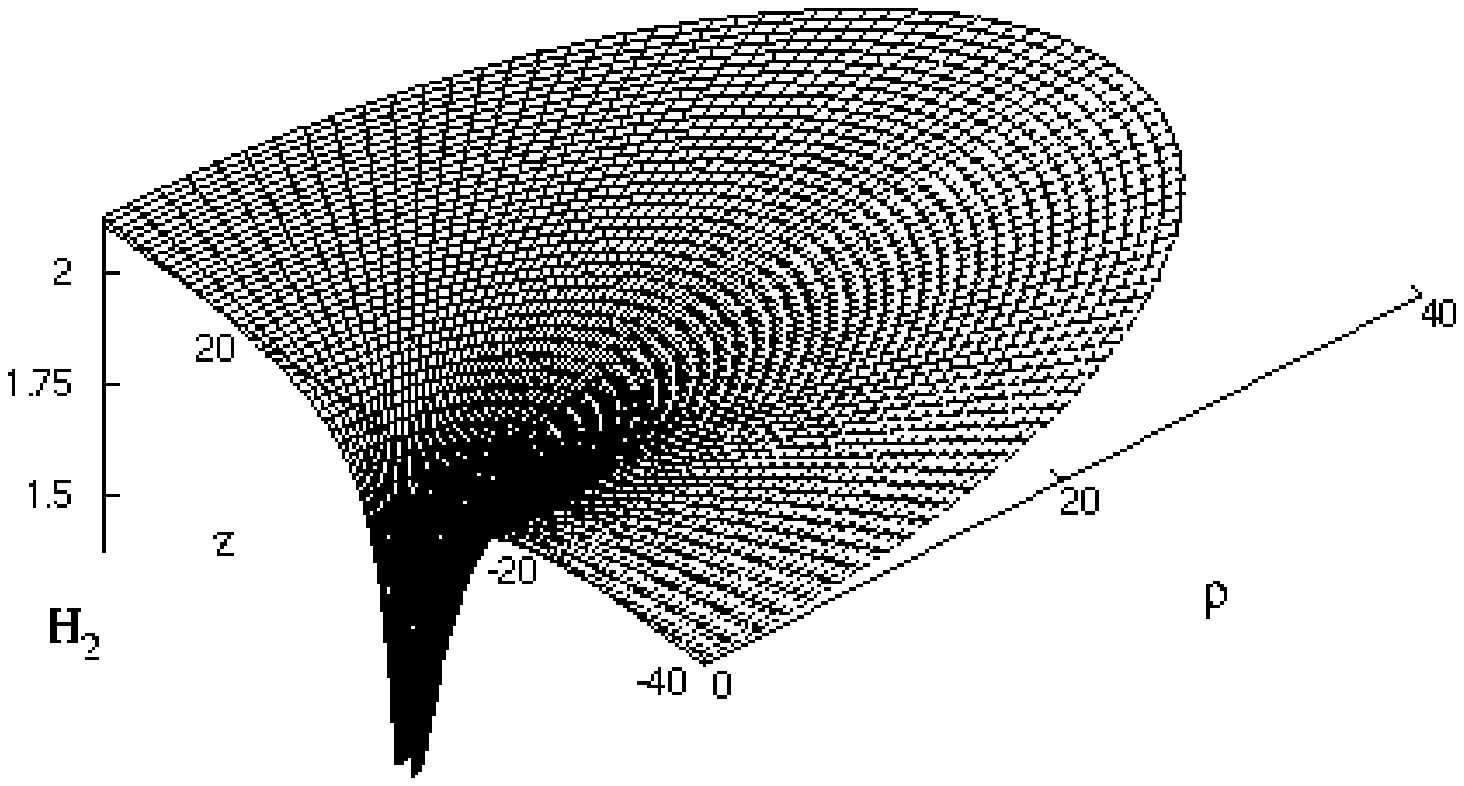}}
\caption
{
Same as Fig.~6a for the gauge function $H_2$.
}
\end{figure}

\begin{figure}
\centering
{\large Fig. 6c} \vspace{0.0cm}
\\
\epsfysize=9.5cm
\mbox{\epsffile{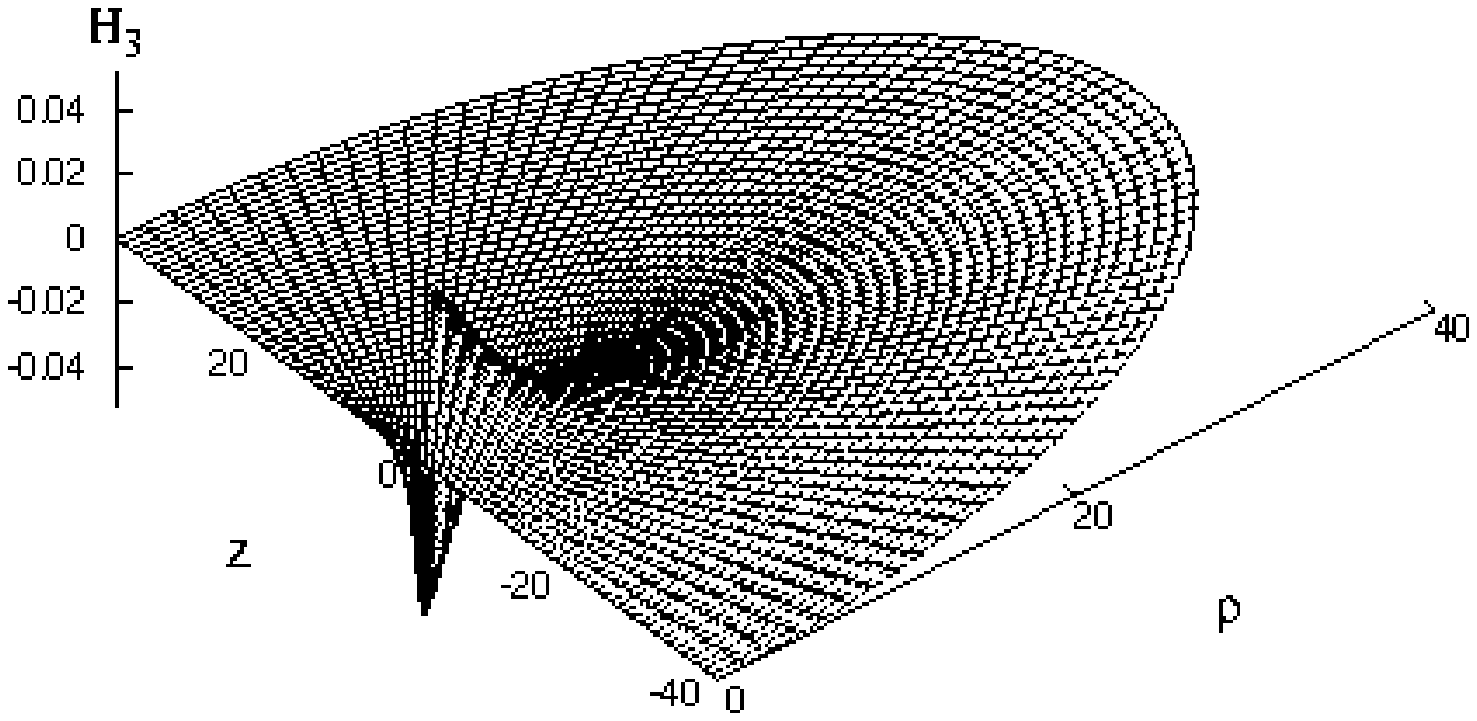}}
\caption
{
Same as Fig.~6a for the gauge function $H_3$.
} 
\end{figure}

\clearpage
\newpage
\begin{figure}
\centering
{\large Fig. 6d} \vspace{0.0cm}
\\
\epsfysize=9.5cm
\mbox{\epsffile{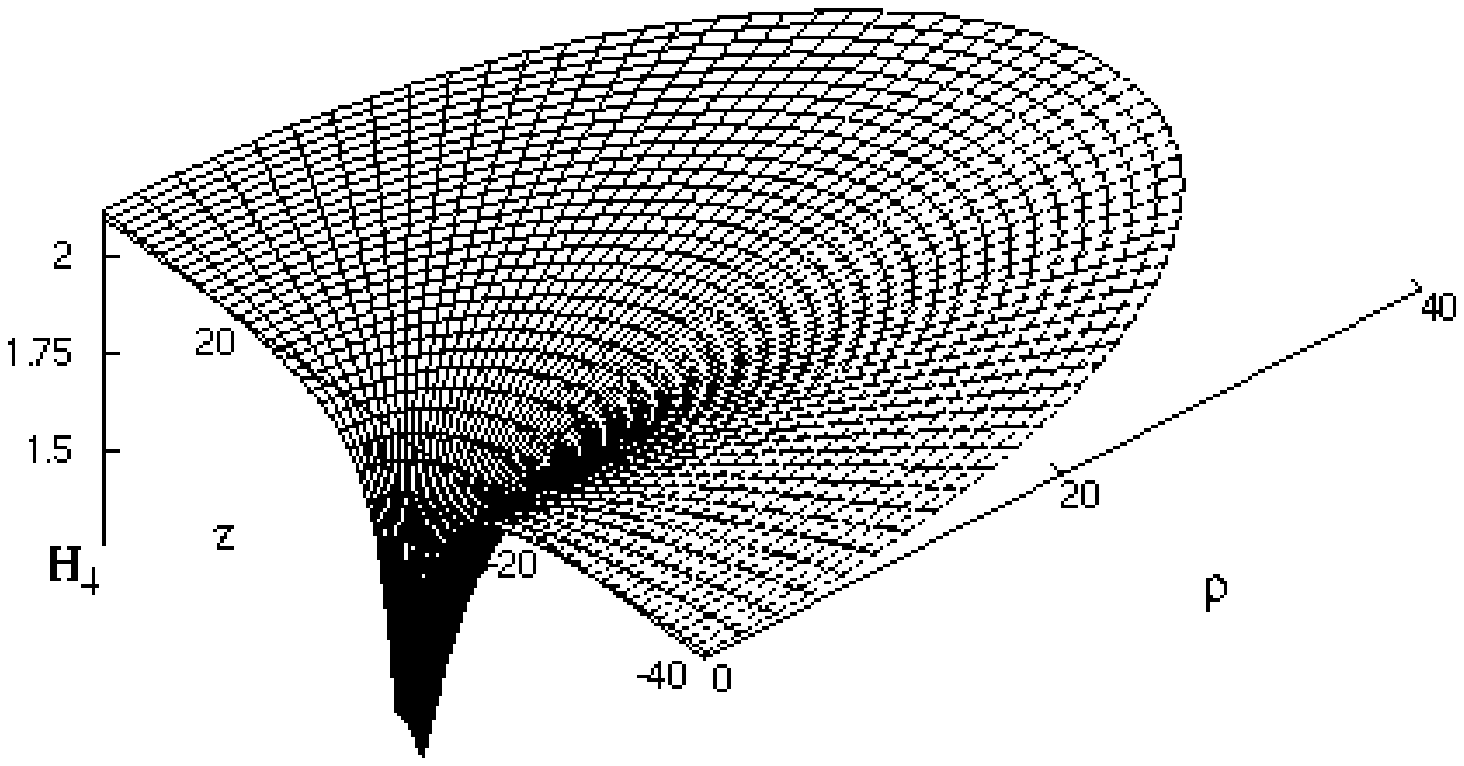}}
\caption
{
Same as Fig.~6a for the gauge function $H_4$.
} 
\end{figure}

\begin{figure}
\centering
{\large Fig. 6e} \vspace{0.0cm}
\\
\epsfysize=9.5cm
\mbox{\epsffile{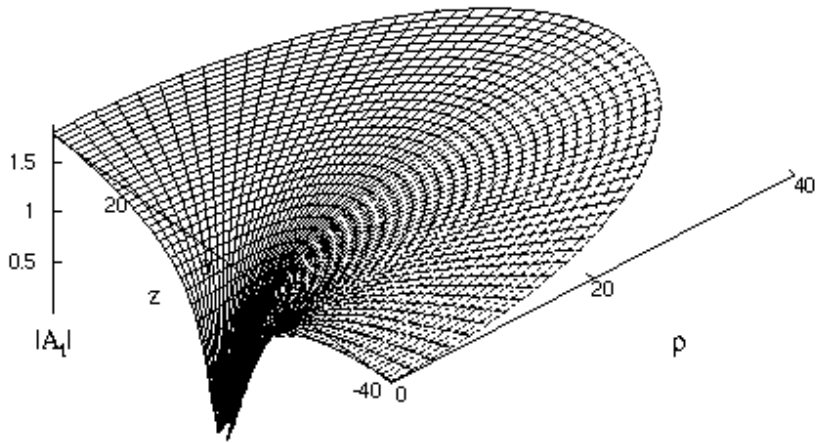}}
\caption
{
Same as Fig.~6a for the magnitude of the electric gauge functions
  $|A_t|$.
} 
\end{figure}
\clearpage
\newpage
\begin{figure}
\centering
{\large Fig. 6f} \vspace{0.0cm}
\\
\epsfysize=9.5cm
\mbox{\epsffile{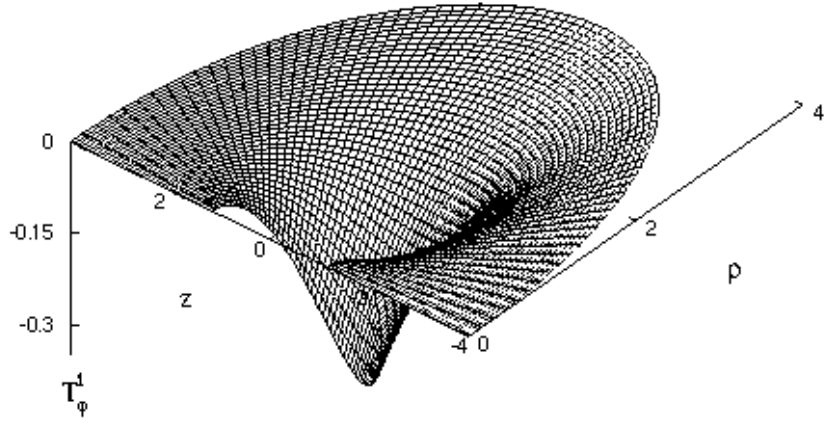}}
\caption
{
Same as Fig.~6a for the component $T_{\varphi}^t$
of the energy momentum tensor associated with rotation. 
}
\end{figure}

\begin{figure}
\centering
{\large Fig. 6g} \vspace{0.0cm}
\\
\epsfysize=9.5cm
\mbox{\epsffile{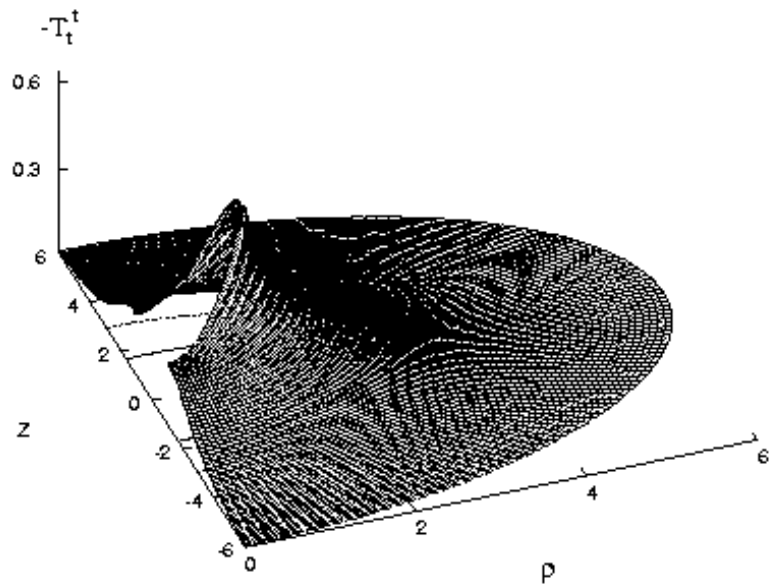}}
\caption
{
Same as Fig.~6a for the component $T_{t}^t$ 
of the energy momentum tensor. 
} 
\end{figure}
\clearpage
\newpage
\begin{figure}
\centering
{\large Fig. 6h} \vspace{0.0cm}
\\
\epsfysize=9.5cm
\mbox{\epsffile{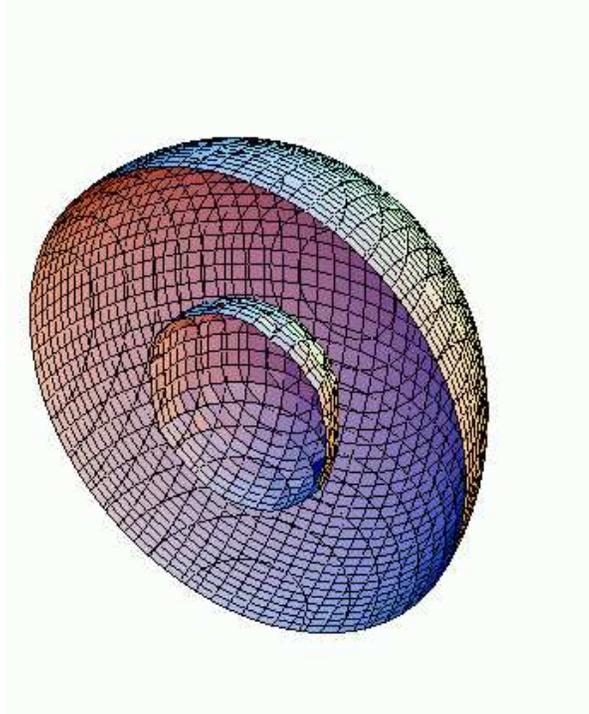}}
\caption
{
Surfaces of constant energy density $\epsilon=-T_t^t=0.16$ 
are plotted for the same solution.
} 
\end{figure}

\begin{figure}
\centering
{\large Fig. 6i} \vspace{0.0cm}
\\
\epsfysize=9.5cm
\mbox{\epsffile{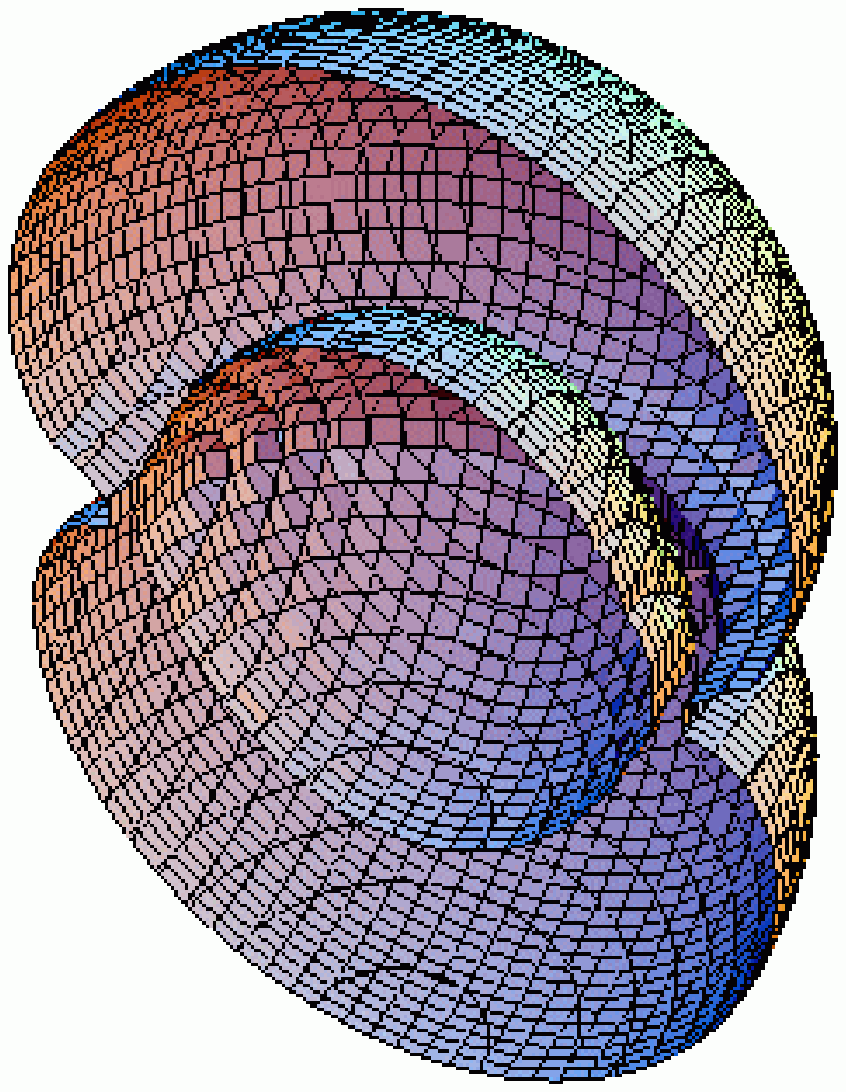}}
\caption
{
Same as Fig.~6h for the  $\epsilon=-T_t^t=0.22$. 
} 
\end{figure}

\clearpage
\newpage

\begin{figure}
\centering
{\large Fig. 6j} \vspace{0.0cm}
\\
\epsfysize=9.5cm
\mbox{\epsffile{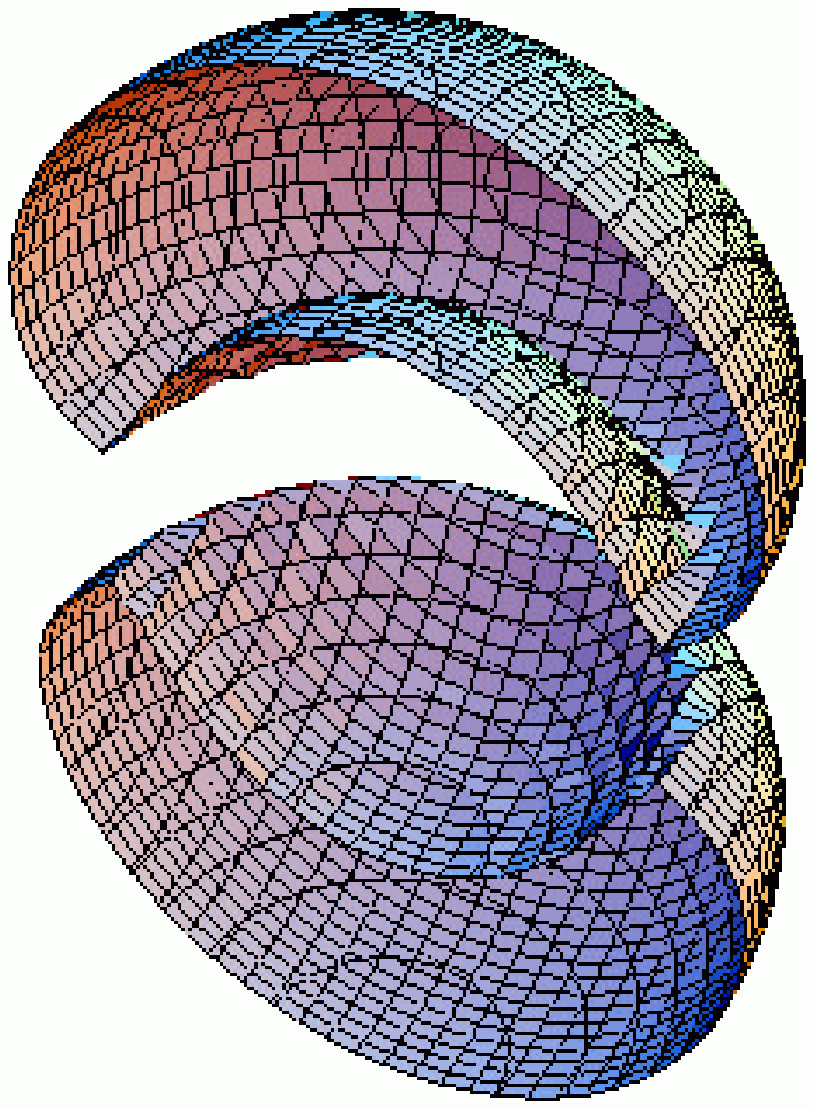}}
\caption
{
Same as Fig.~6a for the  $\epsilon=-T_t^t=0.24$. 
} 
\end{figure}
\end{fixy}

\begin{fixy}{0}
\begin{figure}
\centering
{\large Fig. 7a} \vspace{0.0cm}
\\
\epsfysize=8.5cm
\mbox{\epsffile{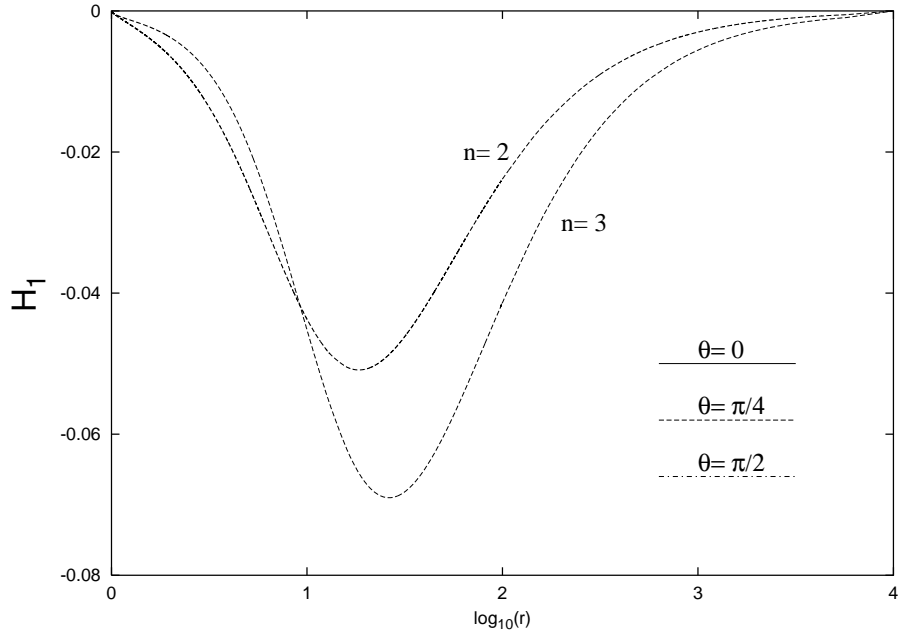}}
\caption
{
The gauge function $H_1$ 
is shown as a function of the radial coordinate
$r$ for the angles $\theta=0$, $\pi/4$ and $\pi/2$.
The parameters of these gravitating solutions are: 
$n=1$,~2, and 3, $k=0$, $r_h=1,~\Lambda=-0.1$, 
the node number $k=0$, $Q_M/n=-2.027$, 
$M(n=1)= 1.541$, $M(n=2)= 1.714$ and  $M(n=3)=2.071$.
}
\end{figure}

\clearpage
\newpage

\begin{figure}
\centering
{\large Fig. 7b} \vspace{0.0cm}
\\
\epsfysize=9.5cm
\mbox{\epsffile{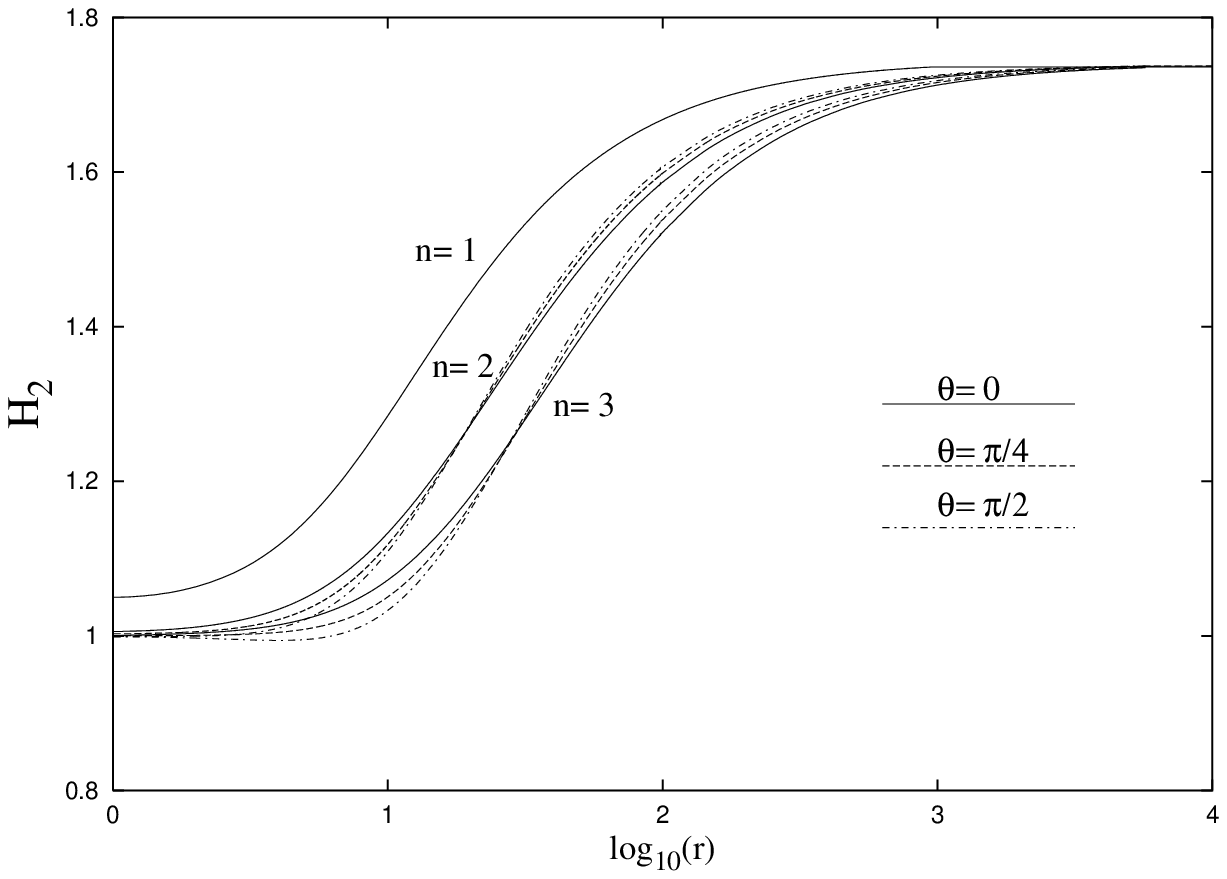}}
\caption
{
Same as Fig.~7a for the  gauge function $H_2$. 
} 
\end{figure}

\begin{figure}
\centering
{\large Fig. 7c} \vspace{0.0cm}
\\
\epsfysize=9.5cm
\mbox{\epsffile{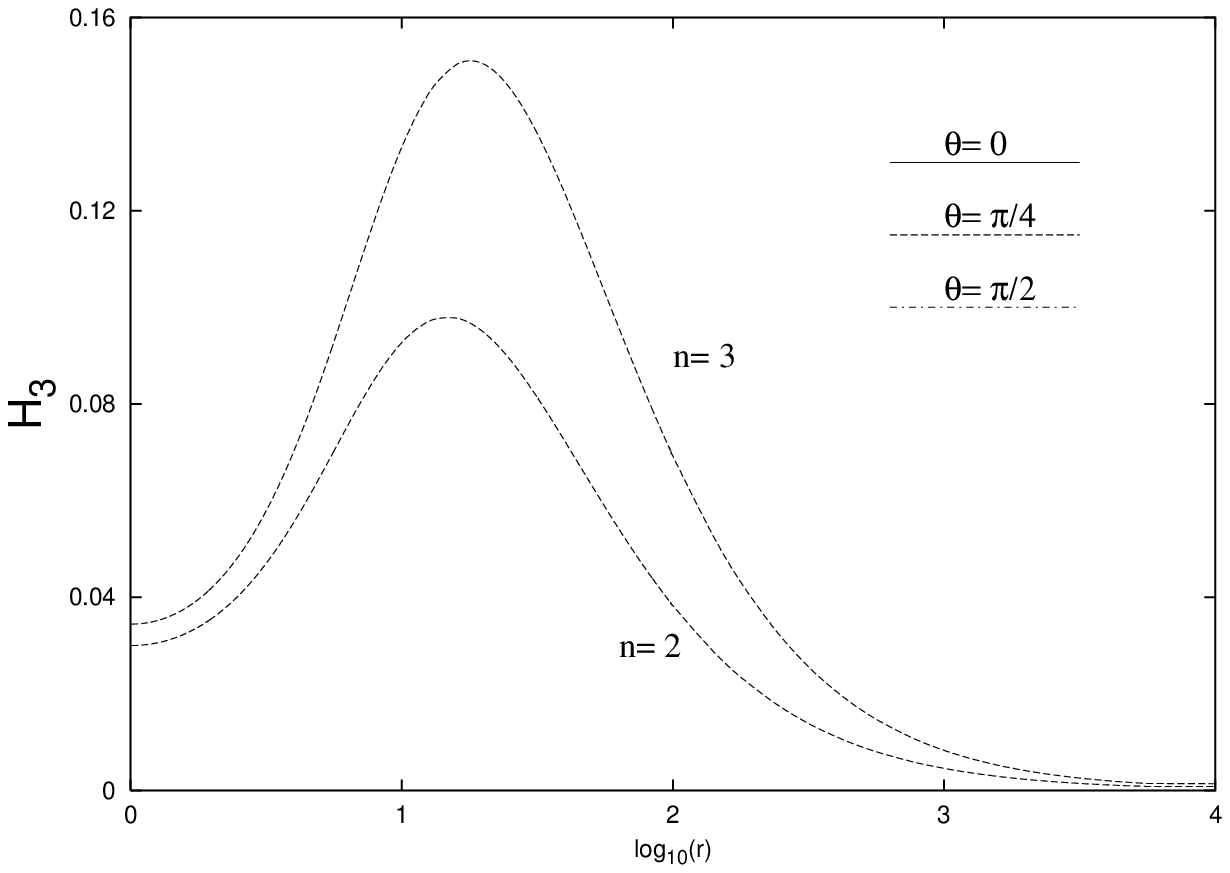}}
\caption
{
Same as Fig.~7a for the  gauge function $H_3$. 
}
\end{figure}

\clearpage
\newpage

\begin{figure}
\centering
{\large Fig. 7d} \vspace{0.0cm}
\\
\epsfysize=9.5cm
\mbox{\epsffile{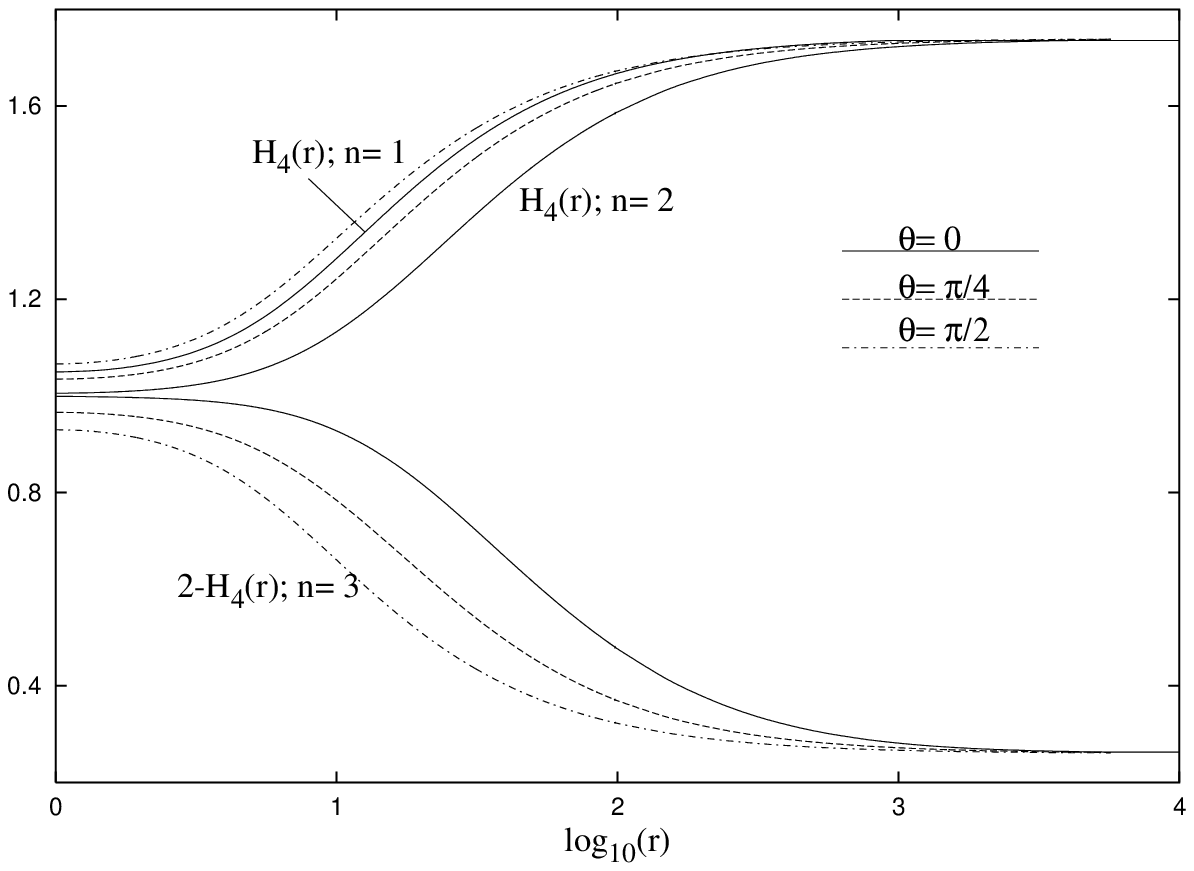}}
\caption
{
Same as Fig.~7a for the  gauge function $H_4$. 
} 
\end{figure}

\begin{figure}
\centering
{\large Fig. 7e} \vspace{0.0cm}
\\
\epsfysize=9.5cm
\mbox{\epsffile{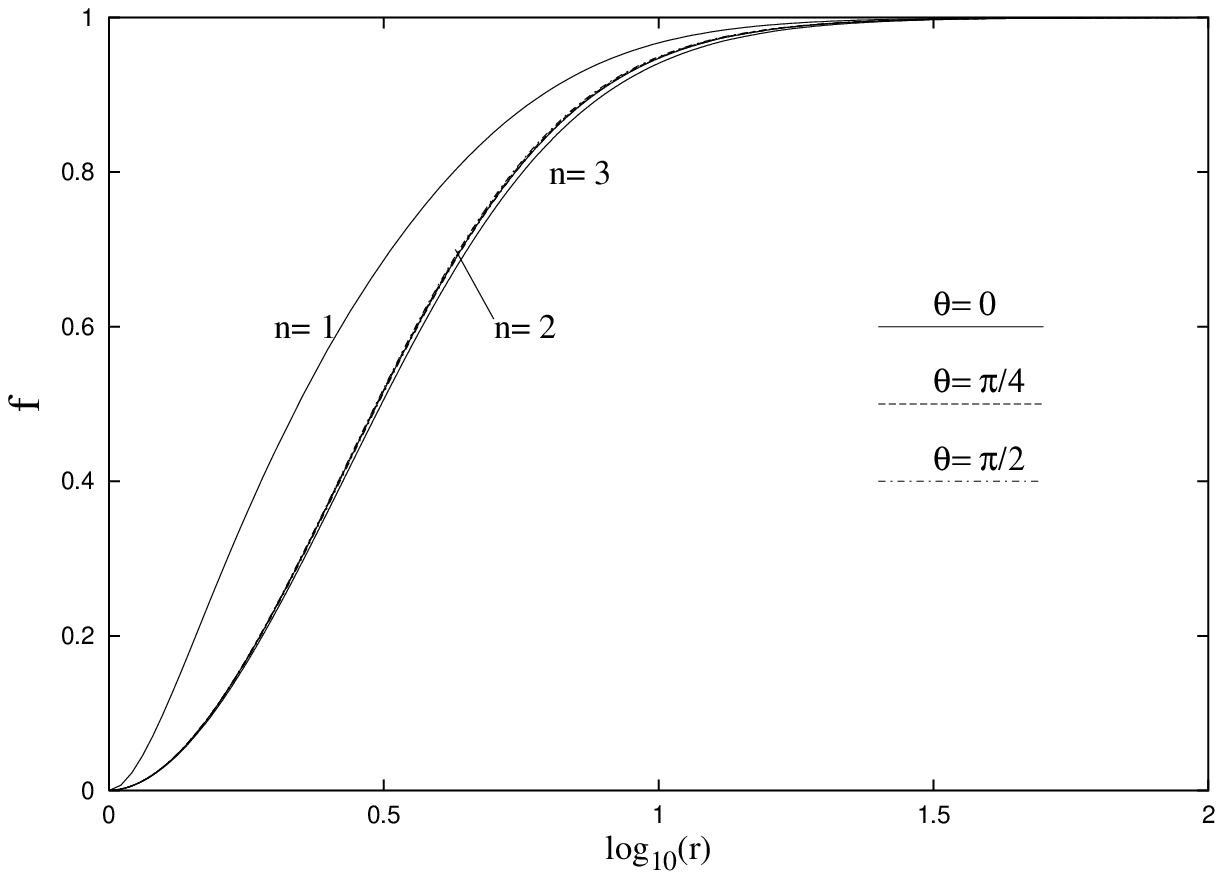}}
\caption
{
Same as Fig.~7a for the  metric function $f$. 
} 
\end{figure}

\clearpage
\newpage

\begin{figure}
\centering
{\large Fig. 7f} \vspace{0.0cm}
\\
\epsfysize=9.5cm
\mbox{\epsffile{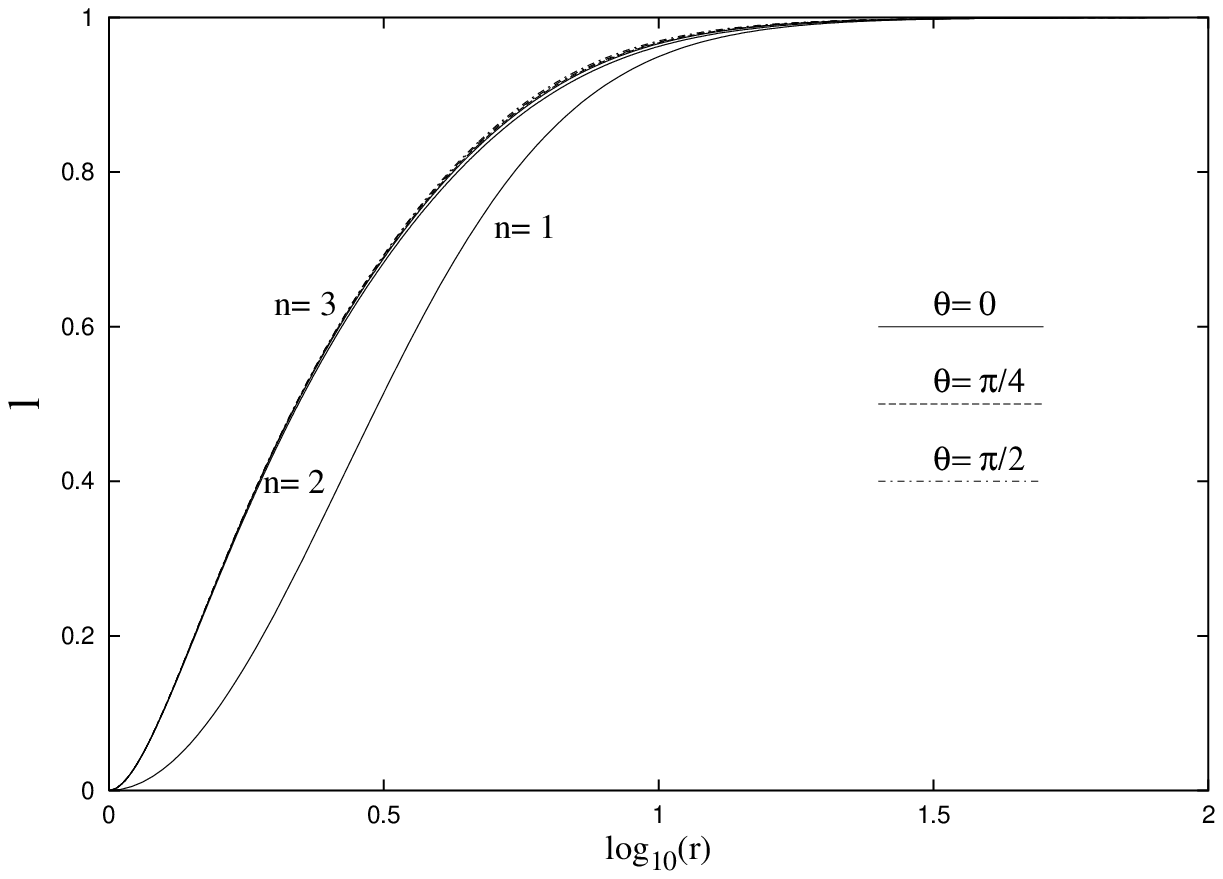}}
\caption
{
Same as Fig.~7a for the metric function $l$.  
} 
\end{figure}

\begin{figure}
\centering
{\large Fig. 7e} \vspace{0.0cm}
\\
\epsfysize=9.5cm
\mbox{\epsffile{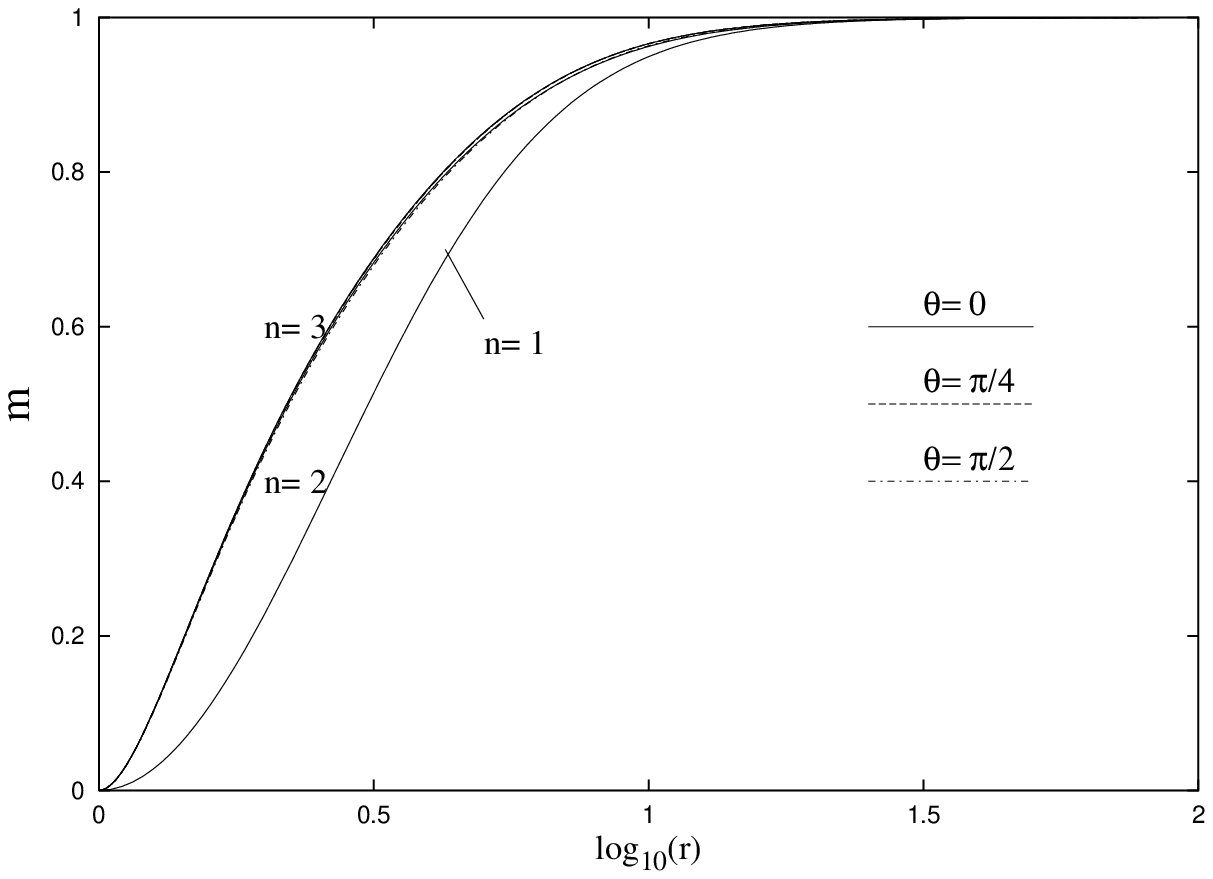}}
\caption
{
Same as Fig.~7a for the  metric function $m$. 
} 
\end{figure}

\clearpage
\newpage

\begin{figure}
\centering
{\large Fig. 7f} \vspace{0.0cm}
\\
\epsfysize=9.5cm
\mbox{\epsffile{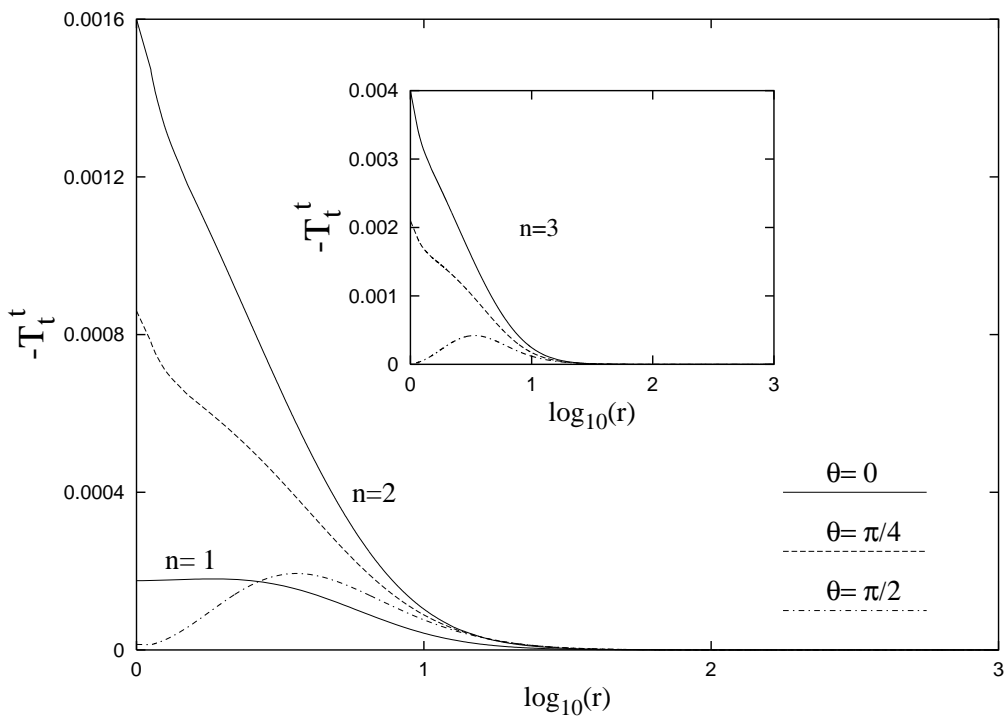}}
\caption
{
Same as Fig.~7a for the energy density $\epsilon=-T_t^t$
of the solutions.  
} 
\end{figure}

\end{fixy}

 \begin{fixy}{-1}
\begin{figure}
\centering
{\large Fig. 8} \vspace{0.0cm}
\\
\epsfysize=9.cm
\mbox{\epsffile{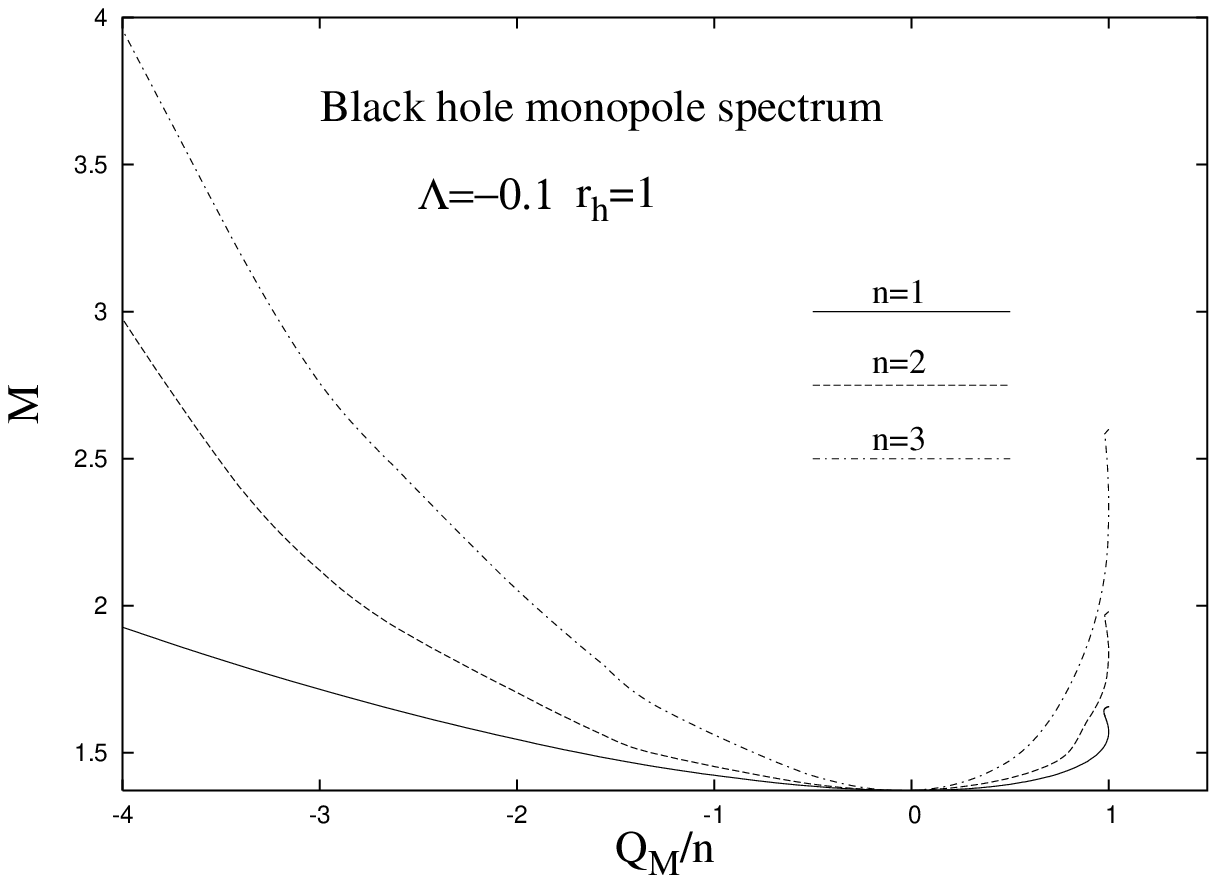}}
\caption
{
The total mass $M$ is plotted as a function of magnetic charge $Q_M$
for black hole gravitating monopole solutions at $\Lambda=-0.1,~r_h=1$.
The winding number $n$ is also  marked.
} 
\end{figure}
\end{fixy}

\clearpage
\newpage

\begin{fixy}{0}

\begin{figure}
\centering
{\large Fig. 9a} \vspace{0.0cm}
\\
\epsfysize=9.cm
\mbox{\epsffile{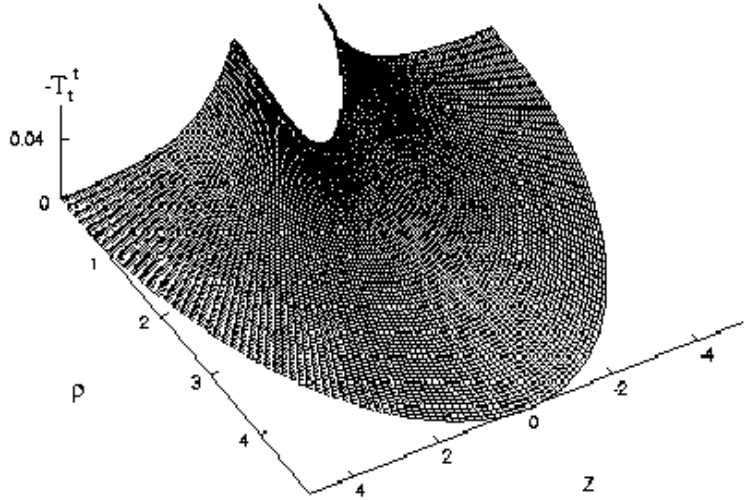}}
\caption
{ The energy density $\epsilon=-T_t^t$ 
is shown as for a gravitating monopole solution
with $\omega_0=0.49,~n=2$. 
Here the cosmological constant is 
$\Lambda=-1$ and the black hole radius is $r_h=1$.
}
\end{figure}

\begin{figure}
\centering
{\large Fig. 9b} \vspace{0.0cm}
\\
\epsfysize=9.cm
\mbox{\epsffile{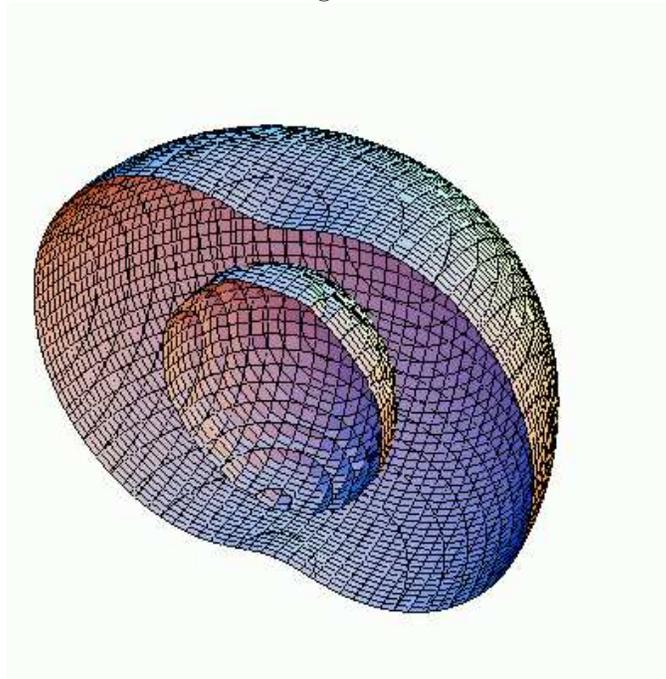}}
\caption
{ A surface  of constant energy density $\epsilon=-T_t^t=0.014$  
are plotted for the solution of Fig. 9a.
} 
\end{figure}

\clearpage
\newpage

\begin{figure}
\centering
{\large Fig. 9c} \vspace{0.0cm}
\\
\epsfysize=9.5cm
\mbox{\epsffile{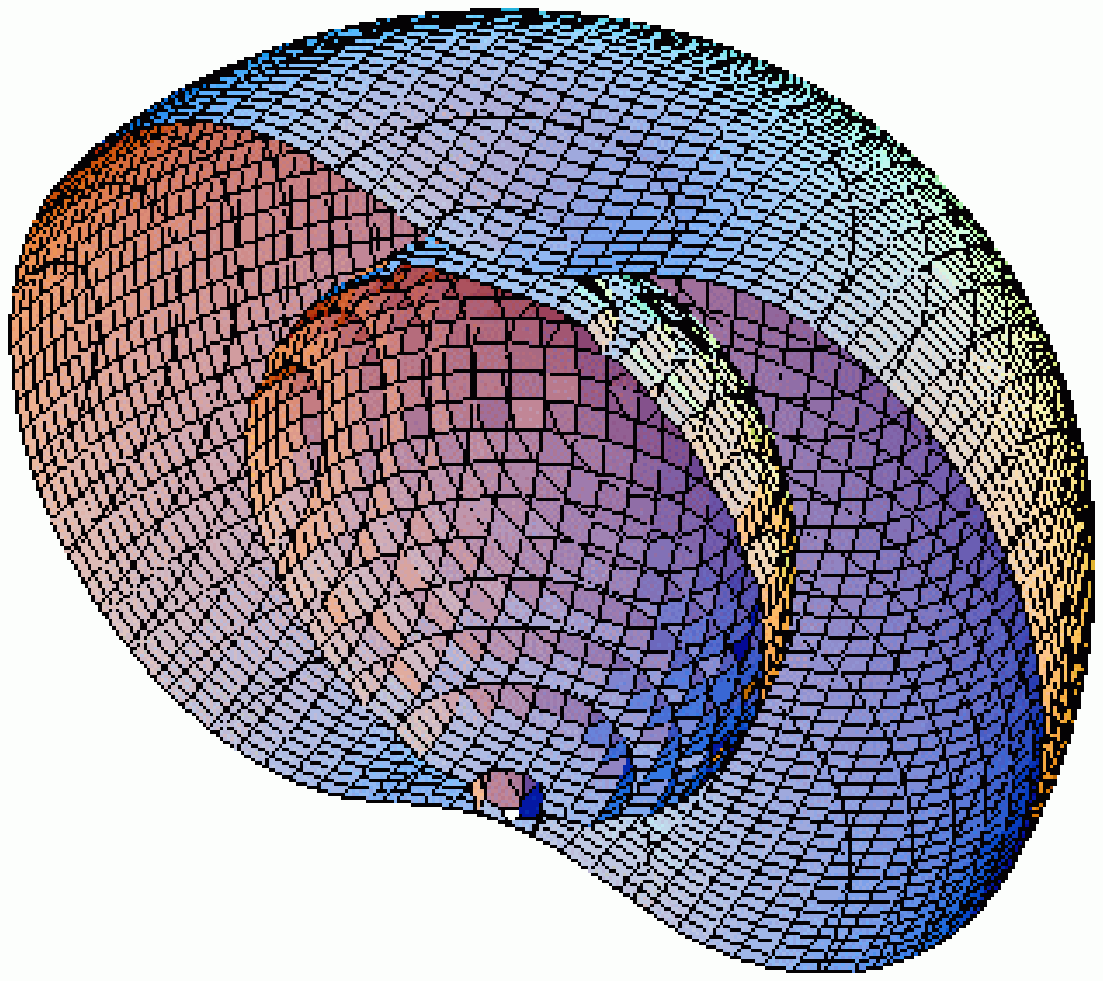}}
\caption
{Same as Fig. 9b for $\epsilon= 0.019$  
} 
\end{figure}
\begin{figure}
\centering
{\large Fig. 9d} \vspace{0.0cm}
\\
\epsfysize=9.5cm
\mbox{\epsffile{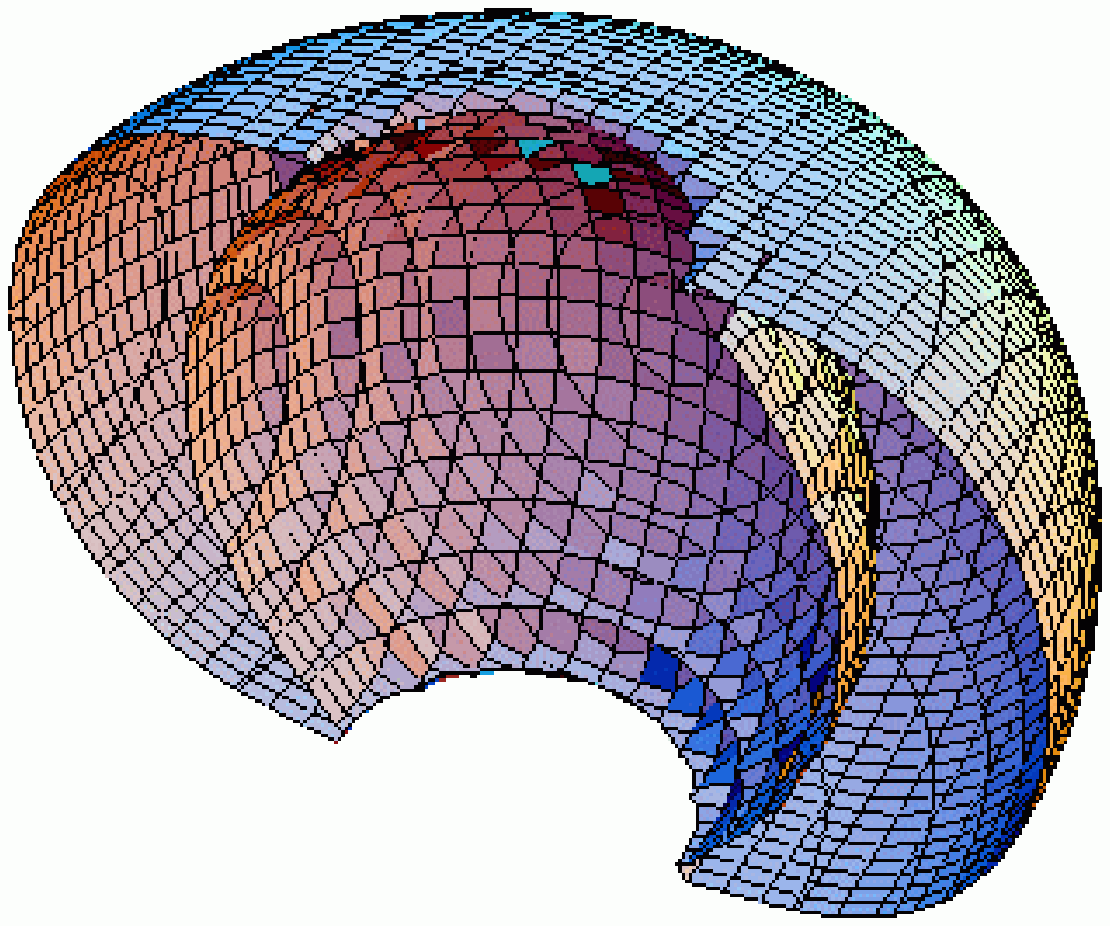}}
\caption
{Same as Fig. 9b for $\epsilon= 0.03 $ 
} 
\end{figure}

\end{fixy}

\clearpage
\newpage

\begin{fixy}{0}

\begin{figure}
\centering
{\large Fig. 10a} \vspace{0.0cm}
\\
\epsfysize=9.cm
\mbox{\epsffile{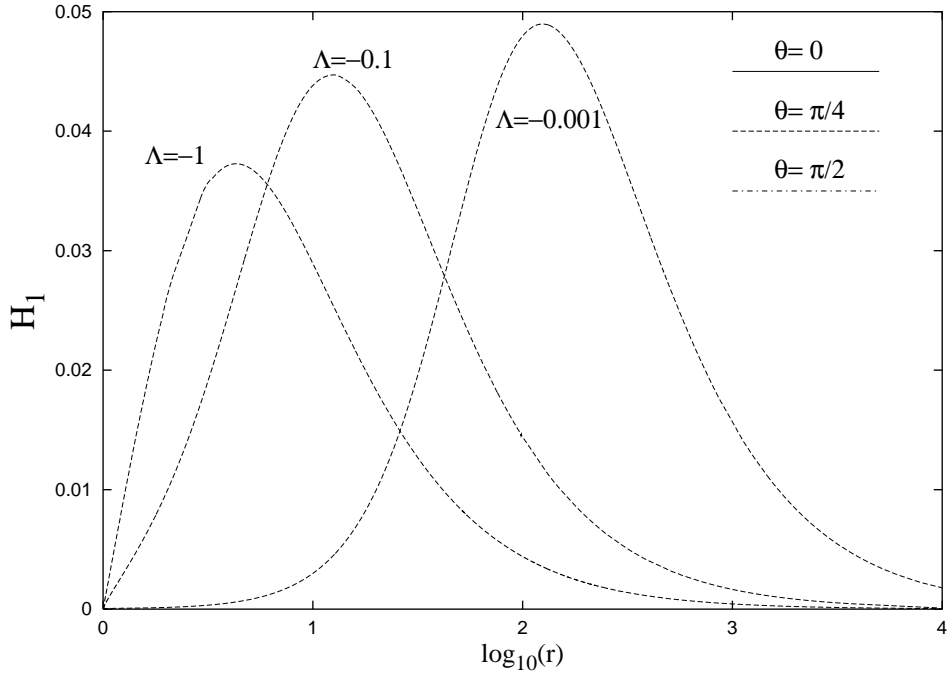}}
\caption
{ The gauge function $H_i$
is shown as a function of 
the radial coordinate
$r$ for the angles $\theta=0$, $\pi/4$ and $\pi/2$.
The results correspond to
EYM black hole solutions with winding number $n=2$,
magnetic charge $Q_M=1.52$, horizon radius $r_h=1$ and cosmological constants
$\Lambda=-1,~-0.1$ and $-0.001$.} 
\end{figure}

\begin{figure}
\centering
{\large Fig. 10b} \vspace{0.0cm}
\\
\epsfysize=9.cm
\mbox{\epsffile{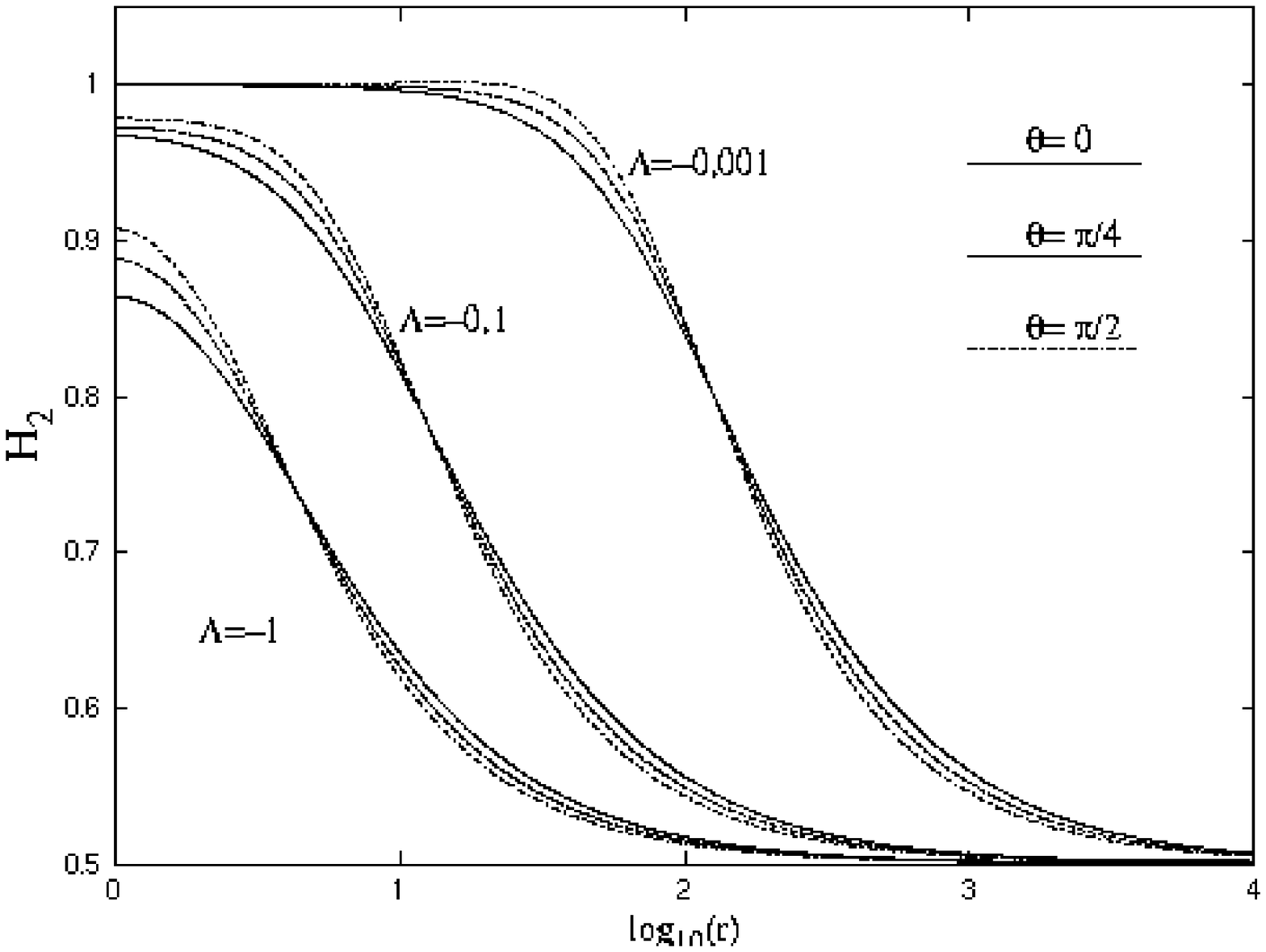}}
\caption
{ Same as Fig.~10a for the  gauge function $H_2$. } 
\end{figure}

\clearpage
\newpage

\begin{figure}
\centering
{\large Fig. 10c} \vspace{0.0cm}
\\
\epsfysize=9.5cm
\mbox{\epsffile{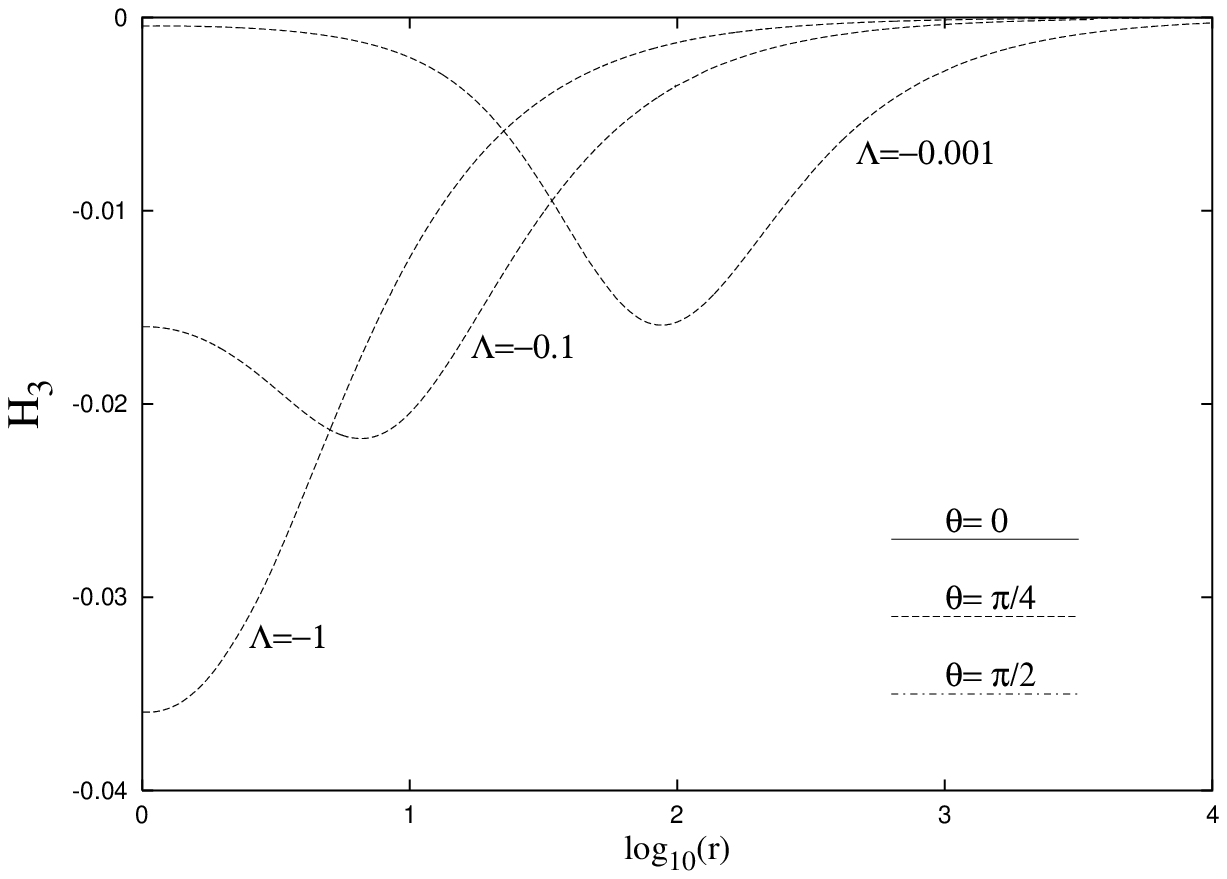}}
\caption
{ Same as Fig.~10a for the  gauge function $H_3$. } 
\end{figure}
\begin{figure}
\centering
{\large Fig. 10d} \vspace{0.0cm}
\\
\epsfysize=9.5cm
\mbox{\epsffile{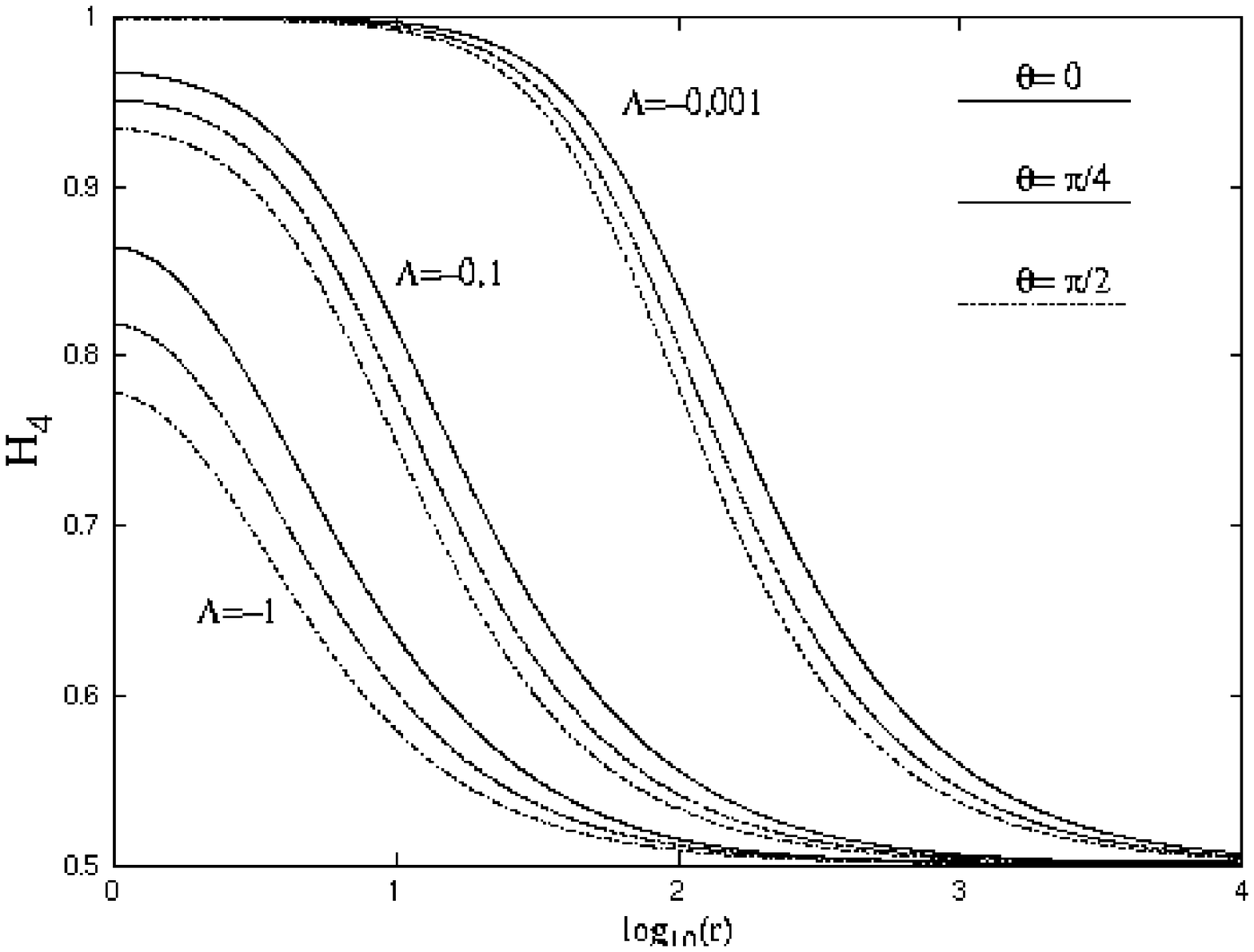}}
\caption
{ Same as Fig.~10a for the  gauge function $H_4$. } 
\end{figure}

\clearpage
\newpage

\begin{figure}
\centering
{\large Fig. 10e} \vspace{0.0cm}
\\
\epsfysize=9.5cm
\mbox{\epsffile{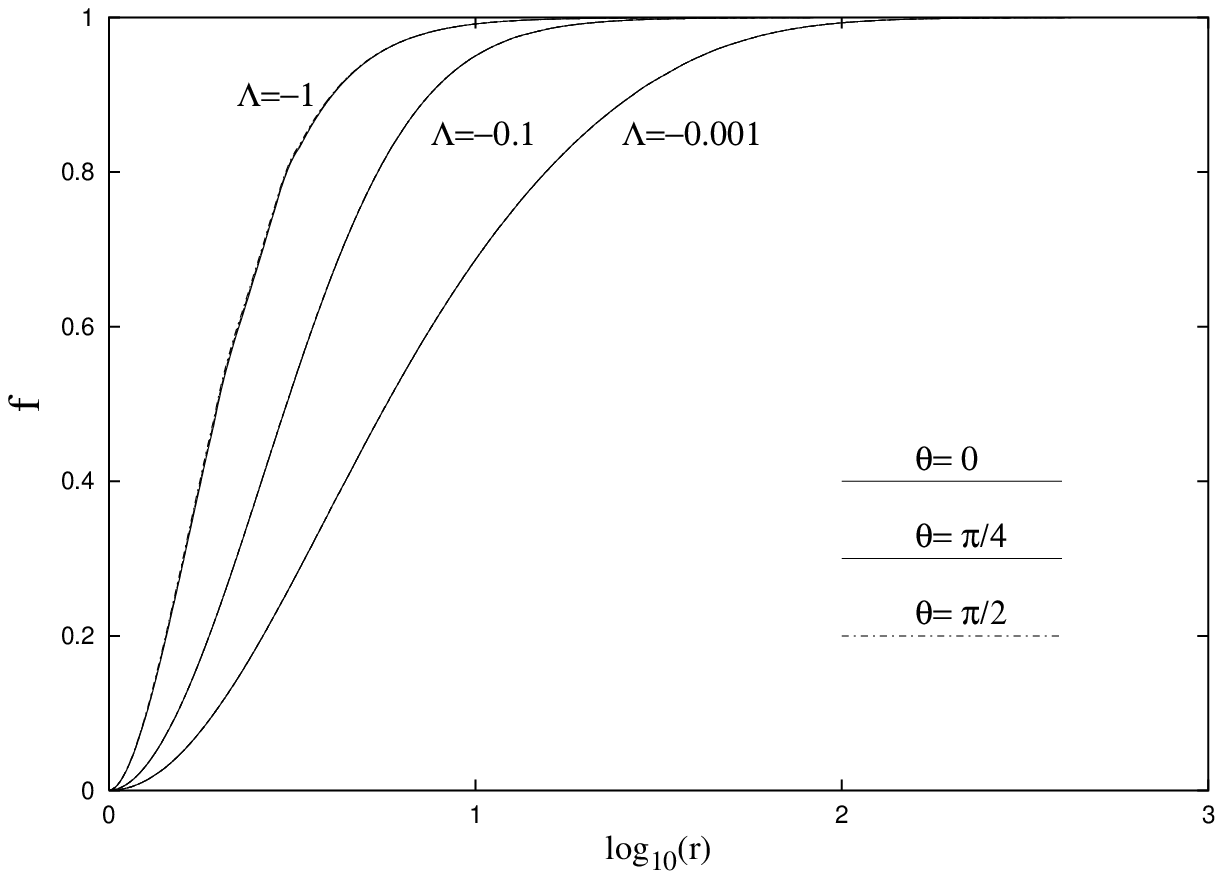}}
\caption
{ Same as Fig.~10a for the  metric function $f$. } 
\end{figure}

\begin{figure}
\centering
{\large Fig. 10f} \vspace{0.0cm}
\\
\epsfysize=9.5cm
\mbox{\epsffile{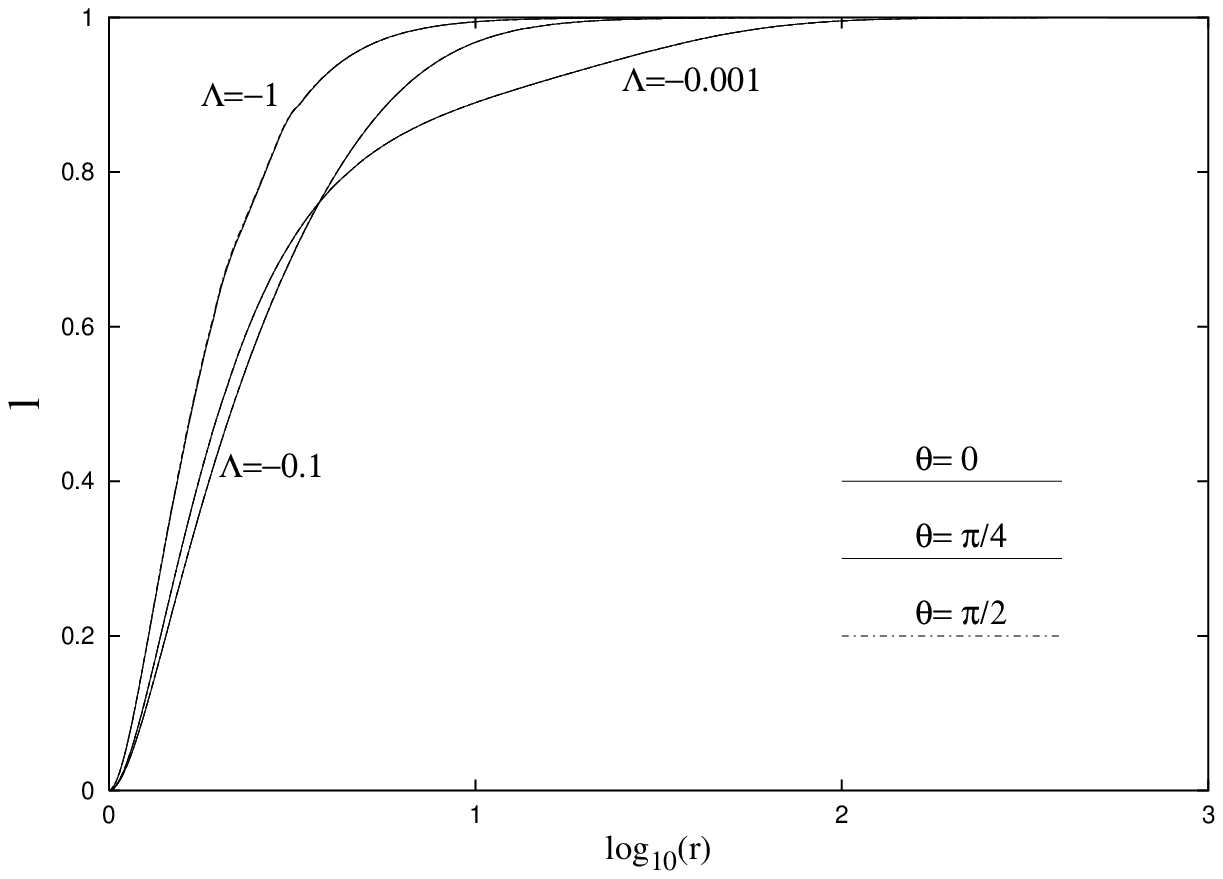}}
\caption
{ Same as Fig.~10a for   the  metric function $l$. } 
\end{figure}

\clearpage
\newpage

\begin{figure}
\centering
{\large Fig. 10g} \vspace{0.0cm}
\\
\epsfysize=9.5cm
\mbox{\epsffile{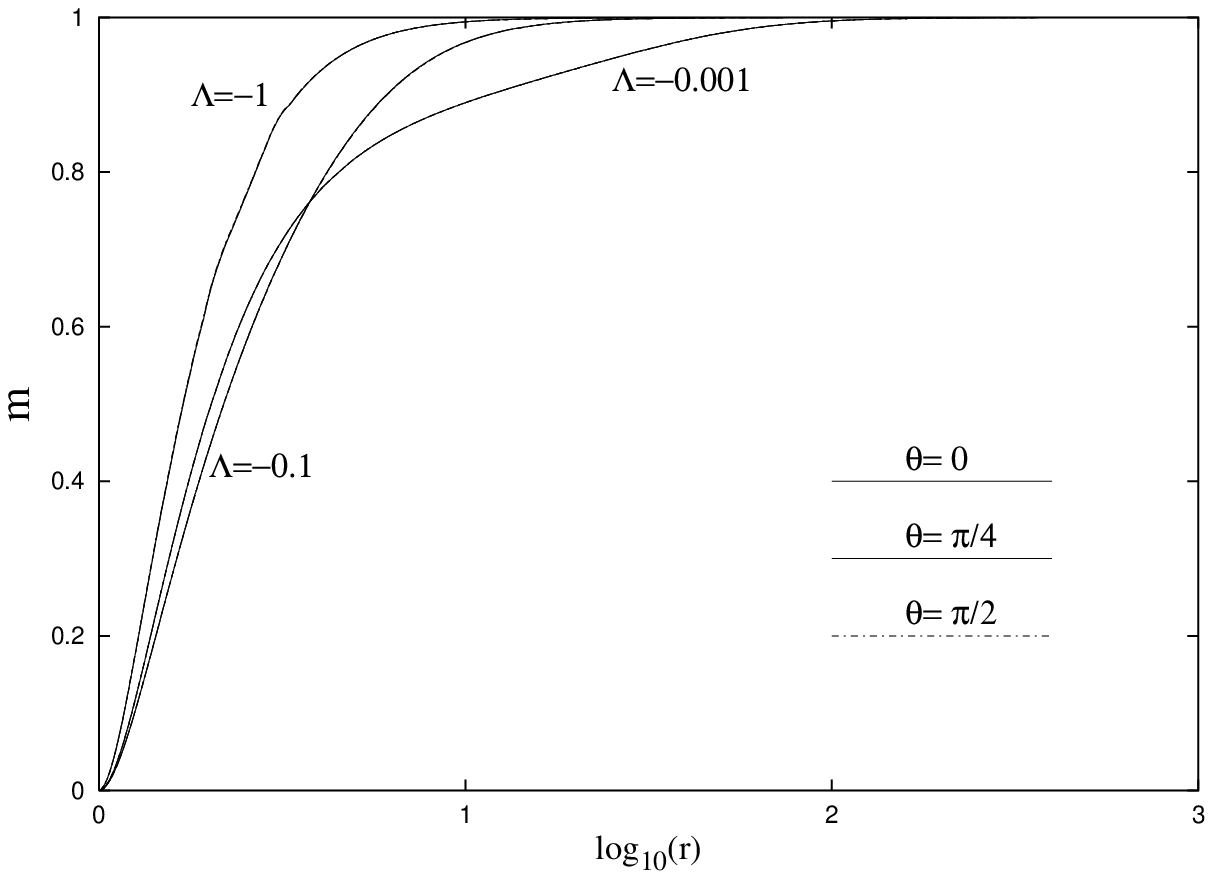}}
\caption
{ Same as Fig.~10a for the  metric function $m$. } 
\end{figure}

\begin{figure}
\centering
{\large Fig. 10h} \vspace{0.0cm}
\\
\epsfysize=9.5cm
\mbox{\epsffile{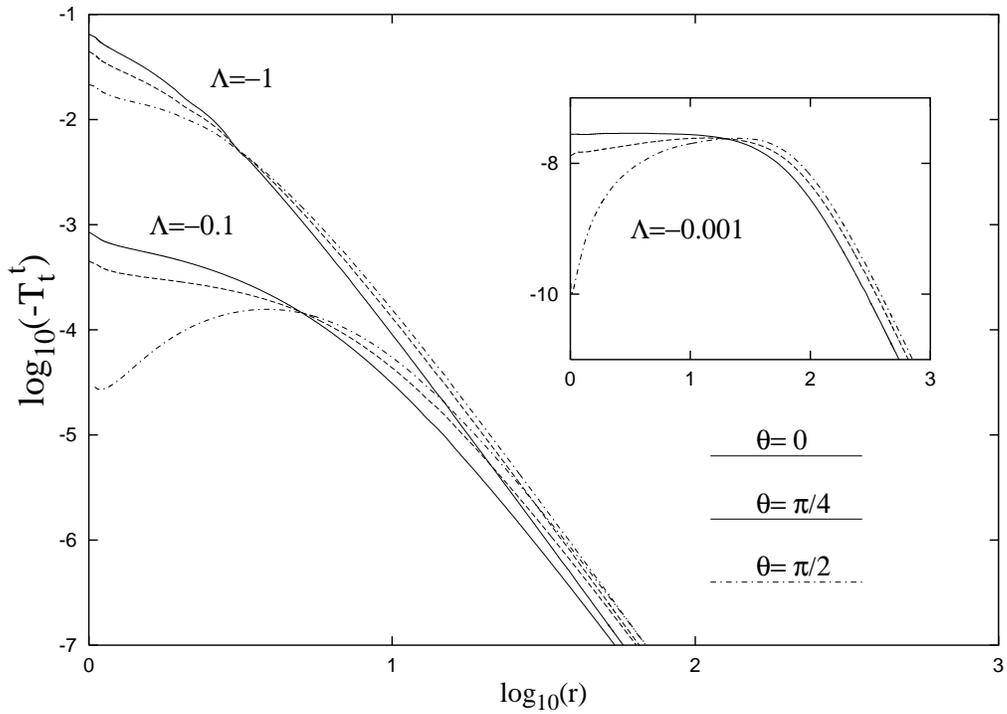}}
\caption
{ Same as Fig.~10a for the energy density $\epsilon=-T_t^t$. }
\end{figure}

\end{fixy}


\clearpage
\newpage

\begin{fixy}{-1}

\begin{figure}
\centering
{\large Fig. 11} \vspace{0.0cm}
\\
\epsfysize=9.cm
\mbox{\epsffile{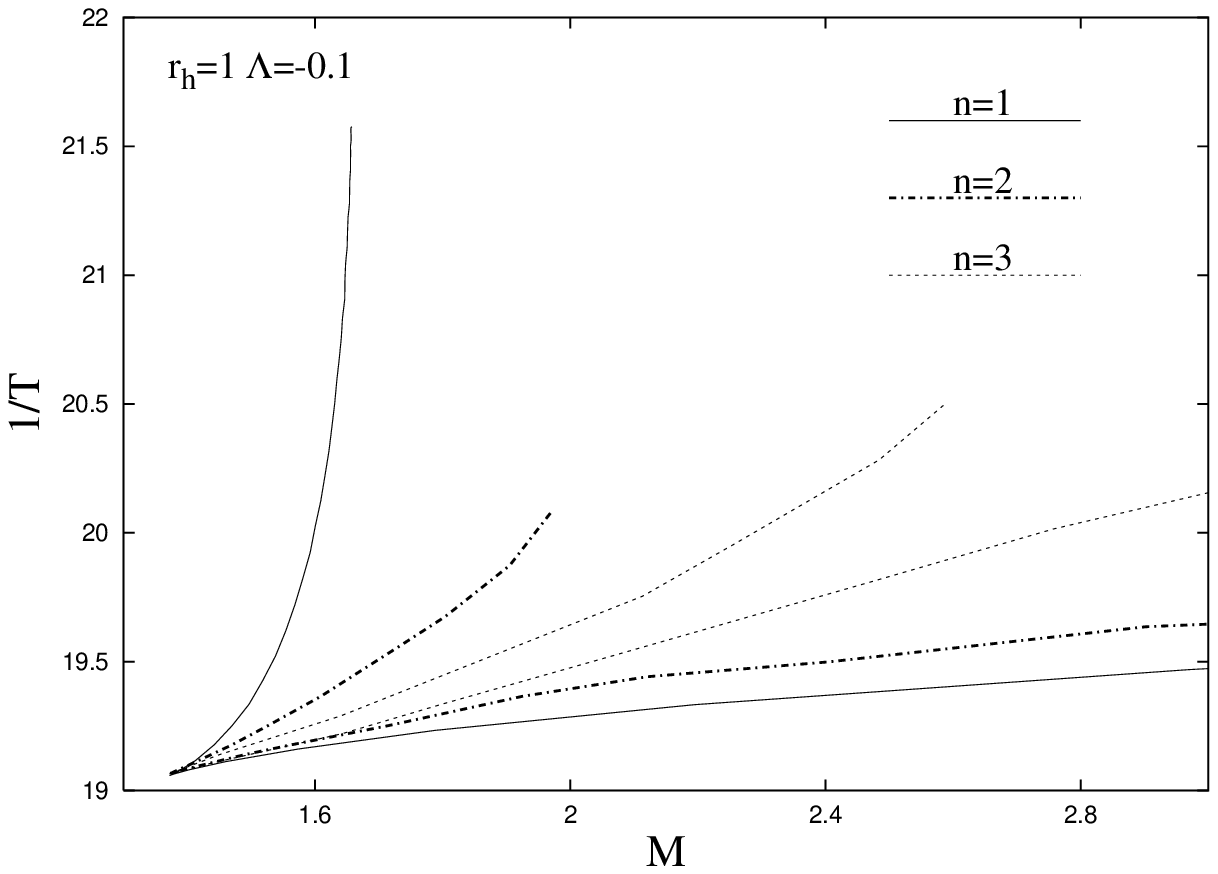}}
\caption
{ The mass-temperature diagrams for black hole monopole solutions at 
$\Lambda=-0.1,~r_h=1$ and three winding numbers.
The value of the magnetic potential at infinity is the control  parameter which 
varies along each curve.} 
\end{figure}

\end{fixy}

\end{document}